%% file: PIP_22b_PlanckXMM_final.tex
\documentclass[a4paper,traditabstract,longauth]{aa}  
\usepackage{graphicx}
\usepackage{txfonts}
\usepackage[breaklinks, colorlinks, citecolor=blue]{hyperref}
\usepackage{url}
\usepackage{natbib}
\usepackage{color}
\usepackage{amssymb}
\usepackage{booktabs}
\usepackage{subfig}

\usepackage{sidecap}

\input{Planck.tex}

\def\xmm{{\it XMM-Newton}}

\def\planck{\Planck}
\def\rosat{{\it ROSAT}}
\def\rass{{\rm RASS}}
\newfont{\gwpfont}{cmssq8 scaled 1000}
\newcommand{\rexcess}{{\gwpfont REXCESS}}

\newcommand{\reflex}{{REFLEX}}
\newcommand{\noras}{{NORAS}}
\newcommand{\bcs}{{BCS}}
\newcommand{\ebcs}{{eBCS}}
\newcommand{\macs}{{MACS}}
\def \sas {\hbox{SAS}}

\def\Mv{M_{500}}
\def\Rv{R_{500}}
\def\Mgv{M_{\rm g,500}}

\def\YX {Y_{\rm X}}
\def\TX {T_{\rm X}}
\def\YSZ {Y_{\rm SZ}}
\def\YSZ {Y_{500}}

\def\kT {kT}
\def\Mv {M_{\rm 500}}
\def\Rv {R_{500}}
\def\keV {\rm keV}
\def\Yv {Y_{500}}
\def\LX {L_{500,[0.1-2.4]}}
\def\msol {{\rm M_{\odot}}}

\def\MYX {$M_{500}$--$Y_{\rm X}$}

\def\Lxz{$L_{\rm X}$--$z$}

\def\YSZYX {$\YSZ$--$\YX$}

\def\ergscm{\rm erg\,s^{-1}\,cm^{-2}}
\def\arcms{\rm arcmin^2}

\def\lesssim{\mathrel{\hbox{\rlap{\hbox{\lower4pt\hbox{$\sim$}}}\hbox{$<$}}}}
\def\gtrsim{\mathrel{\hbox{\rlap{\hbox{\lower4pt\hbox{$\sim$}}}\hbox{$>$}}}}

\newcommand{\propsim}{\lower 3pt \hbox{$\, \buildrel {\textstyle
       \propto}\over {\textstyle \sim}\,$}}

\begin{document} 
\input{PIP_22b_authors_and_institutes.tex}

\title{ \textit{Planck} Intermediate  Results. IV. The \textit{XMM-Newton} validation  programme for new \textit{Planck} galaxy clusters}

\authorrunning{Planck Collaboration}

\date{Received May 2 / Accepted July 29}

\abstract{
We present the final results from the \xmm\ validation follow-up of new \planck\ galaxy cluster candidates.  We observed 15 new candidates, detected with signal-to-noise ratios between 4.0 and 6.1 in the 15.5-month nominal \planck\ survey.  The candidates were selected using ancillary data flags derived from the {\it ROSAT\/} All Sky Survey (\rass) and Digitized Sky Survey all-sky maps, with the aim of pushing into the low SZ flux, high-$z$ regime and testing \rass\ flags as indicators of candidate reliability.  Fourteen new clusters were detected by \xmm, ten single clusters and two double systems.  Redshifts from X-ray spectroscopy lie in the range 0.2 to 0.9, with six clusters at $z>0.5$.  Estimated masses ($\Mv$) range from $2.5 \times 10^{14}$ to $8 \times 10^{14}\,\msol$.  We discuss our results in the context of the full \xmm\ validation programme, in which 51 new clusters have been detected.  This includes four double and two triple systems, some of which are chance projections on the sky of clusters at different redshifts.  We find that  association with a source from the \rass-Bright Source Catalogue is a robust indicator of the reliability of a candidate, whereas association with a source from the \rass-Faint Source Catalogue does not guarantee  that the SZ candidate is a {\it bona fide\/} cluster.   Nevertheless, most \planck\ clusters appear in \rass\ maps, with a significance greater than $2\sigma$ being a good indication that the candidate is a real cluster. Candidate validation from association with SDSS galaxy overdensity at  $z > 0.5$ is also discussed.  The full sample gives a \planck\ sensitivity threshold of $\Yv \sim 4 \times 10^{-4}$ arcmin$^2$, with indication for Malmquist bias in the $\YX$--$\Yv$ relation below this threshold. The corresponding mass threshold depends on redshift. Systems with $\Mv > 5 \times 10^{14}\, \msol$ at $z > 0.5$ are easily detectable with \planck. The newly-detected clusters follow the $\YX$--$\Yv$ relation derived from X-ray selected samples.  Compared to X-ray selected clusters, the new SZ clusters have a lower X-ray luminosity on average for their mass. There is no indication of departure from standard self-similar evolution in the X-ray versus SZ scaling properties. In particular, there is no significant evolution of the $\YX/\Yv$ ratio. }

\authorrunning{Planck Collaboration}
\titlerunning{Validation  of new \planck\ clusters  with \xmm}

\maketitle

%%%%%%%%%%%%%%%%%%%%%%%%%%%%%%%%%%%%%%%%%%%%%%%%%%%%%

\section{Introduction}       
\label{sec:introduction}
The \planck\footnote{\Planck\ (\url{http://www.esa.int/Planck}) is a project of the European Space Agency (ESA) with instruments provided by two scientific consortia funded by ESA member states (in particular the lead countries: France and Italy) with contributions from NASA (USA), and telescope reflectors provided in a collaboration between ESA and a scientific consortium led and funded by Denmark.} satellite has been surveying the millimetre sky since 2009. Its two instruments together cover nine frequency bands: the Low Frequency Instrument \citep[LFI; ][]{mandolesi2010,bersanelli2010,planck2011-1.4} at 30, 44, and 70\,GHz, and the High Frequency Instrument \citep[HFI; ][]{lamarre2010, planck2011-1.5} at 100, 143, 217, 353, 545, and 857 GHz. Before the HFI coolant ran out in January 2012, \planck\ had successfully performed nearly 5 surveys of the entire sky. 

\planck\ allows the detection of galaxy clusters by their imprint on the cosmic microwave background (CMB) via the Sunyaev-Zeldovich (SZ) effect, a characteristic spectral distortion of the CMB due to  inverse Compton scattering of photons by hot electrons in the intra-cluster medium \citep{sun72}. The SZ signal of galaxy clusters is expected to correlate tightly with cluster mass  \citep[e.g.,][]{das04} and its surface brightness is independent of redshift. SZ selected cluster samples are thus particularly well-suited for statistical studies of the galaxy cluster population, either as a probe of the physics of structure formation, or for cosmological studies based on cluster abundance as a function of mass and redshift.  Compared to other SZ surveys, such as those with the Atacama Cosmology Telescope \citep[ACT,][]{mar11} or the South Pole Telescope  \citep[SPT,][]{car09a}, the \planck\ survey covers an exceptionally large volume; indeed, it is the first all-sky survey since the \rosat\  All-Sky Survey (\rass) in the X-ray domain. \planck\ allows the detection of clusters below the flux limit of \rass\ based catalogues at redshifts typically greater than $0.3$ \citep[][Fig.~9]{planck2012-I}.  The first \planck\ SZ catalogue, the Early SZ (ESZ) sample, was published in  \citet{planck2011-5.1a}. It contains 189 clusters and candidates detected at high signal-to-noise ratio (${\rm S /N} > 6$) in the all-sky maps from the first ten months of observations, 20 of which were previously unknown.  At the release of the ESZ sample, 12 of those 20 had been confirmed as new clusters, 11 using \xmm\  validation observations undertaken in Director's Discretionary Time  (DDT) via an agreement between the \xmm\ and \planck\ Project Scientists.

All cluster surveys include false detections.  For \Planck\/, these are mainly due to inhomogeneous, non-isotropic, and highly non-Gaussian fluctuations (galactic dust emission, confusion noise as result of the unsubtracted point sources, etc.) in the complex microwave astrophysical sky.  After identification of known clusters, a follow-up programme is required for cluster confirmation and redshift estimation. It is essential to build as pure as possible an initial candidate sample in order for such a programme to be efficient and manageable.  For this we rely both on internal \planck\ candidate selection and assessment of the SZ signal quality, and on cross-correlation with ancillary data, as described in \citet{planck2011-5.1a}.  Beyond simple confirmation of new clusters, the \xmm\ validation programme aims to refine this validation process and to yield a better understanding of the new objects that \planck\ is detecting.  It consists of snapshot exposures ($\sim10$\,ks), sufficient for unambiguous discrimination between clusters and false candidates \citep{planck2011-5.1b}, for a total allocated time of 500\,ks for 50~candidates.

In the first two follow-up programmes, described by \citet{planck2011-5.1b}, we observed 25 candidates in total and helped to define the selection criteria for the ESZ sample. They yielded the confirmation of 17 single clusters, two double systems, and two triple systems\footnote{These multiple systems, where more than one cluster contribute to the \planck\ signal,  can be either chance association on the sky of clusters at different redshifts, or physically related objects at the same redshift. When referring to double or triple systems in the text, we do not distinguish between the two cases.}. The observations showed that the new clusters are on average less X-ray-luminous and more morphologically disturbed than their X-ray-selected counterparts of similar mass, suggesting that  \planck\ may be revealing a non-negligible population of massive, dynamically perturbed objects that are under-represented in X-ray surveys. However, despite their particular properties, the new clusters appear to follow the $\YSZ$--$\YX$ relation established for X-ray selected objects, where $\YX$,  introduced by \citet{kra06}, is the product of the gas mass and temperature. 

In the third follow-up programme,  described in \citet{planck2012-I}, we observed 11 candidates with lower SZ detection levels  ($4.5\!<\!\textrm{S/N}\!<\!5.3$) than the previous programmes  ($5.1\!<\!{\rm S/N}\!<\!10.6$) in order to investigate the internal SZ quality flags.  Probing lower SZ flux than previous campaigns, the third programme also demonstrated the capability of \planck\ to find new clusters below the \rass\ limit and up to high $z$, including the  blind detection at ${\rm S/N}\sim5$ of PLCK G266.6$-$27.3,  confirmed by \xmm\ to be an  $M_{500}\sim8 \times 10^{14}\, \msol$  cluster at $z\sim1$ \citep{planck2011-5.1c}.  We also detected tentative evidence for Malmquist bias in the $Y_{\rm SZ}$--$Y_{\rm X}$ relation, with a turnover at $Y_{\rm SZ} \sim 4\times10^{-4}$ arcmin$^2$.

In the fourth and last \xmm\ validation programme, presented here, we further probe the low SZ flux, high redshift regime.  The sample includes 15 candidates, detected at signal-to-noise ratios between 4~and 6.1 in the 15.5-month nominal survey data.  We use the results from all \xmm\ validation observations to address the use of  ancillary \rass\ information as an indicator of candidate reliability (Sec.~\ref{sec:rass}).  The evolution of cluster SZ/X-ray properties is discussed in Sec.~\ref{sec:evol}. This paper, together with \citet{planck2011-5.1b,planck2011-5.1c,planck2012-I}, presents our  complete analysis of the DDT \xmm\ validation programme.

We adopt a $\Lambda$CDM cosmology with $H_0=70\,\kmsMpc$, $\Omega_{\rm M}=0.3$, and $\Omega_\Lambda=0.7$. The factor $E(z)= \sqrt{\Omega_{\rm M} (1+z)^3+\Omega_\Lambda}$ is the ratio of the Hubble constant at redshift $z$ to its present-day value. The quantities $\Mv$ and $\Rv$ are the total mass and radius corresponding to a total density contrast $\delta=500$, as compared to $\rho_{\rm c}(z)$, the critical density of the Universe at the cluster redshift;  $\Mv = (4\pi/3)\,500\,\rho_c(z)\,\Rv^3$. The SZ flux is characterised by $\YSZ$, where $\YSZ\,D_{\rm A}^2$ is the spherically integrated Compton parameter within $\Rv$, and $D_{\rm A}$ is the angular-diameter distance to the cluster. Thus, as defined here, $\YSZ$ has units of solid angle and is given in arcmin$^2$ in Table~2.

%==============================================================================================================================================
\begin{table*}[t]
\begingroup
\caption{\footnotesize{Summary of ancillary information used in selecting candidates for XMM observations, and log of the \xmm\ observations.  Column~(1): \planck\ source name. Columns~(2) \& (3): Right ascension and declination of the Planck source (J2000). Columns (4) \& (5): Signal-to-noise ratio of the \planck\ cluster candidate detection with the MMF3 algorithm in the \planck-maps, and number of methods blindly detecting the candidate. Column~(6): quality grade of the SZ detection (A is best). Column~(7)--(10): \xmm\ observation identification number, filter used, on-source exposure time with the EPN camera, and fraction of useful time after cleaning for periods of high background due to soft proton flares (EMOS and EPN camera, respectively). Column~(11): category resulting from the pre-selection of the candidates. (12) Confirmed clusters are flagged. $\dagger$ indicates double projected systems. 
} }
\resizebox{\textwidth}{!} {
\begin{tabular}{l r r c c c c c r c c c}
\toprule
\toprule
\multicolumn{1}{c}{Name} &
\multicolumn{1}{c}{RA$_{\rm SZ}$} & \multicolumn{1}{c}{Dec$_{\rm SZ}$} & 
\multicolumn{1}{c}{$\textrm{S/N}$} & \multicolumn{1}{c}{$\textrm{$N_{\rm det}$}$} & \multicolumn{1}{c}{$\textrm{$Q_{\rm SZ}$}$}  
 & \multicolumn{1}{c}{OBSID} & 
\multicolumn{1}{c}{Filter} & \multicolumn{1}{c}{$t_{\rm exp}$} &
\multicolumn{1}{c}{Clean fraction} &   \multicolumn{1}{c}{Category} & \multicolumn{1}{c}{Confirmed}\\
\noalign{\smallskip}
\multicolumn{1}{c}{} &  \multicolumn{1}{c}{(deg)} & 
\multicolumn{1}{c}{(deg)}  & \multicolumn{1}{c}{} &  \multicolumn{1}{c}{}&  \multicolumn{1}{c}{} & \multicolumn{1}{c}{}  &  
\multicolumn{1}{c}{} & \multicolumn{1}{c}{(ks EPN)} &
\multicolumn{1}{c}{(EMOS/EPN)} & \multicolumn{1}{c}{} \\
\midrule
PLCK\,G348.4$-$25.5  & 291.257  &$-$49.426  & 6.12 &       3  & A  & 0679180101  &  t t t  & 10.6  & 1.0 $/$0.9  & PHZ  &  Y \\
PLCK\,G329.5$-$22.7$^{\dagger}$  & 278.270  &$-$65.570  & 5.84 &       3  & B  & 0679181501  &  m m t  &  8.5  & 1.0 $/$1.0  &   \ldots  &  Y \\
PLCK\,G219.9$-$34.4  &  73.680  &$-$20.269  & 5.74 &       2  & A  & 0679180501  &  t t t  &  9.5  & 1.0 $/$0.9  & PHZ  &  Y \\
PLCK\,G352.1$-$24.0  & 290.233  &$-$45.842  & 5.63 &       2  & C  & 0679180201  &  m m t  &  8.5  & 1.0 $/$1.0  &   PHZ  &  Y \\
PLCK\,G305.9$-$44.6  &   5.946  &$-$72.393  & 5.40 &       3  & B  & 0679180301  &  t t t  & 10.1  & 0.4 $/$0.2  &    \ldots  &  Y \\
PLCK\,G196.7$-$45.5$^{\dagger}$  &  55.759  & $-$8.704  & 5.21 &       3  & B  & 0679180401  &  m m m  &  9.0  & 1.0 $/$0.8  &    \ldots  &  Y \\
PLCK\,G208.6$-$74.4  &  30.044  &$-$24.897  & 5.01 &       3  & B  & 0679180601  &  t t t  &  2.9  & 0.8 $/$0.8  &    \ldots  &  Y \\
PLCK\,G130.1$-$17.0  &  22.678  &   45.288  & 4.93 &       3  & C  & 0679180801  &  t t t  &  7.5  & 1.0 $/$1.0  &    \ldots  &  Y \\
PLCK\,G239.9$-$40.0  &  71.683  &$-$37.029  & 4.76 &       3  & B  & 0679181001  &  t t t  &  9.9  & 1.0 $/$1.0  &   \ldots  &  Y \\
PLCK\,G310.5$+$27.1  & 201.148  &$-$35.245  & 4.77 &       2  & B  & 0679180901  &  t t t  & 12.9  & 0.9 $/$0.8  & PHZ  & \ldots \\
PLCK\,G196.4$-$68.3  &  34.921  &$-$19.263  & 4.73 &       2  & B  & 0679181101  &  t t t  & 11.8  & 0.7 $/$0.4  & PHZ  & \ldots \\
PLCK\,G204.7$+$15.9  & 113.614  &   14.295  & 4.57 &       3  & A  & 0679180701  &  t t t  &  9.4  & 1.0 $/$1.0  &   \ldots  &  Y \\
PLCK\,G011.2$-$40.4  & 315.233  &$-$33.107  & 4.47 &       1  & C  & 0679181201  &  t t t  &  8.5  & 1.0 $/$1.0  & PHZ  &  Y \\
PLCK\,G147.3$-$16.6  &  44.099  &   40.291  & 4.41 &       3  & B  & 0679181301  &  t t t  & 10.7  & 0.9 $/$0.6  & PHZ  &  Y \\
PLCK\,G210.6$+$20.4  & 120.218  &   11.093  & 4.01 &       1  & C  & 0679181401  &  t t t  &  8.5  & 1.0 $/$1.0  &  SDSS  & \ldots \\
\bottomrule
\end{tabular}
}
\label{tab:sample}
\endPlancktable 
 \endgroup
\end{table*}
%==============================================================================================================================================

\section{Sample selection}
\label{sec:sample}

\subsection{\textit{Planck}  catalogue} 
In this paper,  candidates were chosen from the catalogue derived from the first 15.5\,months of data (the ``nominal'' mission).  The processing status, calibration, and map versions were those of March 2011. The detection and quality assessment of the cluster candidates followed the general procedure described in \citet{planck2011-5.1a}.  Briefly, a blind cluster search was performed with three methods: the matched multi-frequency filter ``MMF3'' developed by \citet[][]{mel06};  an independent matched multi-frequency filter ``MMF1''; and the PowellSnakes algorithm \citep[PWS;][]{car09b,car11}.  Candidates then underwent internal SZ quality checks, removing spurious detections (e.g., association with artefacts or galactic sources), and assessment of the SZ signal detection.  The signal assessment included quantitative criteria such as the signal-to-noise ratio and the number of methods blindly detecting the candidate, $N_{\rm det}$, as well as a qualitative assessment based on visual inspection of the frequency maps, reconstructed SZ images, and the frequency spectra for each cluster.  The latter procedure is summarised in an SZ quality grade, $Q_{\rm SZ}$, as described in \citet{planck2012-I}.  

Previously known clusters were identified via cross-correlation with catalogues and NED/Simbad queries. Possible counterparts were searched for within a 5\arcm\ radius of the \planck\ position, allowing us to assign two further external reliability flags:

\begin{itemize}
\item association of a FSC (Faint Source Catalogue) or a BSC (Bright Source Catalogue) \rass\ source \citep{vog99,vog00} or an excess of counts (with corresponding signal-to-noise ratio) in the RASS [0.5--2]\,\keV\ image.
\item  galaxy over-density in the Digitized Sky Survey (DSS) red plates\footnote{\url{http://stdatu.stsci.edu/dss}}, from a  visual check.  In the Sloan Digital Sky Survey (SDSS) area\footnote{\url{http://www.sdss.org}}, two independent galaxy detection algorithms were applied to the DR7 galaxy catalogues (Fromenteau et al., in preparation; Li \& White, in preparation).  Both algorithms use photometric redshift information.  Quality match criteria were assigned based on cluster richness or the over-density signal-to-noise ratio.
\end{itemize}

\subsection{\xmm\ target selection} 

The resulting targets are listed together with their SZ quality flags in Table~\ref{tab:sample}. The  range of signal-to-noise ratios, $4\!<\!{\rm S/N}\!<\!6.1$, is wide, with nearly uniform coverage, so that the validation results can be useful for defining the final signal-to-noise ratio for the {\it Planck\/} Cluster  Catalogue.  We considered lower signal-to-noise ratios than the previous validation programme, with 9 targets at ${\rm S/N}<5$ and a median S/N of $4.9$, as compared to 5.1 previously (for 10.5\,months of survey data).  A priori, this allows us to reach lower flux or higher redshift. To further push the sample towards high redshift, we discarded candidates with estimated $\Rv$ size greater than $5\arcmin$.  Although the large positional uncertainty of \planck\ candidates makes the search for a DSS counterpart non-trivial, the brightest galaxies of clusters at $z < 0.5$ are generally visible in DSS \citep[e.g.,][]{fas11}. We thus also used DSS images to select high-$z$ clusters.  Half the targets, labelled PHZ (potentially at high z) in Table~\ref{tab:sample}, have no visible counterpart in DSS red plates. These are obviously riskier candidates, particularly those with low $N_{\rm det}$ or $Q_{\rm SZ}$.

As previous validation observations have shown, the association of a SZ candidate with a \rass\ FSC or BSC source is not in itself sufficient to confirm the candidate, as chance association with a point source is always a possibility.  Conversely, a candidate with no counterpart in the \rass\ catalogue may well be a {\it bona fide} cluster.  With this campaign, in combination with the previous observations,  we also aim to address the use of \rass\ data as an indicator of candidate reliability.  In the sample of 36 candidates observed previously, thirteen candidates were associated with a BSC source and seventeen candidates with an FSC source.  Only six SZ candidates had no FSC/BSC counterpart, of which the three confirmed candidates were detected in \rass\ at a signal to noise ratio of $1.7\!<\!{\rm S/N}\!<\!2.8$. To better span the range of external \rass\ flags, we chose ten candidates with no FSC or BSC counterpart, six of which correspond to a \rass\ ${\rm S/N} < 1.5$.  Of the remaining five candidates, only one is  associated with a BSC source and four are associated with an FSC source. The \rass\ association for all \xmm\ validation targets is summarised in Table.~\ref{tab:RASS}. 

%==============================================================================================================================================
\begin{figure*}
\centering
\begin{minipage}[t]{0.9\textwidth}
\resizebox{\hsize}{!} {
\includegraphics[height=5cm,clip]{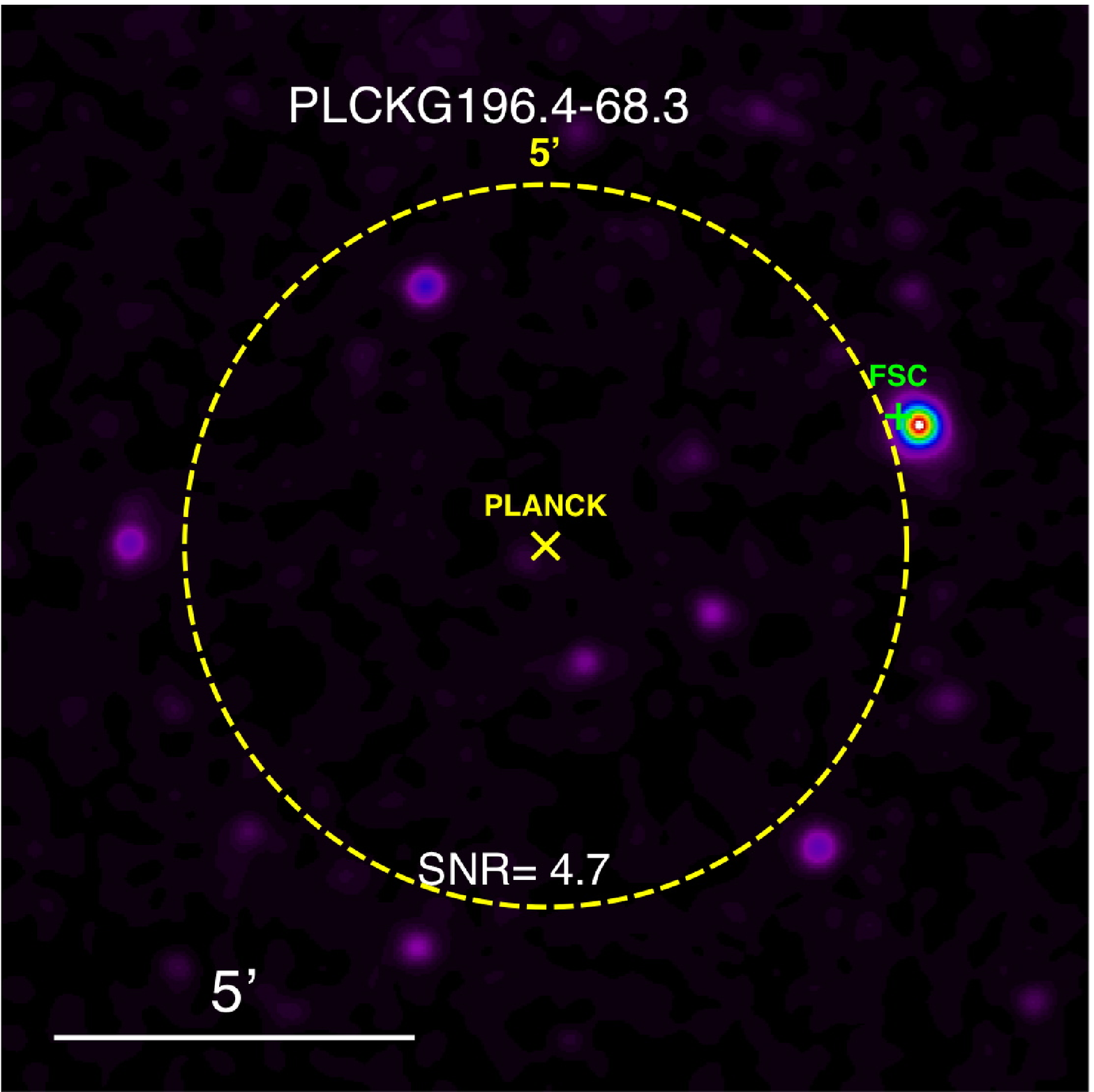} 
\hspace{0.5mm}
\includegraphics[height=5cm,clip]{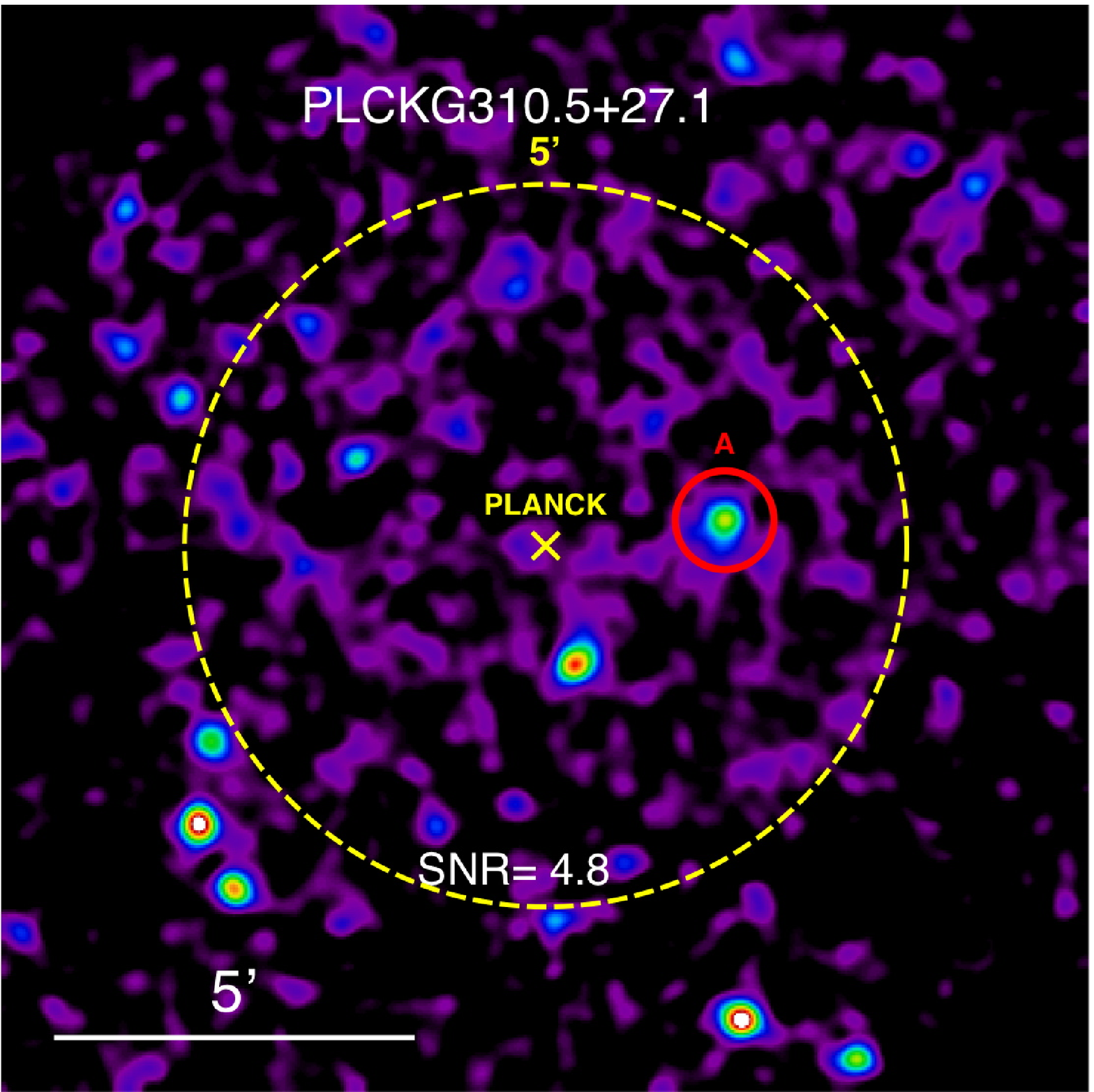}  
\hspace{0.5mm}
\includegraphics[height=5cm,clip]{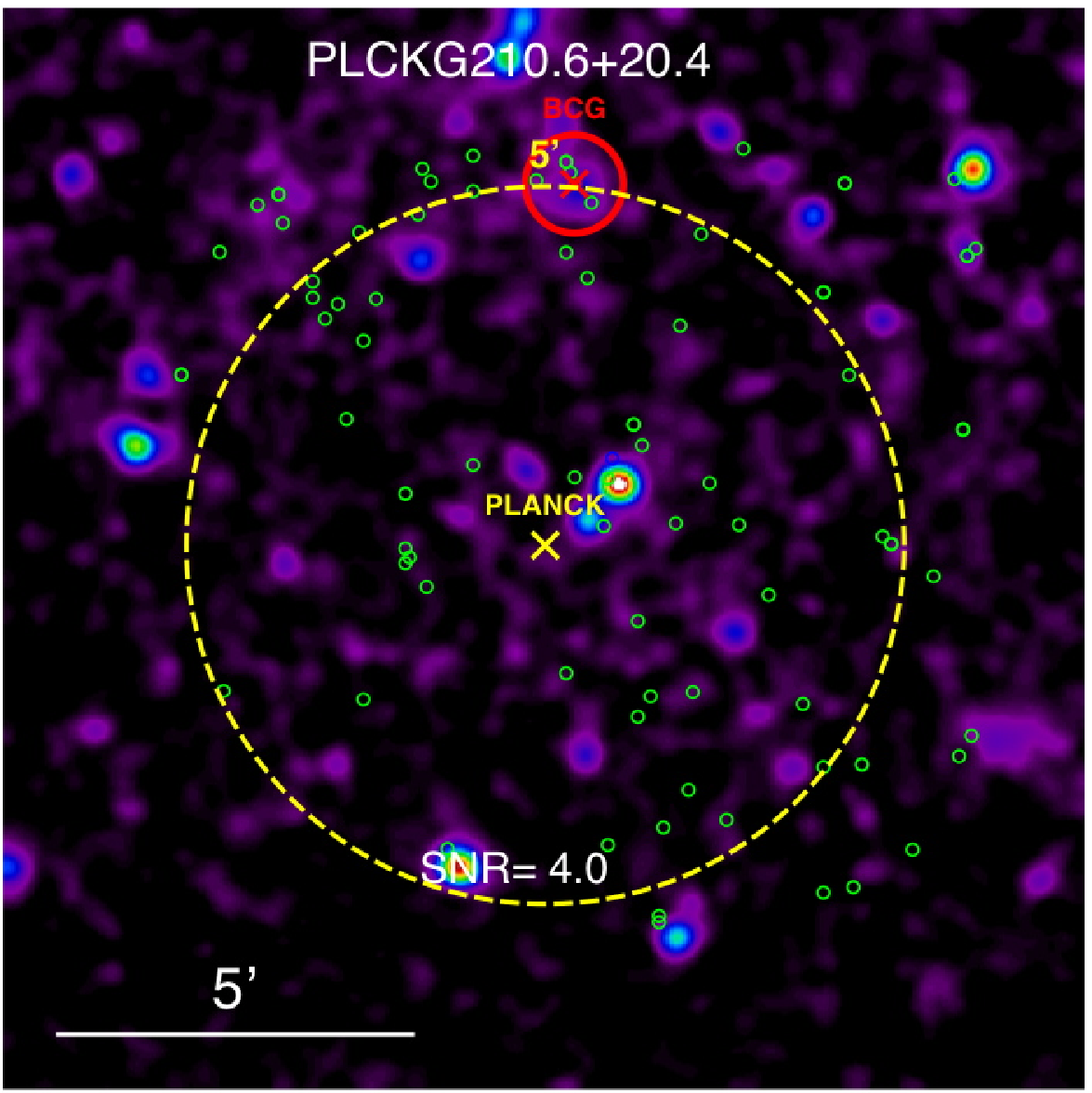}  
}
\end{minipage}
\caption{{\footnotesize \xmm\ [0.3--2]\,\keV\ energy band images of the three unconfirmed cluster candidates centred on the SZ position (yellow cross). The red circles indicate the presence of an extended source. Green squares in the right panel are positions of galaxies in the SDSS over-density. }}
\label{fig:false_gal}
\end{figure*}
%==============================================================================================================================================

Finally one candidate, PLCK\,G210.6$+$20.4, was specifically chosen to further test our SDSS-based confirmation of very poor SZ candidates.  PLCK\,G210.6$+$20.4  is the lowest SZ signal-to-noise candidate, detected at ${\rm S/N}=4$ by one method only, with a $Q_{\rm SZ}={\rm C}$ grade and no significant signal in \rass\ data.  However, the galaxy-detection algorithms (Sec.~2.1) that we used indicated that the candidate is associated with an SDSS  galaxy over-density at $z=0.5$.

\section{\textit{XMM-Newton} observations and data analysis}
\label{sec:analysis}

%==============================================================================================================================================
\begin{figure*}[tbp]
\begin{minipage}[t]{0.95\textwidth}
\centering
\includegraphics[height=0.23\hsize]{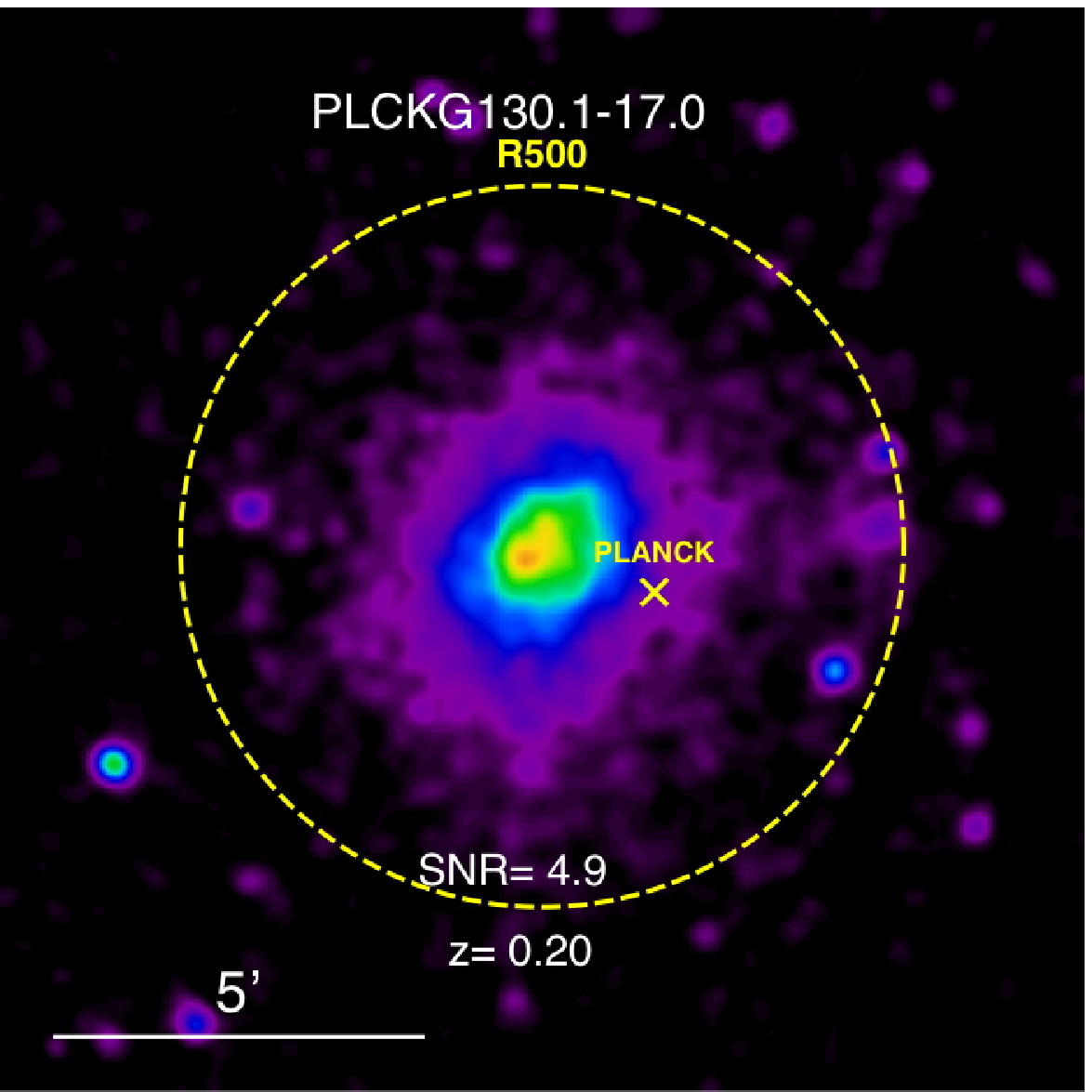} 
\hspace{0.2mm}
\includegraphics[height=0.23\hsize]{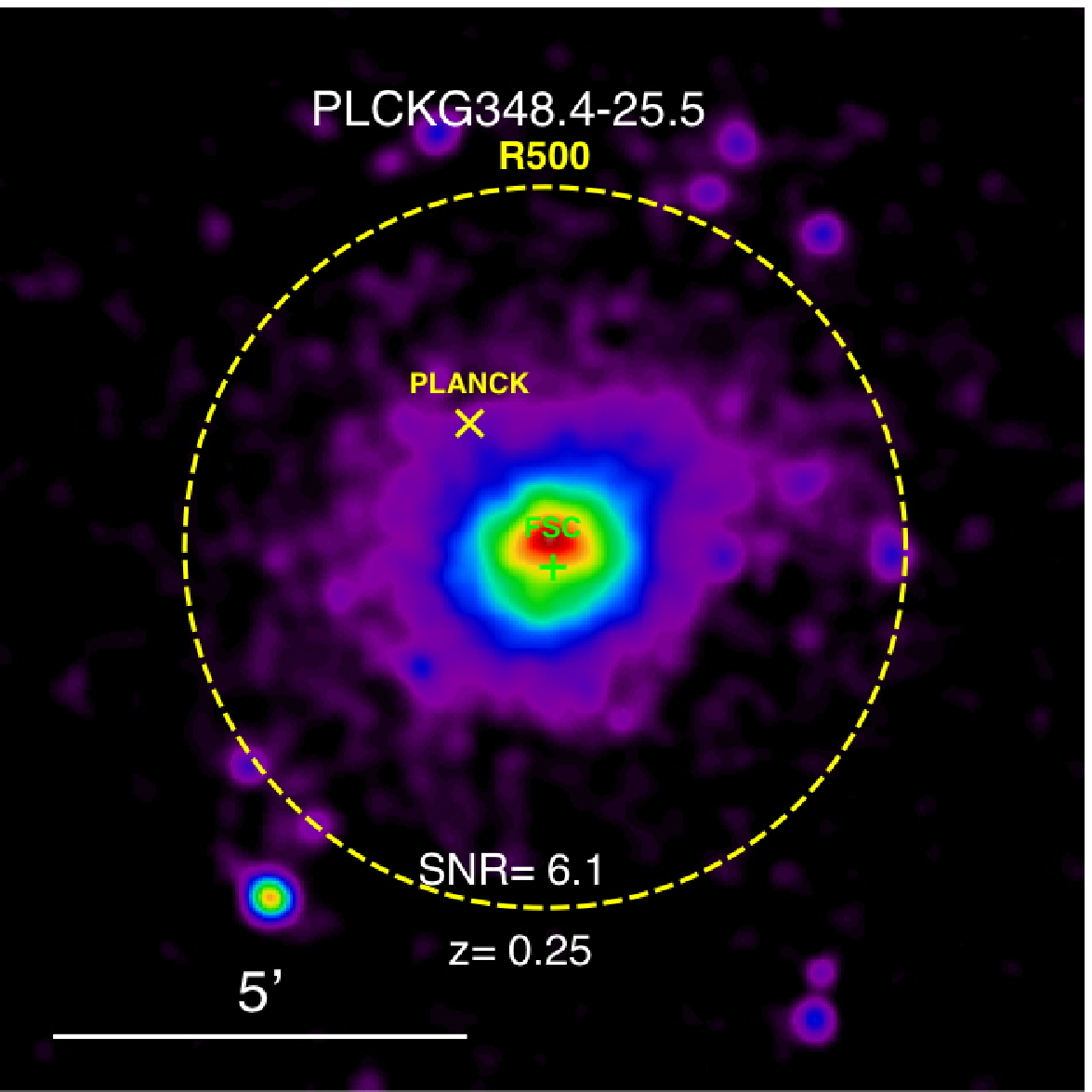}  
\hspace{0.2mm}
\includegraphics[height=0.23\hsize]{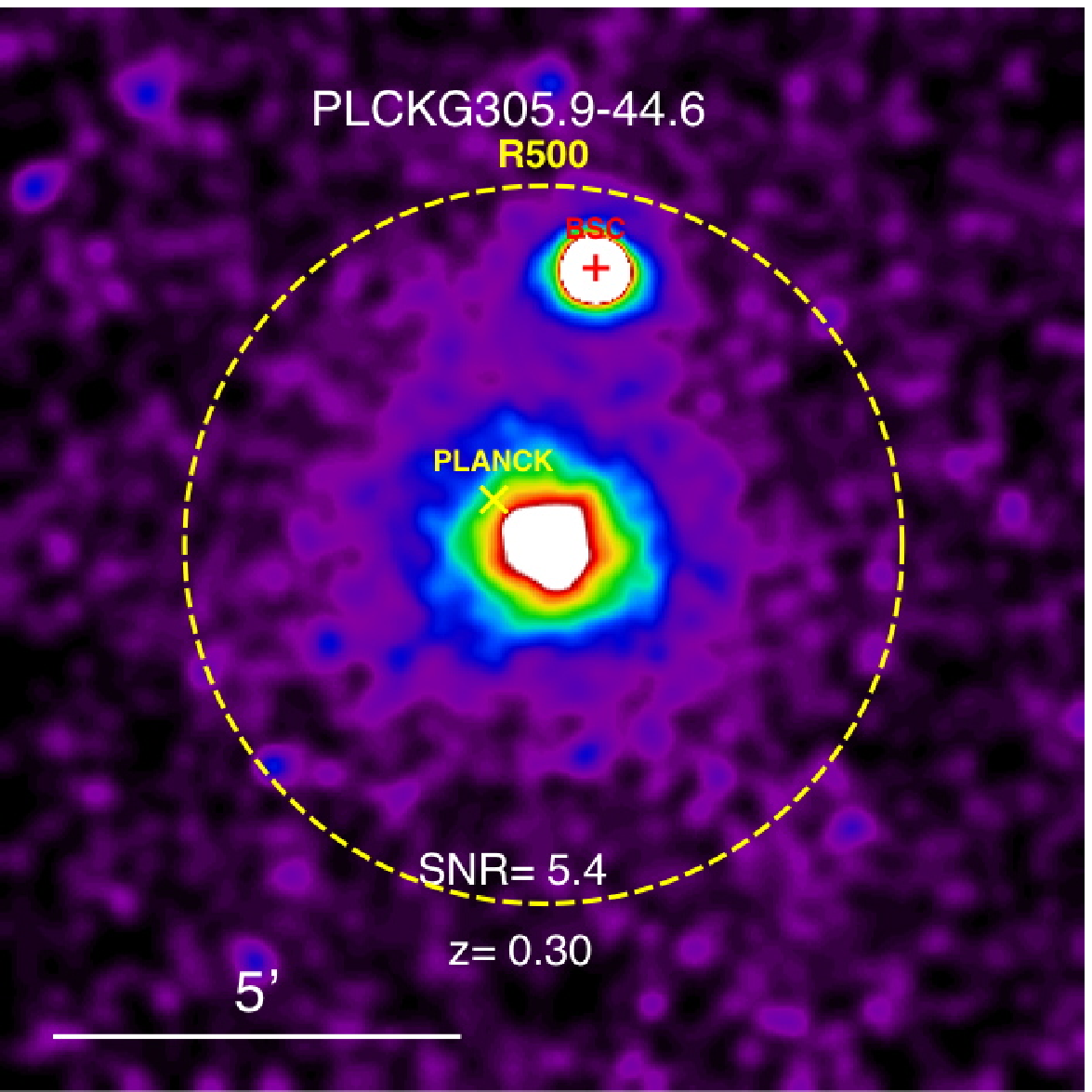} 
\hspace{0.2mm}
\includegraphics[height=0.23\hsize]{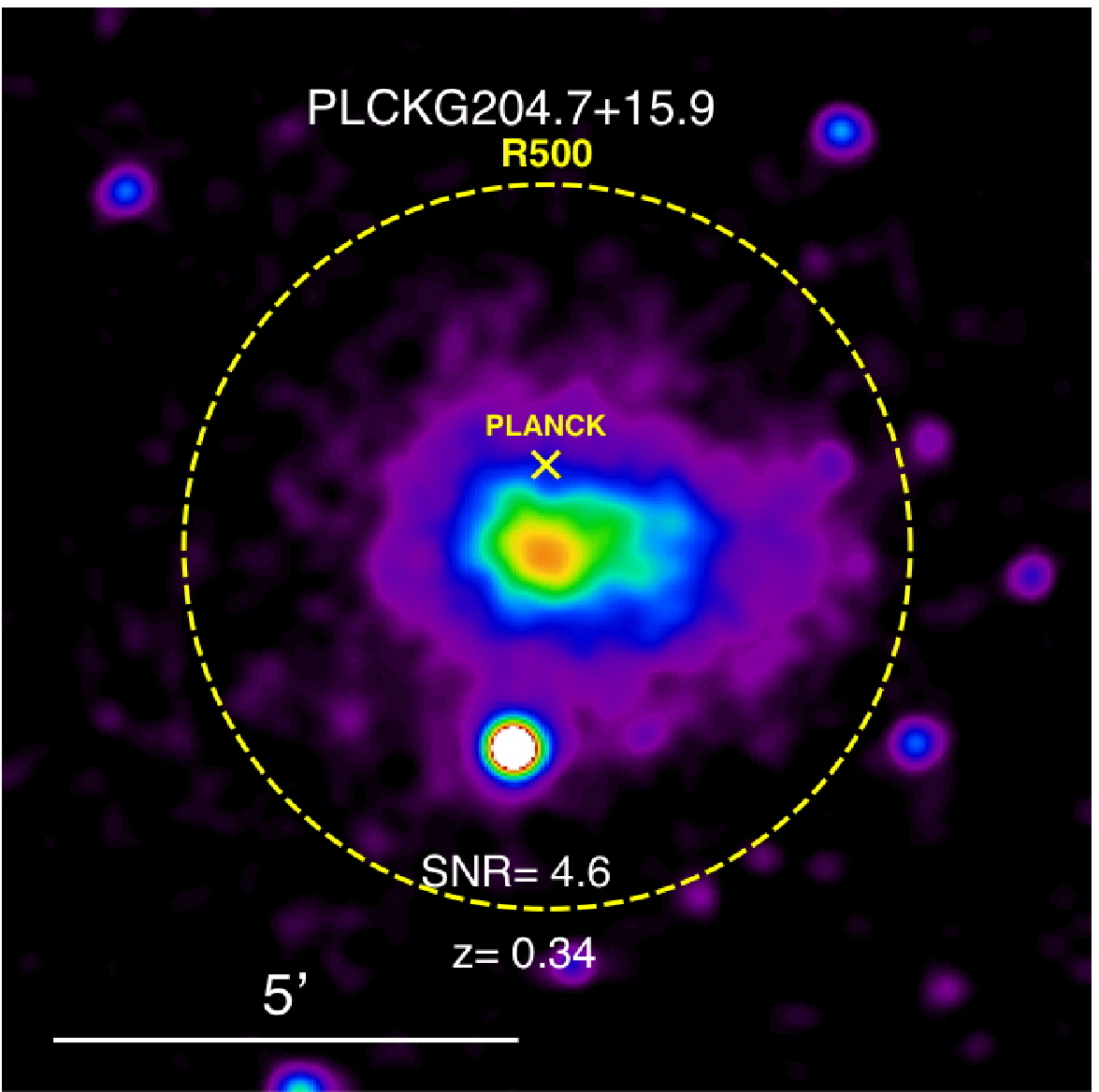}  
\end{minipage}\\[1.6mm]
\begin{minipage}[t]{0.95\textwidth}
\centering
\includegraphics[height =0.23\hsize]{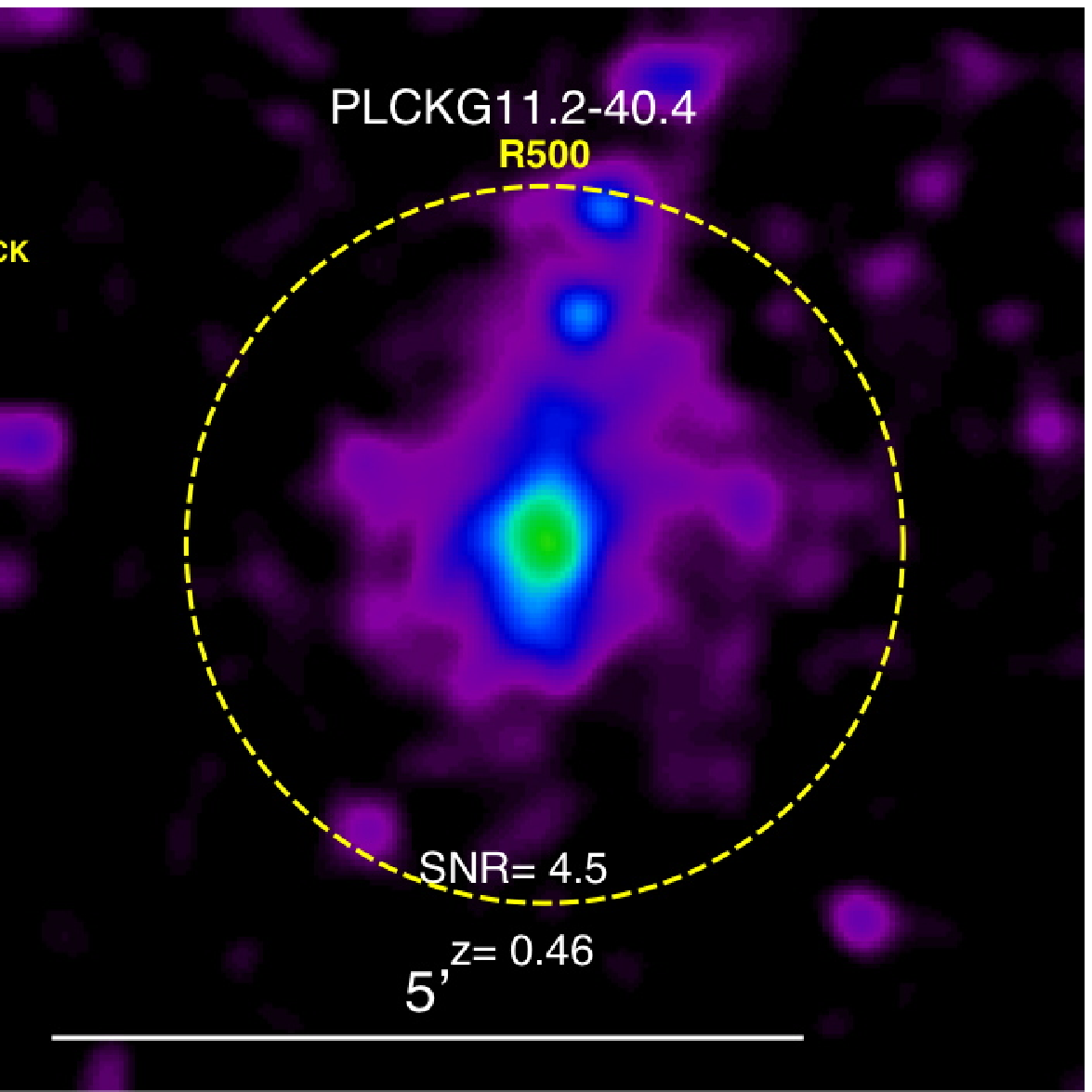} 
\hspace{0.2mm}
\includegraphics[height =0.23\hsize]{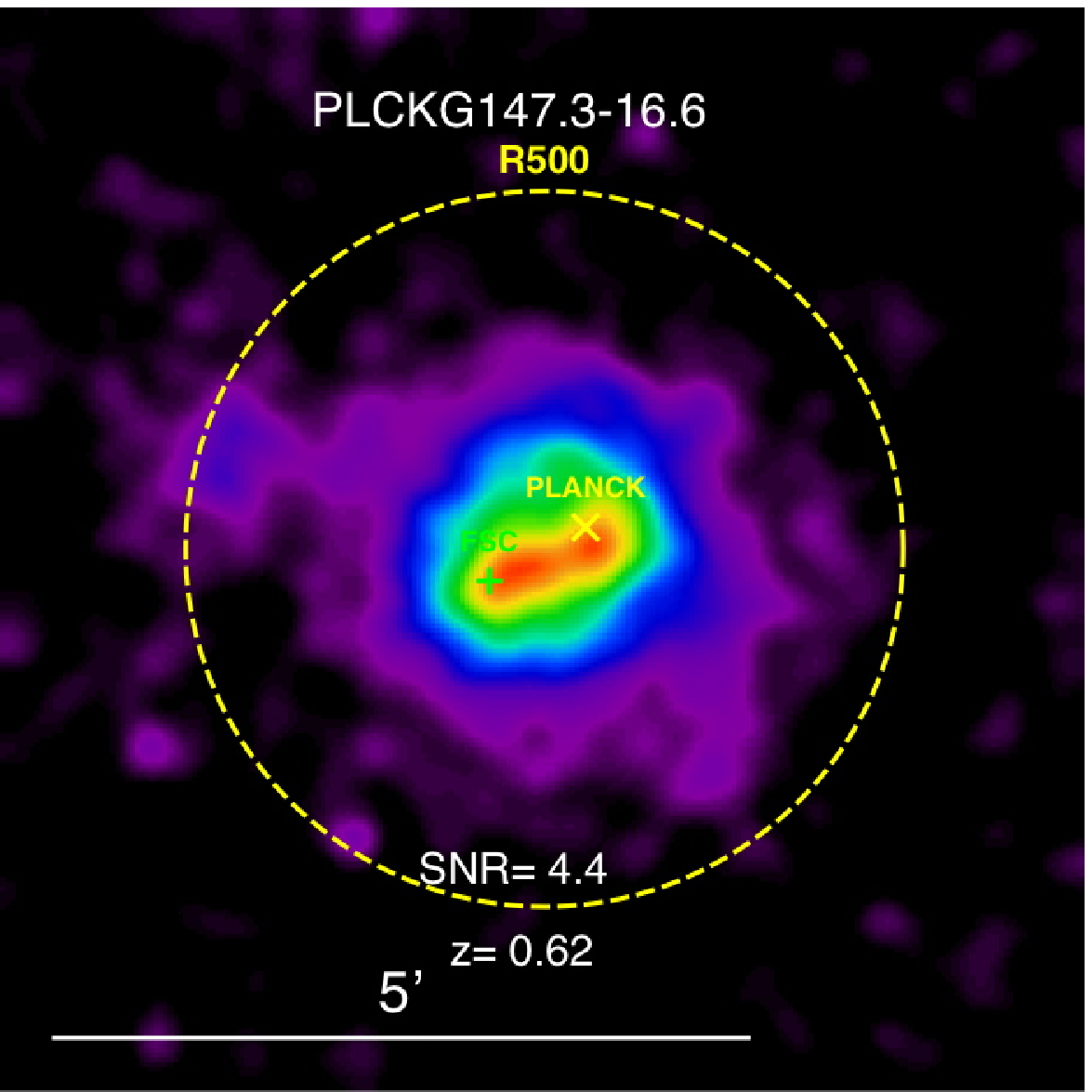}  
\hspace{0.2mm}
\includegraphics[height =0.23\hsize]{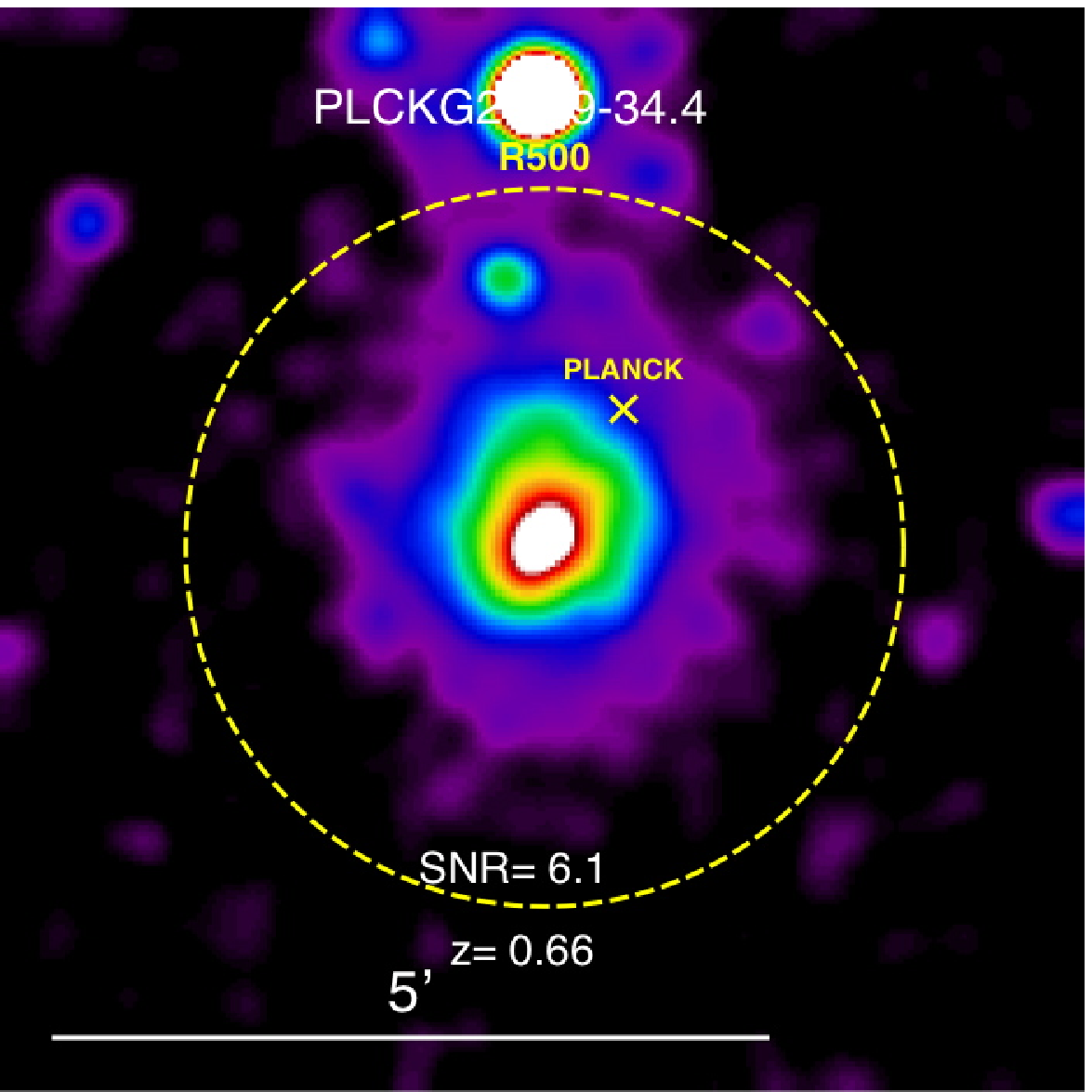} 
\hspace{0.2mm}
\includegraphics[height =0.23\hsize]{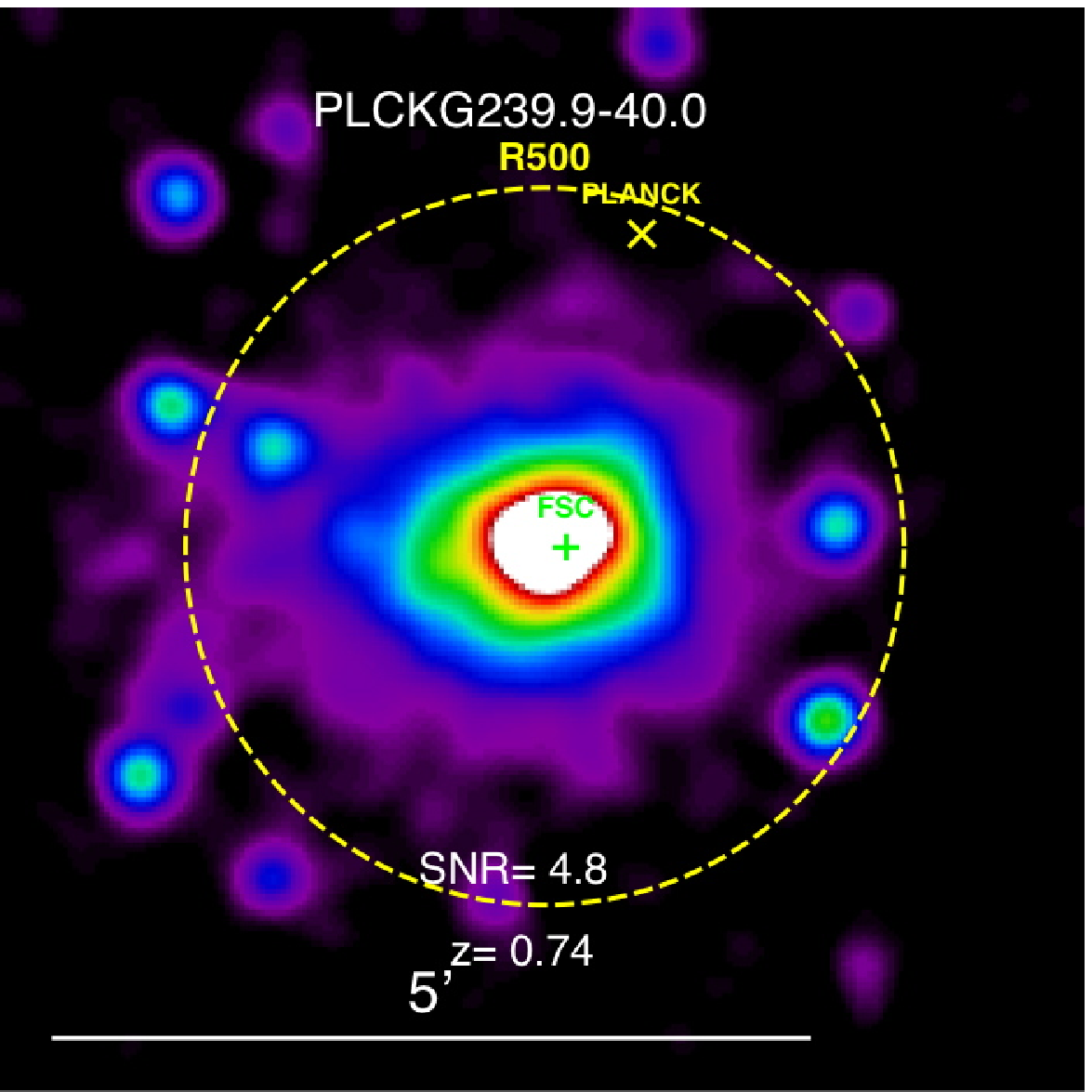} 
\end{minipage}\\[1.6mm]
\begin{minipage}[t]{0.95\textwidth}
\centering
\includegraphics[height =0.23\hsize]{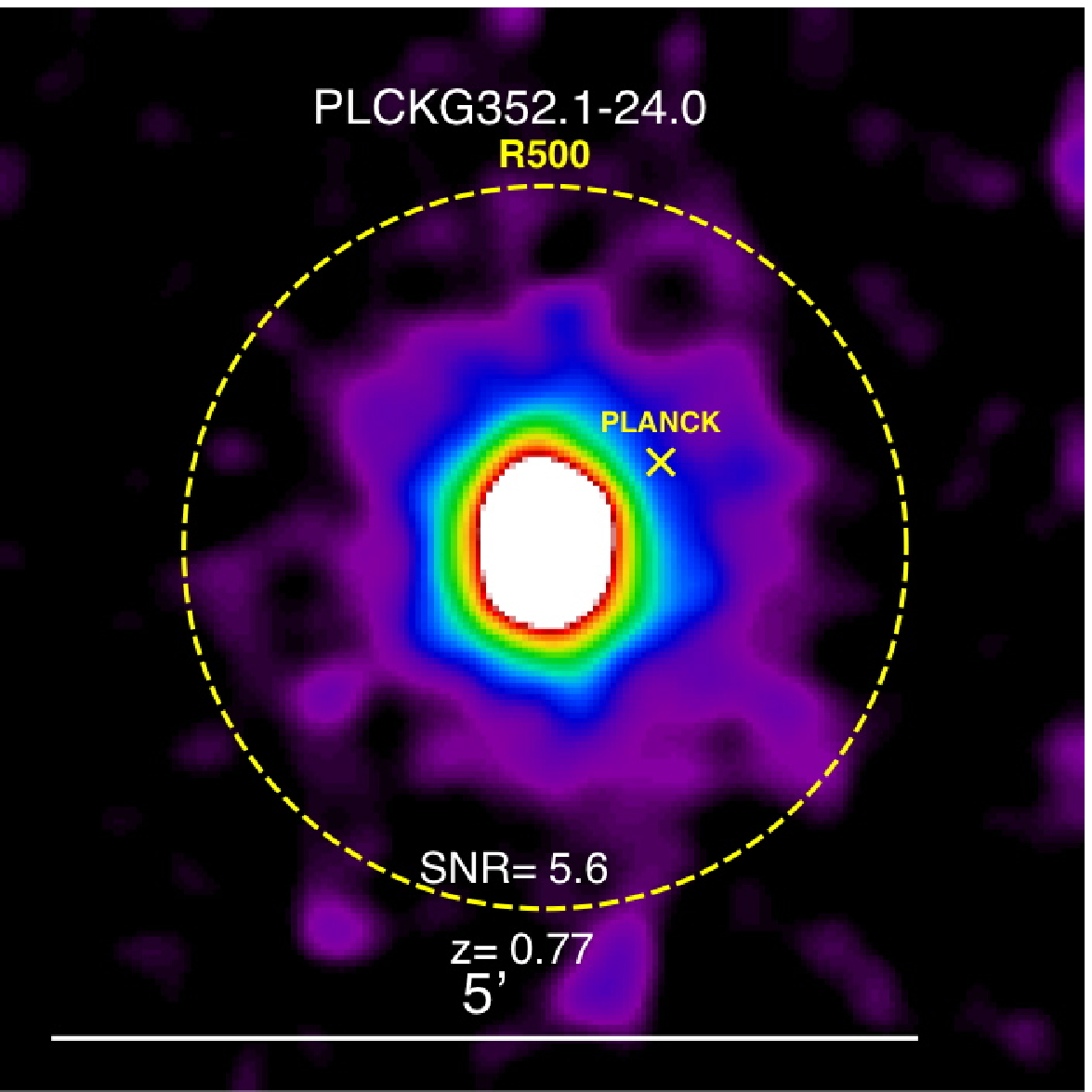} 
\hspace{0.2mm}
\includegraphics[height =0.23\hsize]{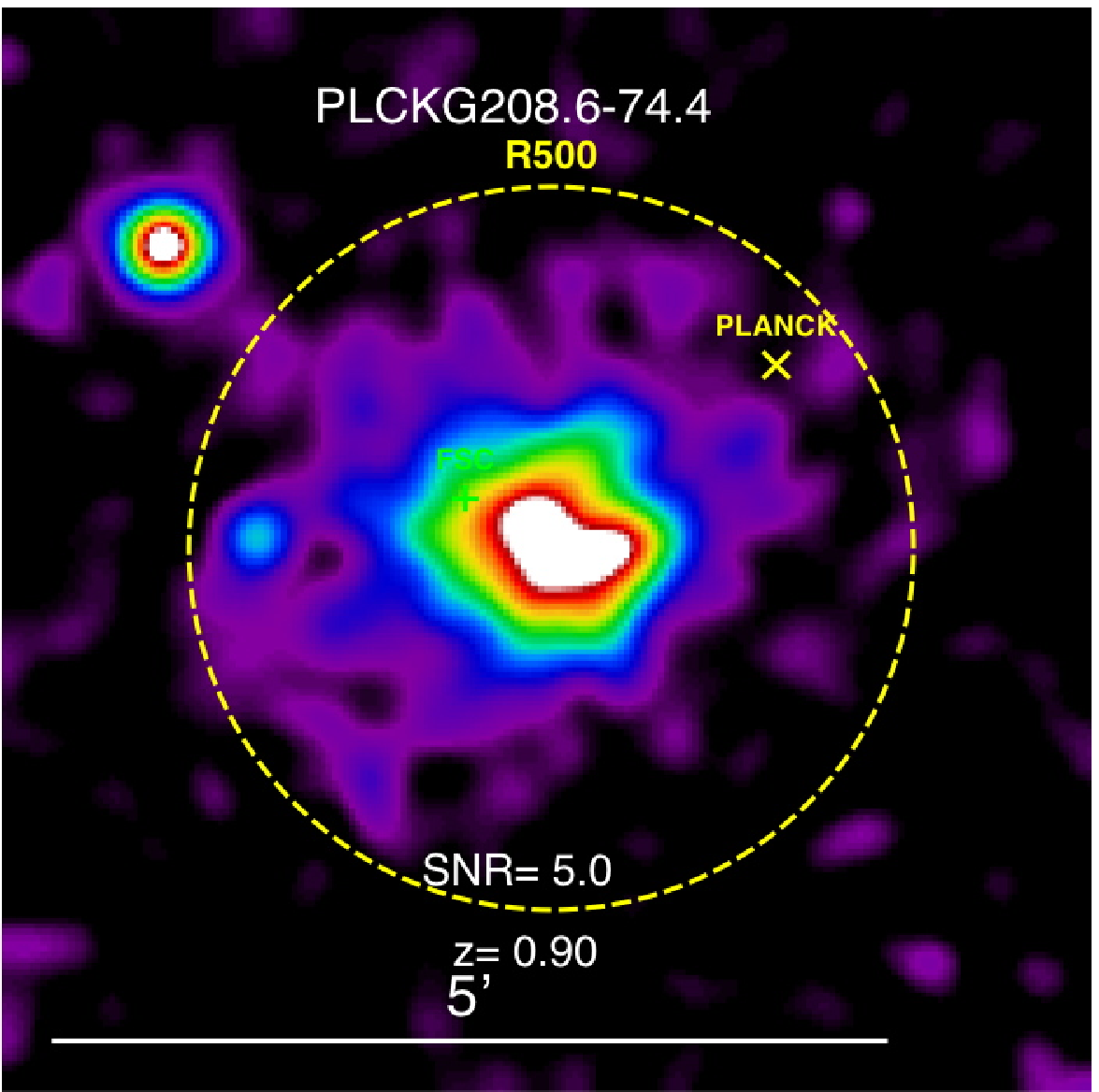}  
\end{minipage}\\[1.6mm]
\begin{minipage}[t]{0.95\textwidth}
\centering
\includegraphics[height =0.23\hsize]{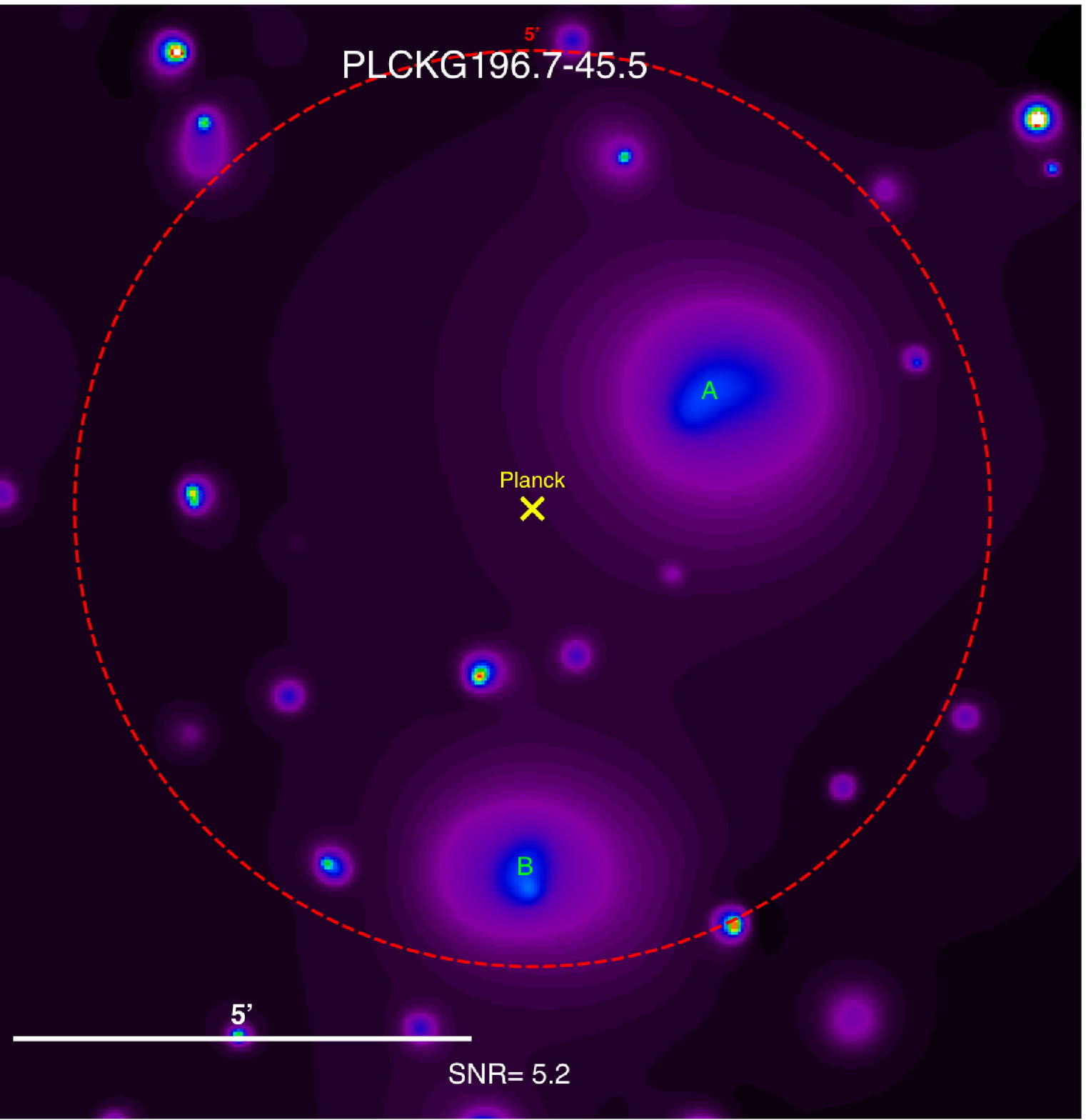}  
\hspace{0.2mm}
\includegraphics[height =0.23\hsize]{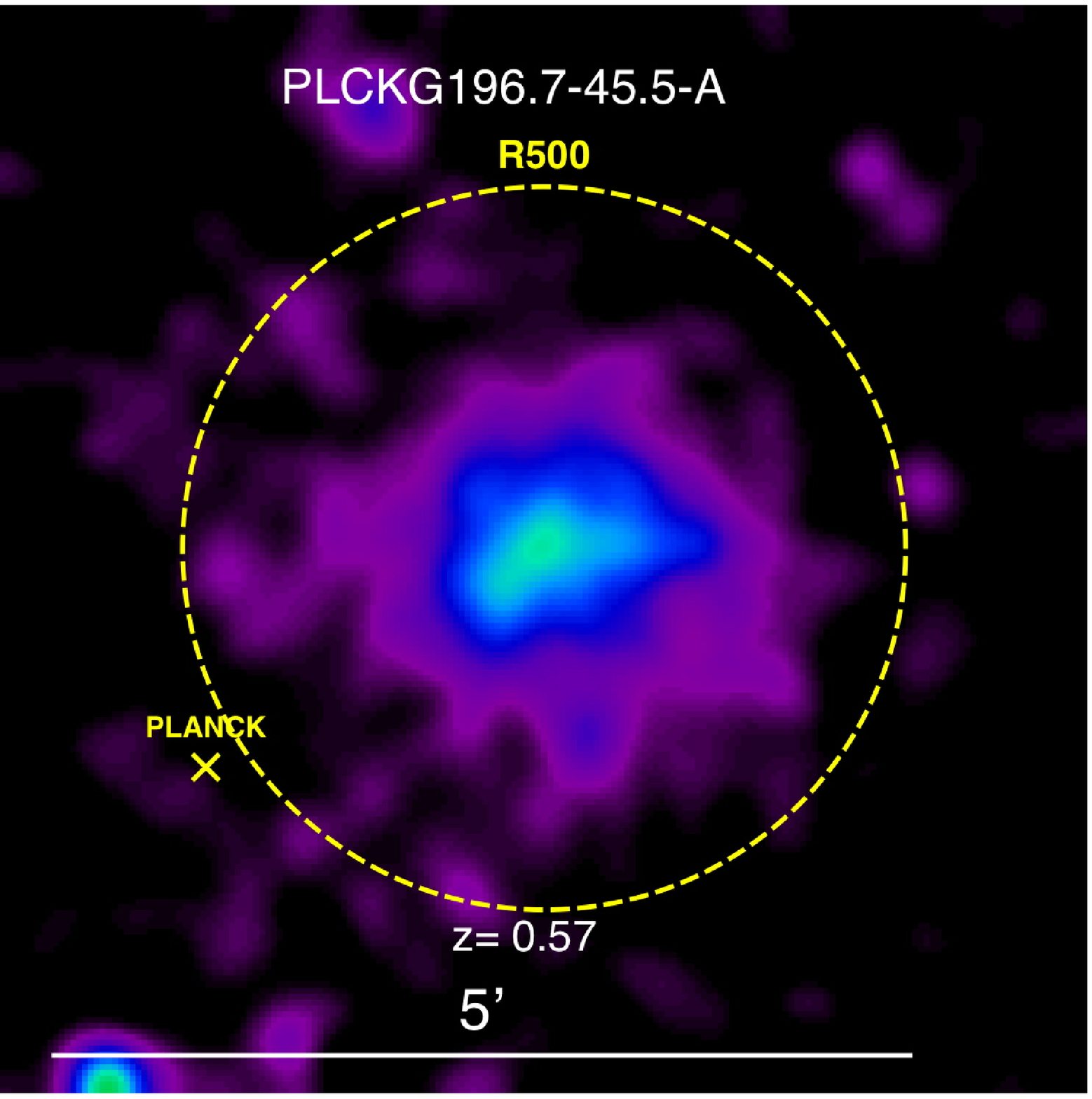}  
\hspace{0.2mm}
\includegraphics[height =0.23\hsize]{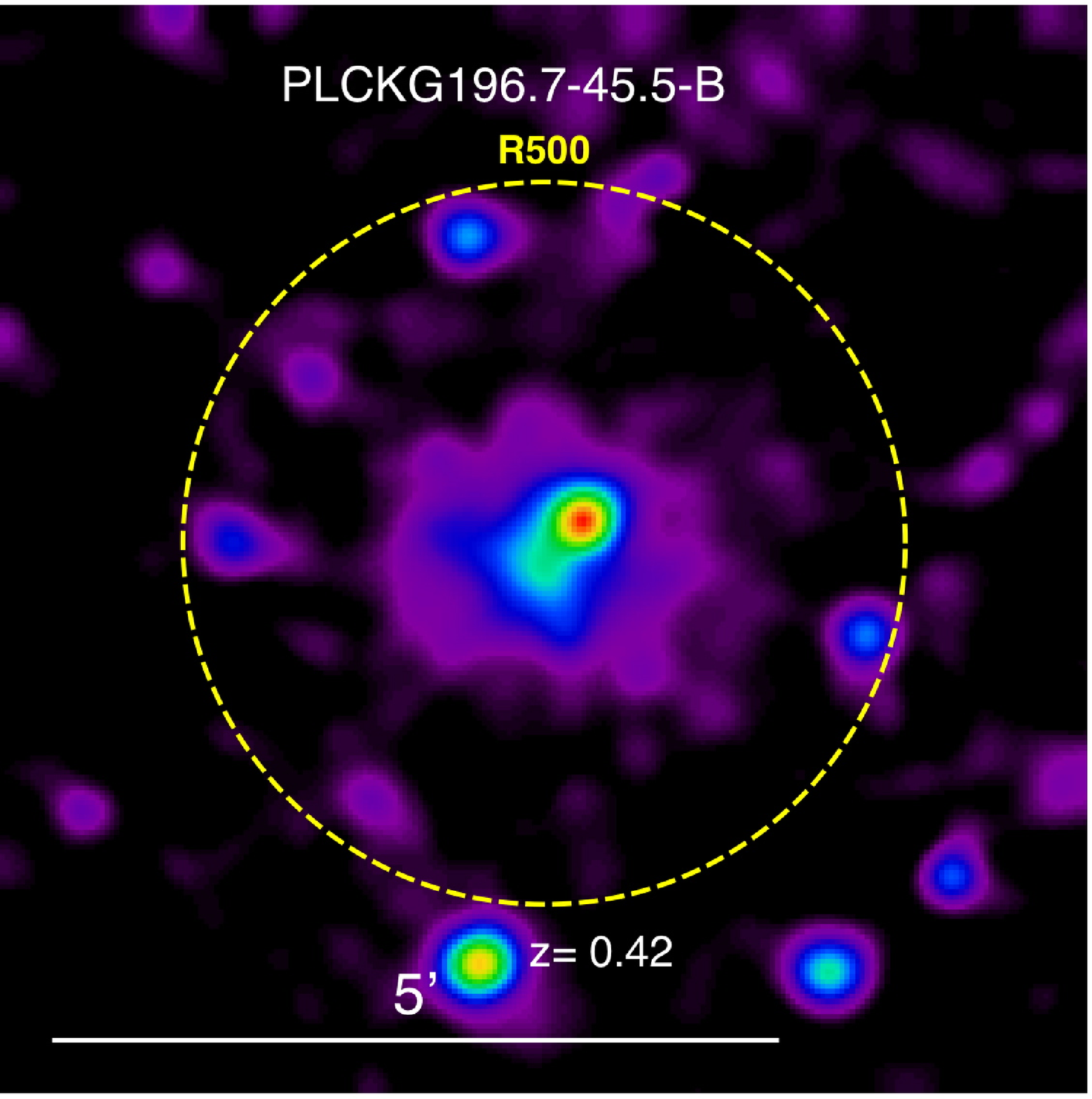} 
\end{minipage}\\[1.6mm]
\begin{minipage}[t]{0.95\textwidth}
\centering
\includegraphics[height=0.23\hsize]{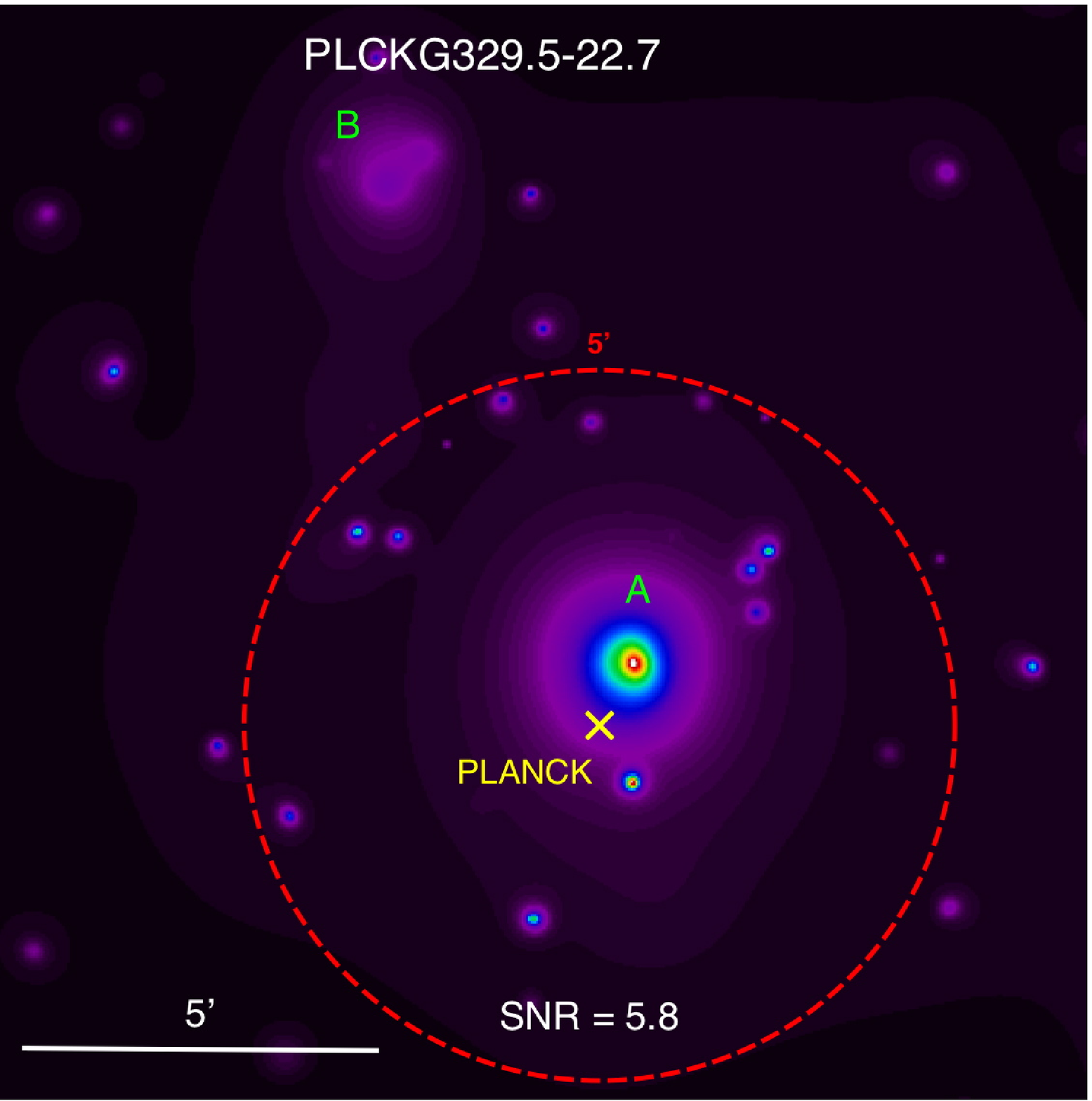}  
\hspace{0.2mm}
\includegraphics[height =0.23\hsize]{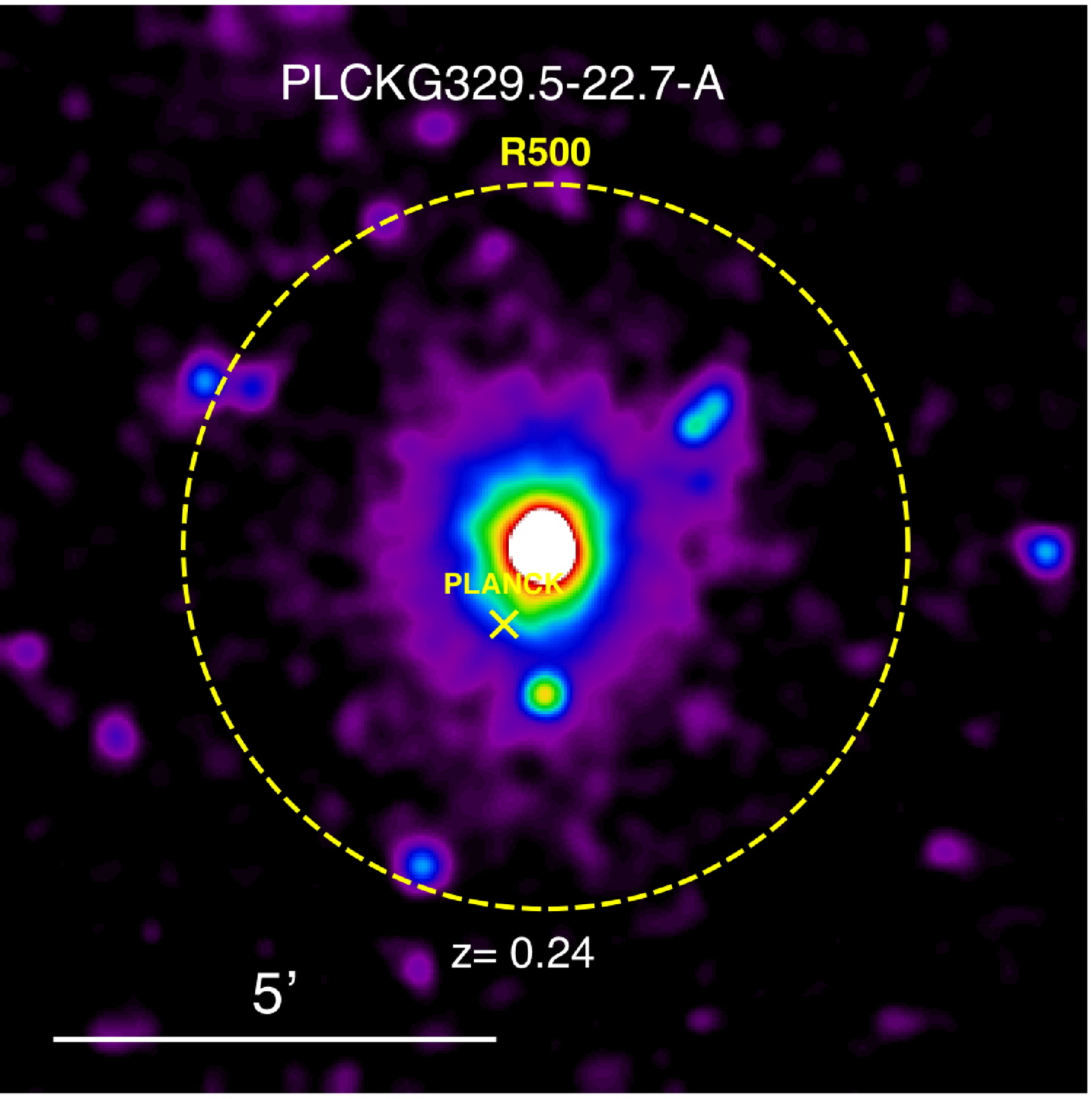} 
\hspace{0.2mm}
\includegraphics[height=0.23\hsize]{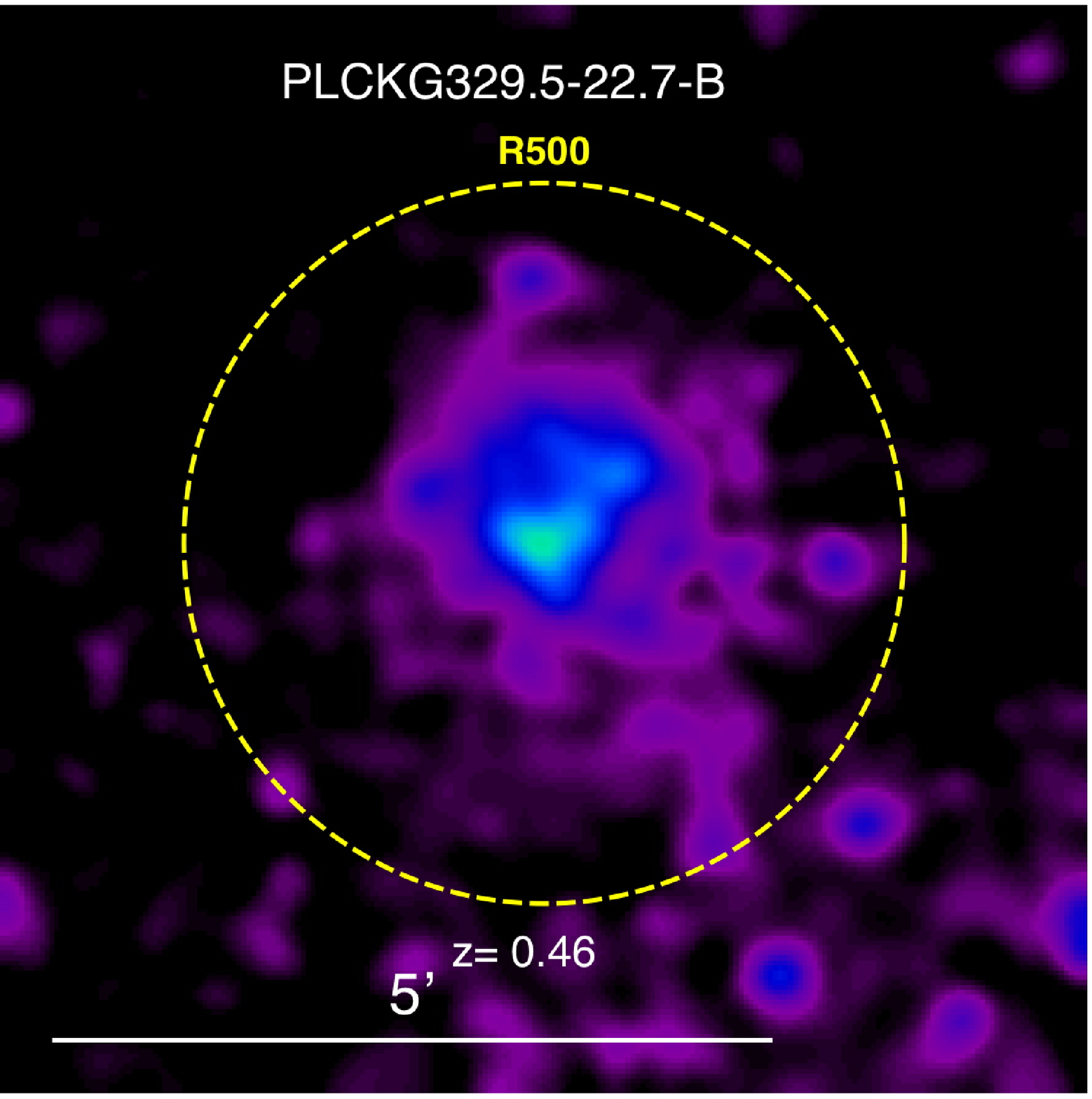}  
\end{minipage}\\[1.6mm]
\caption{{\footnotesize {\it XMM-Newton\/} [0.3-2] keV energy band images of confirmed cluster candidates. North is up and East is to the left. Image sizes are $3\theta_{500}$ on a side, where $\theta_{500}$ is estimated from the $M_{500}-Y_X$ relation of \citet{arn10} assuming standard evolution. Images are corrected for surface brightness dimming with $z$, divided by the emissivity in the energy band, taking into account galactic absorption and instrument response, and scaled according to the self-similar model. The colour table is the same for all clusters, so that the images would be identical if clusters obeyed strict self-similarity. A yellow cross indicates the\planck\ position and a red/green plus sign the position of a \rass-BSC/FSC source. The clusters are sorted according their estimated redshift. For the double systems (last two rows) the middle and right panels show the two components and the left panel the wavelet-filtered  overall image.  }  }
\label{fig:confirmed_gal}
\end{figure*}
%==============================================================================================================================================

Candidates were observed between 31 July 2011 and 13 October 2011. The observation identification number and observation setup are given in Table~\ref{tab:sample}.  Due to a slew failure in the satellite revolution 2132, the PLCK\,G208.6$-$74.4 observation was incomplete, with an EPN exposure time of $3.4$\,ks. The target was observed initially at the end of its summer visibility window, and could only be reobserved five months later.  It was replaced with an additional visible candidate, PLCK\,G329.5$-$22.7. 

Calibrated event lists were produced with v11.0 of XMM-\sas. Data that were affected by periods of high background due to soft proton flares were omitted from the analysis \citep{pra07}; clean observing time after flare removal is given in Table~\ref{tab:sample}. The status of each SZ candidate is also given in Table~\ref{tab:sample}: 12 of the 15 candidates are confirmed to be real clusters, among which two are double systems.  \xmm\ images of unconfirmed candidates are shown in Fig.~\ref{fig:false_gal}; confirmed candidates are shown in Fig.~\ref{fig:confirmed_gal}. 

We derived redshifts and physical parameters of the confirmed candidates as described in \citet{planck2011-5.1a, planck2012-I}.  Cleaned \xmm\ data were {\sc pattern}-selected.  Each photon was then assigned a weight equivalent to the ratio of the effective area at the photon energy and position to the central effective area, computed with \sas\ task {\sc evigweight}. Images and spectra were extracted using this weight,  assuring full vignetting correction \citep[see][]{arn01}.
Bright point sources were excised from the data and the background was handled as described in \citet{pra10}.  The particle-induced background (PB) was estimated using a stacked event list built from observations obtained with the filter wheel in closed position. The cosmic X-ray background was modeled using a PB-subtracted spectrum of an annular region external to the cluster emission. 

In the spectroscopic analysis, the hydrogen column density was fixed at the 21-cm value of \citet{lab05}. The redshift was estimated  by fitting an absorbed redshifted thermal model to the spectrum extracted within a circular region corresponding to the maximum X-ray detection significance. 
 The quality of the $z$ estimate was characterised by the quality flag  $Q_{\rm z}$ as introduced in \citet{planck2011-5.1b}.  $Q_{\rm z}$ was set to $Q_{\rm z}\!\!=\!\!0$ when the redshift could not be constrained due to the lack of  line detection. $Q_{\rm z}\!\!=\!\!1$
  corresponds to ambiguous $z_{\rm Fe}$ estimate, when the spectral fit as a function of $z$ exhibited several $\chi^2$ minima that could not be distinguished at the $90\%$ confidence level.  $Q_{\rm z}\!\!=\!\!2$ corresponds to a well constrained redshift (i.e., a single $\chi^2$ minimum). 
  
Surface brightness profiles centred on the X-ray peak were extracted from $3\farcs3$  bins in the [0.3--2]\,\keV\ band for each instrument independently, background subtracted, co-added and rebinned to $3\sigma$ per bin. 3D gas density profile were obtained using the regularised non-parametric method of direct deprojection and PSF deconvolution of the surface brightness profile developed by \citet{cro06}.
Global cluster parameters are  estimated self-consistently within $\Rv$ via iteration about the $\Mv$--$\YX$ relation of \citet{arn10}, assuming standard evolution,  
$$E(z)^{2/5}\Mv = 10^{14.567 \pm 0.010} \left[\frac{\YX}{2\times10^{14}\,{\msol}\,\keV}\right]^{0.561 \pm 0.018}\,\msol.$$ 
The quantity $\YX$, is defined as the product of $\Mgv$, the gas mass within $\Rv$, and $\TX$, the spectroscopic temperature measured in the [0.15--0.75]\,$\Rv$ aperture.  In addition, $L_{500}$, the X-ray luminosity inside $\Rv$, is calculated as described in \citet{pra09}. The SZ flux was then re-extracted, $\Yv$ being calculated with the X-ray position and size $\Rv$ fixed to the refined values derived from the high-quality \xmm\ observation. The X-ray properties of the clusters and resulting refined $\Yv$ values are listed in Table~\ref{tab:phys}. 

\section{\xmm\ validation outcome}

\subsection{False cluster candidates}

For the three candidates shown in Fig.~\ref{fig:false_gal}, no obvious extended X-ray sources were found within $5\arcmin$ of the \planck\ position. We followed the maximum likelihood procedure described by \citet{planck2011-5.1a} to find all extended sources in the field detected at the $\gtrsim 3 \sigma$ level.  We then assessed whether they could be the counterpart of the \planck\ candidate  from their position and X-ray flux, using the relation between the X-ray flux in the [0.1--2.4]\,\keV\ band, $F_{\rm X}$, and the SZ flux $\YSZ$ established by \citet{planck2012-I}:
\begin{equation}
\frac{F_{\rm X}/{\rm 10^{-12}\, erg\,s^{-1}\,cm^2}} {\YSZ/{\rm 10^{-3}\,arcmin^2}}=   4.95\, E(z)^{5/3}\, (1+z)^{-4}\, K(z).
\label{eq:SXY500}
\end{equation}
Here $K = K(z)$ is the $K$ correction, neglecting its temperature dependance. 

%==============================================================================================================================================
\begin{table*}[t]
\centering
\caption{{\footnotesize X-ray and SZ properties of the confirmed \Planck\ sources. Columns~(2) \& (3): Right ascension and declination of the peak of the X-ray emission (J2000). Column~(4): redshift from X-ray spectral fitting. Column~(5): Quality flag for the X-ray redshift measurement (see Sec.~\ref{sec:analysis})}. Column~(6): Total EPIC count rates in the [0.3-2]\,keV band, within the maximum radius of detection given in Column~(7). Columns~(8)--(14): $R_{500}$ is the radius corresponding to a density contrast of 500, estimated iteratively from the $M_{500}-Y_X$ relation, $Y_X = M_{g,500}T_X$ is the product of the gas mass within $R_{500}$ and the spectroscopic temperature $T_X$, and $M_{500}$ is the total mass within $R_{500}$.  $L_{500, [0.1-2.4]}$ is the luminosity within $R_{500}$ in the [0.1-2.4]\,keV band. $Y_{500}$ is the spherically integrated Compton parameter measured with Planck, centred on the X-ray peak, interior to the $R_{500}$ estimated with the X-ray observations.}
\label{tab:phys}
\resizebox{\textwidth}{!} {
\begin{tabular}{lrrllrrrrrcccc}
\toprule
\toprule
\multicolumn{1}{c}{Name} & 
\multicolumn{1}{c}{RA$_{\rm X}$} & \multicolumn{1}{c}{Dec$_{\rm X}$} & 
\multicolumn{1}{c}{$z_{\rm Fe}$} & \multicolumn{1}{c}{$Q_{\rm z}$} &
\multicolumn{1}{c}{$R$} & \multicolumn{1}{c}{$\theta_{\rm det}$} &
\multicolumn{1}{c}{$R_{500}$} & \multicolumn{1}{c}{$\TX$} &
\multicolumn{1}{c}{$M_{\rm gas,500}$} & \multicolumn{1}{c}{$\YX$} &
\multicolumn{1}{c}{$Y_{500}$} & \multicolumn{1}{c}{$M_{500}$} &
\multicolumn{1}{c}{$\LX$} \\
\noalign{\smallskip}
\multicolumn{1}{c}{} &
\multicolumn{1}{c}{[h:m:s]} &\multicolumn{1}{c}{[d:m:s]} & 
\multicolumn{1}{c}{} &
\multicolumn{1}{c}{} & \multicolumn{1}{c}{$[{\rm cts\,s^{-1}}]$} & \multicolumn{1}{c}{$[\arcmin]$} &
\multicolumn{1}{c}{[kpc]} & \multicolumn{1}{c}{[keV]} &
\multicolumn{1}{c}{$[10^{14}\,{\rm M_{\odot}}]$} & \multicolumn{1}{c}{$[10^{14}\,{\rm M_{\odot}\,\keV}]$} &
\multicolumn{1}{c}{$[10^{-4}\,{\rm arcmin^2}]$} & \multicolumn{1}{c}{$[10^{14}\,{\rm M_{\odot}}]$} &
\multicolumn{1}{c}{$[10^{44}\,{\rm erg\,s^{-1}}]$} \\
\midrule
PLCK\,G219.9$-$34.4&{04:54:45.4}&{$-$20:17:06.6}&0.66& 2			&$0.37\pm0.01$& 2.7&1048&$ 9.4\pm1.0$&$ 0.82\pm0.03$&$ 7.7\pm1.1$&$ 6.3\pm1.5$&$ 6.8\pm0.5$&$ 7.3\pm0.3$\\
PLCK\,G348.4$-$25.5&{19:24:56.1}&{$-$49:27:02.1}&0.25& 2			&$1.24\pm0.02$& 6.4&1020&$ 5.1\pm0.2$&$ 0.47\pm0.01$&$ 2.4\pm0.2$&$ 8.5\pm2.0$&$ 3.9\pm0.1$&$ 2.92\pm0.04$\\
PLCK\,G352.1$-$24.0&{19:20:59.7}&{$-$45:51:02.2}&0.77& 1$^\dagger$	&$0.33\pm0.01$& 2.0& 925 &$ 7.8\pm0.8$&$ 0.67\pm0.04$&$ 5.3\pm0.9$&$ 4.0\pm1.4$&$ 5.3\pm0.4$&$10.4\pm0.3$\\
PLCK\,G305.9$-$44.6&{00:23:38.9}&{$-$72:24:06.1}&0.30& 2			&$1.61\pm0.04$& 6.7&1178 &$ 7.4\pm0.6$&$ 0.79\pm0.02$&$ 5.8\pm0.6$&$8.3\pm2.4$&$ 6.3\pm0.4$&$ 5.9\pm0.1$\\
PLCK\,G208.6$-$74.4&{02:00:16.4}&{$-$24:54:54.4}&0.90& 1			&$0.38\pm0.02$& 3.2&1012&$11.5\pm2.4$&$ 1.02\pm0.08$&$12.\pm3.$&$5.6\pm1.4$&$ 8.1\pm1.3$&$12.6\pm0.6$\\
PLCK\,G130.1$-$17.0&{01:30:51.3}&{$+$45:17:54.9}&0.20& 2			&$1.25\pm0.02$& 6.3& 963&$ 4.2\pm0.2$&$ 0.37\pm0.01$&$ 1.6\pm0.1$&$8.8\pm2.1$&$ 3.1\pm0.1$&$ 2.04\pm0.03$\\
PLCK\,G239.9$-$40.0&{04:46:47.2}&{$-$37:03:49.7}&0.74& 1$^\ast$ 		&$0.56\pm0.01$& 4.4&1033 &$ 8.4\pm0.6$&$ 1.04\pm0.03$&$ 8.7\pm0.9$&$4.4\pm1.2$&$ 7.1\pm0.4$&$11.8\pm0.2$\\
PLCK\,G204.7$+$15.9&{07:34:27.3}&{$+$14:16:50.2}&0.34& 2			&$0.89\pm0.01$& 4.6&1121 &$ 7.1\pm0.4$&$ 0.68\pm0.02$&$ 4.9\pm0.4$&$7.8\pm2.0$&$ 5.7\pm0.2$&$ 4.20\pm0.06$\\
PLCK\,G011.2$-$40.4&{21:00:37.6}&{$-$33:08:05.7}&0.46& 2			&$0.18\pm0.01$& 2.9& 833 &$ 4.8\pm0.5$&$ 0.28\pm0.02$&$ 1.4\pm0.2$&$2.9\pm1.4$&$ 2.7\pm0.2$&$ 1.78\pm0.06$\\
PLCK\,G147.3$-$16.6&{02:56:25.3}&{$+$40:17:18.7}&0.62& 1$^\clubsuit$	&$0.34\pm0.01$& 2.4&1042 &$ 8.8\pm0.8$&$ 0.76\pm0.03$&$ 6.7\pm0.9$&$5.2\pm1.7$&$ 6.3\pm0.4$&$ 7.15\pm0.66$\\
\midrule
PLCK\,G329.5$-$22.7~A&{18:33:00.3}&{$-$65:33:20.0}&0.24& 2			&$0.91\pm0.02$& 4.9& 917&$ 4.4\pm0.2$&$ 0.30\pm0.01$&$ 1.3\pm0.1$&\ldots &$2.8\pm0.1$&$ 2.16\pm0.04$\\
PLCK\,G329.5$-$22.7~B&{18:33:33.7}&{$-$65:26:39.2}&0.46& 2			&$0.25\pm0.01$& 2.6& 872&$ 4.9\pm0.7$&$ 0.35\pm0.02$&$ 1.7\pm0.3$&\ldots &$3.1\pm0.3$&$ 2.6\pm0.1$\\
PLCK\,G196.7$-$45.5~A&{03:42:54.2}&{$-$08:40:58.2}&0.57& 1$^\ddagger$&$0.29\pm0.02$& 5.6& 820&$ 4.4\pm0.4$&$ 0.37\pm0.02$&$ 1.7\pm0.2$&\ldots &$2.9\pm0.2$&$ 2.8\pm0.1$\\
PLCK\,G196.7$-$45.5~B&{03:43:02.4}&{$-$08:46:09.8}&0.42& 1$^\ddagger$&$0.15\pm0.01$& 2.4& 826&$ 4.8\pm0.8$&$ 0.24\pm0.02$&$ 1.2\pm0.3$&\ldots &$2.5\pm0.3$&$ 1.30\pm0.07$\\
\bottomrule
\end{tabular}
}
\tablefoot{ Other possible $z_{\rm Fe}$: $^\dagger$ $z_{\rm Fe}=0.12, 0.40$; the former solution is excluded from the X-ray versus SZ properties and the latter is unlikely (see Sec~\ref{ap:zx}); $^\ast$ $z_{\rm Fe}=0.26,0.46$. The $z_{\rm Fe}=0.26$ is unlikely in view of the X-ray versus SZ properties (see Sec~\ref{ap:zx}); $^\clubsuit$ $z_{\rm Fe}=0.40,1.03$. The given solution, $z_{\rm Fe}=0.62$ is that consistent with the optical redshift $z_{\rm spec}=0.66\pm0.05$ (Sec.~\ref{ap:zopt}).   $^\ddagger$  $z_{\rm Fe}= 0.87$ (excluded from DSS red image, see Sec.~\ref{ap:zx}) and  $z_{\rm Fe}= 0.10$ for the A and B components, respectively.}
\\
\normalsize
\end{table*}
%==============================================================================================================================================

\subsubsection{PLCK\,G196.4$-$68.3 and PLCK\,G310.5$+$27.1}

PLCK\,G196.4$-$68.3 was classified as PHZ (potentially at high $z$). Analysis of the \xmm\ data on PLCK\,G196.4$-$68.3 revealed two extended sources at 9\parcm9 and 11\parcm8 from the SZ position. The former corresponds to a \rass-FSC source.  Both sources are too far away to be the X-ray counterpart of the \planck\ candidate.  A \rass-FSC source is located at 5\parcm2 from the SZ position  and likely contributes to the ${\rm S/N}=1.7$ signal derived from  \rass\  data at the \planck\ source location. However, the comparison of its surface brightness profile with the \xmm\ PSF shows that it is consistent with a point source. We thus conclude that PLCK\,G196.4$-$68.3 is a false detection.

PLCK\,G310.5$+$27.1 was also classified as PHZ. Two extended X-ray sources were detected at 10\parcm5 and  2\parcm5  from the SZ position, respectively. The former is too far away to be the X-ray counterpart, while the latter is very weak. Analysis of the surface brightness profile confirmed that  it is extended. The detection radius is small, $\theta_{\rm det} = 0\parcm44$ and the spectrum extracted from this region is too poor to put robust constraints on the redshift or the temperature. However, using the $F_{\rm X}$--$\YSZ$ relation (Eq.~\ref{eq:SXY500}) and the measured X-ray flux, we can put an upper limit on $\YSZ$ assuming a redshift as high as $z = 2$ and taking into account a factor of two dispersion around the relation.
For a temperature of $\kT=4\,\keV$ and $z = 2$, we derive a flux within the detection radius of  $F_{\rm X} = 2.8\times10^{-14}$\,erg\,s\mo\,cm$^{-2}$.  Assuming that this flux is close to the total, this gives an upper limit on the SZ flux of $\YSZ\sim 9\times 10^{-5}\,\mathrm{arcmin^2}$, nearly an order of magnitude smaller than the \planck\ value  $\YSZ\sim 6.7\pm1.5\times 10^{-4}\,\mathrm{arcmin^2}$.  Moreover, the SZ significance drops under 2$\sigma$ when the flux is re-extracted at the X-ray position. We conclude that this candidate is also a false detection.

Both of these false candidates were detected by two methods, with a medium quality grade of $Q_{\rm SZ}={\rm B}$ and at ${\rm S/N}=4.7$ and ${\rm S/N}=4.8$, respectively.  A $Q_{\rm SZ}={\rm B}$ quality grade is thus not sufficient to ensure candidate validity at these signal-to-noise ratios.  On the other hand, all $Q_{\rm SZ}={\rm A}$ candidates down to ${\rm S/N}=4.6$ that have been followed up by \xmm\ have been confirmed.

\subsubsection{PLCK\,G210.6$+$20.4}

PLCK\,G210.6$+$20.4 is associated with an SDSS cluster. The SDSS search algorithm identified a galaxy over-density of 77 members at a photometric redshift of $z\sim0.57$, consistent with  the spectroscopic redshift of the brightest cluster galaxy (BCG) at  $z=0.52$. The barycentre of the concentration and  the BCG are located 1\parcm5 and 5\arcm\ from  the \planck\ position (see Fig.~\ref{fig:false_gal}), respectively. The X-ray analysis revealed the presence of an extended source, centred on the BCG, detected at 3.3$\sigma$ in the [0.3--2]\,keV image. However, the source is very faint and more reminiscent of a group of galaxies than of a rich cluster. This is confirmed by the X-ray spectroscopic analysis.  Extracting and fitting the spectrum with an absorbed thermal model at $z=0.52$, we measured a temperature within the detection radius $\theta_{\rm det} = 0\parcm77$ of $T_{\rm R_{\rm det}} = 1.5\pm0.5\,\keV$ and  a flux of $F_{\rm X} = 2.31\times 10^{-14}\,\ergscm$.   Using Eq.~\ref{eq:SXY500} as above, the upper limit on the corresponding SZ flux is $\Yv \sim 2.6\times 10^{-5}\, \mathrm{arcmin^2}$, more than 10 times lower than the \planck\ value of $4.9\pm1.2\times 10^{-4}\, \mathrm{arcmin^2}$.  The X-ray source is too weak to be the \planck\ counterpart and we conclude that the candidate is not a cluster. 

In the previous \xmm\ validation run, the two candidates potentially associated with $z>0.5$ SDSS clusters were confirmed, including PLCK\,G193.3$-$46.1 at $z\sim0.6$. This showed that  SDSS can robustly confirm candidates up to such high $z$.  It  is instructive to compare PLCK\,G210.6$+$20.4 with  PLCK\,G193.3$-$46.1.  In both cases the search algorithm found a rich concentration of galaxies, as expected for \planck\ clusters. The masses, reconstructed from the luminosity function, are $\sim 3\times 10^{14}\,\msol$ and  $9\times 10^{14}\,\msol$, respectively, i.e., the false candidate has a larger mass. In both cases, the galaxy distribution appears rather loose (compare Fig.~\ref{fig:false_gal} right panel and \citealt[][Fig.~5]{planck2012-I}).   The \xmm\ observation revealed that PLCK\,G193.3$-$46.1 is a double peaked cluster, i.e., a dynamically perturbed cluster with an ICM distribution consistent with the galaxy morphology.  In view of the \xmm\ image, the galaxy concentration at the location of PLCK\,G210.6$+$20.4 is likely a filamentary structure where only the part around the BCG is virialised and contains gas that is hot enough to emit in X-rays. This would also explain the large offset between the BCG position and the galaxy concentration barycentre, which is much larger than in the case of  PLCK\,G193.3$-$46.1.  These two cases illustrate the difficulty of distinguishing between massive clusters and  pre-virialised structures with rather shallow SDSS data at high $z$.  Beyond  luminosity and mass estimates, important diagnostics include the offset between the SZ, BCG, and  barycentre, as well as  the galaxy distribution morphology, if available,  and other ancillary data, such as significant \rass\ emission.  These factors must all be considered for firm confirmation of low signal-to-noise-ratio SZ detections.  On the other hand, we cannot be sure that the apparent SZ signal is purely due to noise, and cannot exclude a contribution from the pre-virialised structure itself, especially if it corresponds to a warm  filament along the line of sight.  

\subsection{Confirmed candidates}

Twelve of the 15 candidates are confirmed as real clusters, of which two are double systems as shown in Fig.~\ref{fig:confirmed_gal}. Physical parameters are given in Table~\ref{tab:phys}. For the two double systems, the cluster closest to the \planck\ position is labelled A and the other is labelled B in Table~\ref{tab:phys}.

\subsubsection{Single clusters}

The redshifts of eight clusters are well constrained by the \xmm\ spectrum (quality flag  of $Q_{\rm z}\!\!=\!\!2$). Three of these clusters,  PLCK\,G219.9$-$34.4, PLCK\,G011.2$-$40.4  and PLCK\,G348.4$-$25.5  were classified as \hbox{PHZ}. The first two are indeed at $z=0.46$ and $z=0.66$, respectively, but PLCK\,G348.4$-$25.5 is at  $z=0.25$.  Knowing the precise cluster location with \xmm, we re-examined the DSS image.  A bright galaxy is indeed located exactly at the position of the X-ray peak; however, the field is crowded and there is no obvious galaxy concentration around that \hbox{BCG}.  This explains our initial mis-classification.  
%==============================================================================================================================================
 \begin{figure}[btp]
\centering
\includegraphics[width=0.9\columnwidth, clip]{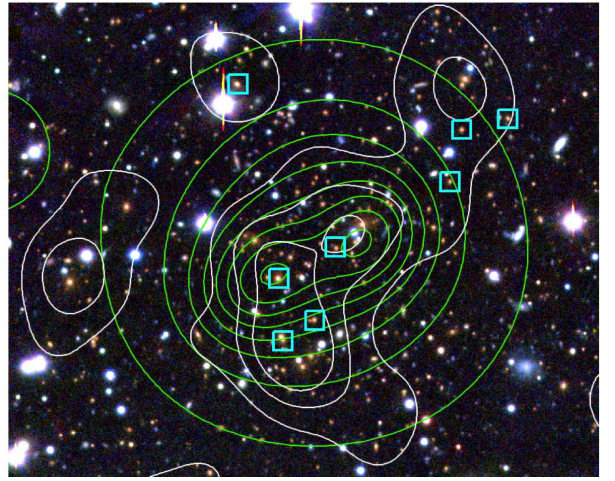}
\caption{ {\footnotesize A $gri$ composite image of the central $5\farcm 5 \times 3\farcm 4$ of PLCK G147.3$-$16.6, based on imaging data from NOT/MOSCA ($g$ and $i$) and TNG/DOLORES ($r$ and $i$).  Boxes: cluster galaxies spectroscopically confirmed  with Gemini (excluding the two galaxies at $z=0.68$).  
North is up and East to the left. The green contours are isocontours of 
the wavelet filtered \xmm\  image. The white contours show the 
luminosity distribution of the red sequence galaxies indicated by red symbols   
in Fig.~\ref{fig:plckg147rs}, smoothed with a $\sigma = 14\arcsec$ Gaussian filter. 
The plotted contour levels are at (10,20,30) times the rms variation in the 
luminosity distribution. 
}}
\label{fig:plckg147im}
\end{figure}
%==============================================================================================================================================

The redshift determination for three single clusters is more uncertain. There are several $\chi^2$ minima that cannot be distinguished at the $68\,\%$ confidence level ($Q_{\rm z}\!\!=\!\!1$).  As proposed by \citet{planck2012-I}, we  used the X-ray versus SZ properties to eliminate unphysical solutions, as well as DSS data. This is  detailed in Appendix~\ref{ap:zx}.  The \xmm\  analysis gives three possible redshifts for  PLCK\,G147.3$-$16.6:  0.4, 0.62, and 1.1, the last being the best-fitting value.  The cluster has an interesting double-peaked morphology. It is likely an on-going merger of two nearly equal mass systems (Fig.~\ref{fig:plckg147im}). The analysis of imaging data obtained with the Telescopio Nazionale Galileo La Palma (TNG) telescope and the Nordic Optical Telescope, as well as spectroscopic data obtained at Gemini, are detailed in Appendix~\ref{ap:zopt}. We confirm a redshift of $z=0.66 \pm 0.05$.

The spectral analysis of PLCK\,G208.6$-$74.4 gives a single $\chi^2$ minimum at $z=0.9\pm 0.04$,  in very good agreement with SZ versus X-ray properties. However we assign a quality flag of  $Q_{\rm z}\!\!=\!\!1$ since the statistical quality of  the spectrum  is poor due to the short exposure time.  Furthermore the DSS image is ambiguous: although there is no visible galaxy at the X-ray maximum, the centroid of the large scale X-ray emission  is close to a bright DSS galaxy.  

In summary, of the seven candidates we classified  as PHZ, two are false, four are indeed at $z\gtrsim 0.5$, and one is at a low redshift of $z=0.25$.  In addition to those clusters which were classified as PHZ, two further $Q_{\rm z}\!\!=\!\!1$ clusters, PLCK\,G239.9$-$40.0  and  PLCK\,G208.6$-$74.4, are most likely at high $z$. 

\subsubsection{Multiple systems}

In PLCK\,G196.7$-$45.5, two clusters, separated by $\approx 5.5$ arcmin, lie within the \planck\ position error box:  PLCK\,G196.7$-$45.5A at 2\parcm34 and PLCK\,G196.7$-$45.5\,B at 3\parcm9 from the SZ position.  In view of the \planck\ resolution, $5\arcmin$ to $30\arcmin$ depending on frequency \citep{planck2011-1.4,planck2011-1.5}, both clusters certainly contribute to the SZ signal.  It is likely a chance association, although given the uncertainty in the redshifts, a binary system cannot be ruled out (see Appendix~\ref{ap:zx}). 
  
  %==============================================================================================================================================
\begin{figure}[btp]
\centering
\includegraphics[width=0.8\columnwidth]{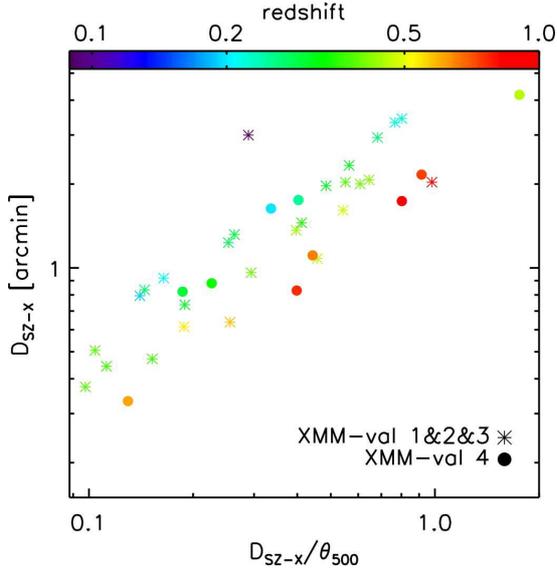} 
\caption{{\footnotesize Distance of blind SZ position to X-ray position, $D_{\rm SZ-X}$, as a function of $D_{\rm SZ-X}$, normalised to the cluster size $\theta_{500,X}$ for single confirmed systems. The clusters are colour-coded according to redshift.}}
\label{fig:dist}
\end{figure}
%==============================================================================================================================================
%==============================================================================================================================================
\begin{figure}[t]
 \centering
\includegraphics[width=0.8\columnwidth]{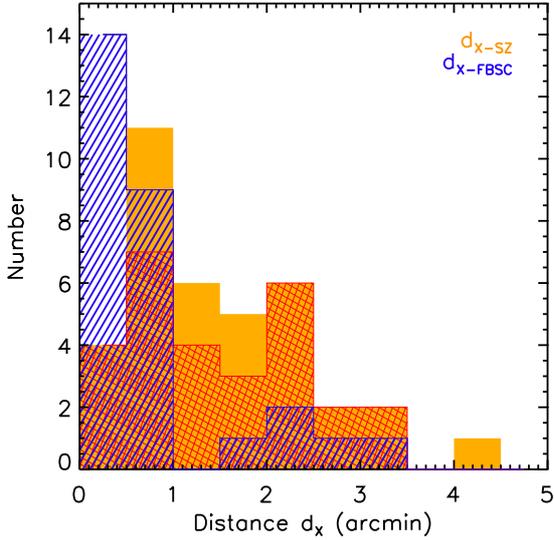}
\caption{{\footnotesize Histogram of the distance between the X-ray peak determined from the \xmm\ validation observations and the  \planck\ SZ position for all clusters (orange filled) and those associated with a source from the \rass\ Faint Source Catalogue  or  Bright Source Catalogue (red hatched). The histogram of the distance between the X-ray peak and  the \rass\  source  position is plotted for comparison (blue hatched). }}
\label{fig:disthisto}
\end{figure}
%==============================================================================================================================================

In PLCK\,G329.5$-$22.7, the cluster PLCK\,G329.5$-$22.7\,A lies about 1\arcm\ from the \planck\ position, while the second object is about 8\arcm\ away. From the $\YX$ values and redshift estimates, cluster B is expected to have a $\YSZ$ flux 1.8 times smaller than that of cluster A, thus contributing $36\%$ to the total flux. Its contribution to the blind signal may differ, as the blind signal is extracted using a single component model found roughly peaked at cluster A. Indeed comparison of such a single component extraction with that using a double component model (with flux ratio fixed to the X-ray constraint)  suggests a contamination from cluster B of about $\sim20\%$. In summary, PLCK\,G329.5$-$22.7\,A  is the main contributor to the SZ detection,  although PLCK\,G329.5$-$22.7\,B certainly contributes. The redshifts of the two clusters are well determined,  $z=0.24$ and $z=0.46$, respectively, showing that they are not physically related. This double system is thus a chance association on the sky.  
 
Overall, we have found four double systems and two triple systems among the 43 \planck\ candidates confirmed by \xmm, i.e., 14\,\% multiple systems.  Since the \xmm\ validation follow-up observations are neither representative nor complete, this fraction of multiple systems cannot be extrapolated to the population at large; however, it is more than five times larger than the fraction of cluster pairs separated by less than 10\arcm\ ($63/1882$ objects) in the whole MCXC X-ray catalogue compilation \citep{pif11}. This is clearly a selection effect due to confusion in the large \planck\ beam, which it might  be necessary to take into account for a precise  estimate of the selection function. 
 %==============================================================================================================================================
 \begin{figure}[btp]
 \centering
\includegraphics[width=0.8\columnwidth]{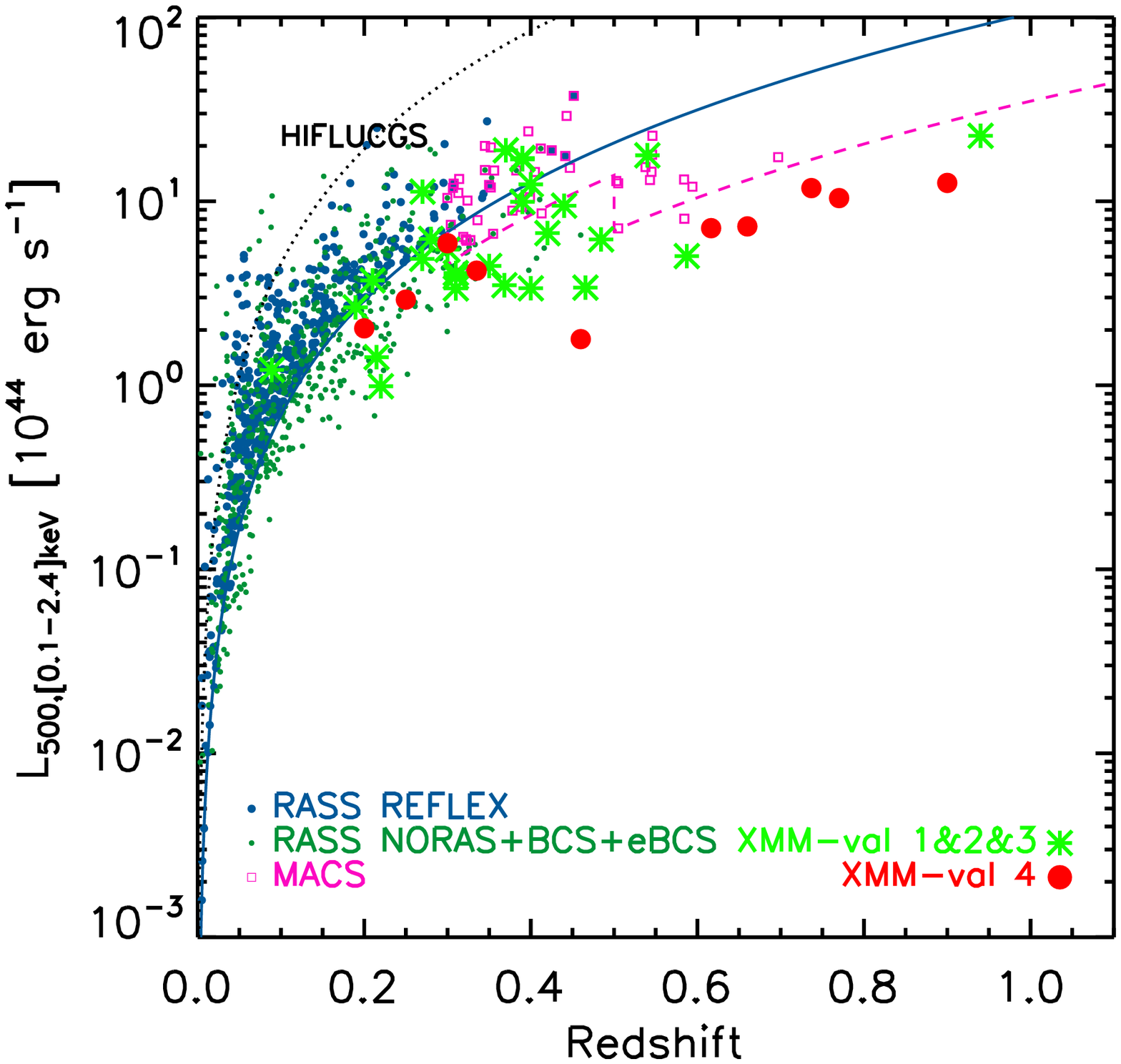}
\caption{ {\footnotesize The new SZ-discovered \planck\ single objects compared to clusters from the \rosat\  All-Sky Survey catalogues in the \Lxz\ plane. Green points represent \planck\ clusters previously confirmed with \xmm\  \citep{planck2011-5.1b, planck2012-I}  and red points are the newly confirmed single clusters.
 The X-ray luminosity  is calculated  in the [0.1--2.4]\,\keV\ band. Catalogues shown are \reflex\ \citep{boe04}, \noras\ \citep{boe00}, \bcs\ \citep{ebe98}, \ebcs\ \citep{ebe00} and \macs\  \citep{ebe07}. The solid line is the REFLEX flux limit, the dotted line is the HIFLUCGS flux limit of $2 \times 10^{-11}\,\ergscm$ and the dashed line is from  the \macs\ flux limits. }}
\label{fig:zlx}
\end{figure}
%==============================================================================================================================================

\subsection{\planck\  position reconstruction uncertainty}
\label{sec:pos}
The  \planck\ position reconstruction uncertainty is  driven by the spatial resolution of the instruments. The  positions determined by the \planck\ detection algorithm  are compared to the precise \xmm\ positions in Fig.~\ref{fig:dist} and Fig.~\ref{fig:disthisto},  where we put together all validation observations of single systems. The mean offset between  the \planck\ and the \xmm\  position is $1\farcm5$, with a median value of $1\farcm3$, as  expected from \planck\ sky  simulations \citep[][Fig.~7 left]{planck2011-5.1a}.  For 70 and 86\,\% of the clusters, this offset is less than $2\arcmin$ and $2\farcm5$, respectively. The assumed positional uncertainty of up to $5\arcmin$ is certainly conservative and  an offset of $5\arcmin$ is actually very unlikely. This needs to be taken into account when searching for possible counterparts in ancillary data or follow-up observations.

The offsets of five sources are greater than $2\farcm5$. Three of those objects are very diffuse,  likely dynamically  unrelaxed systems,  at relatively low $z$, including the prominent outlier PLCK\,G18.7+23.6 at $z=0.09$ (Fig.~\ref{fig:dist}, purple point). As noted by \citet{planck2011-5.1b} a real, physical  offset between the X-ray and SZ peak may contribute to the overall offset for this type of cluster. In all cases but one, the offset remains smaller than the cluster size $\Rv$.  The notable exception is PLCK\,G11.2$-$40.4 (Fig.~\ref{fig:dist}). The \xmm\ position of this cluster  is  $4\farcm2$ or $\sim\,1.8\,\Rv$ from the \planck\ position.  The peak in the SZ reconstructed map is also $\sim 3\arcmin$ away from from the \planck\ position.   This cluster is detected by only one method and has a low quality grade $Q_{\rm SZ}={\rm C}$, being located in a particularly noisy region of the \planck\ map. This is likely to complicate the estimate of the cluster position. 

Finally, we note that the position reconstruction uncertainty is on average smaller than for the ESZ sample that peaks at $\sim2\arcmin$ \citep[][Fig.~7 right]{planck2011-5.1a}.  This is likely the result of the higher redshift range considered here. Indeed, at this redshift the sources are more compact and their position is easier to reconstruct. Furthermore, possible physical offsets are expected to become negligible as they become unresolved. 
%==============================================================================================================================================
\begin{figure}[btp]
\centering
\includegraphics[width=0.8\columnwidth]{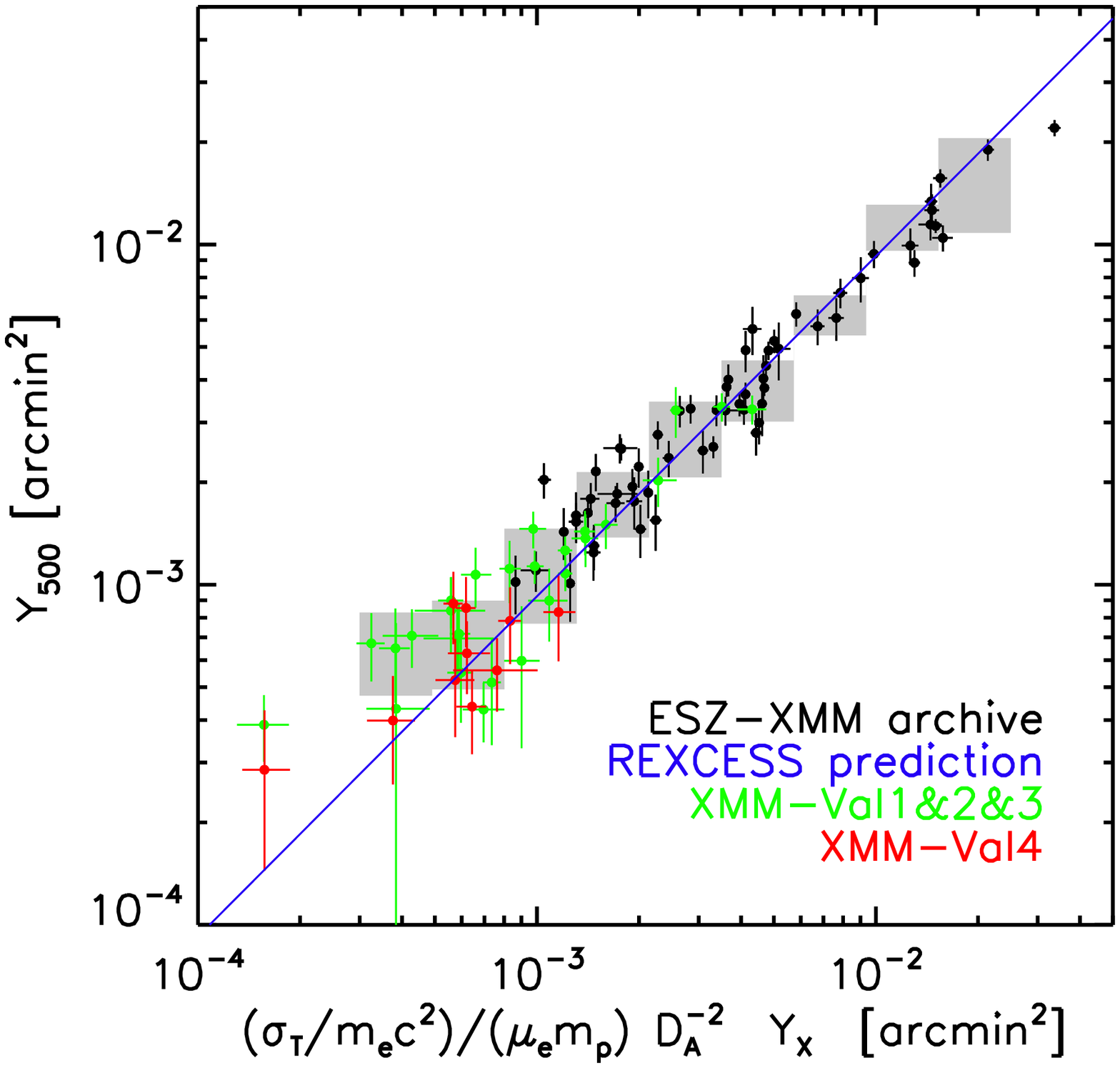}
\caption{{\footnotesize 
 Relation between apparent SZ signal ($\YSZ$) and the corresponding normalised $\YX$ parameter for single systems confirmed with \xmm\ (green and red points).   Black points show clusters in the  \planck-ESZ sample with \xmm\ archival data as presented in \citet{planck2011-5.2b}; The blue lines denote the $\YSZ$ scaling relations predicted from the \rexcess\ X-ray observations \citep{arn10}. The grey area corresponds to median $\YSZ$ values in $\YX$ bins with $\pm1\sigma$ standard deviation. } 
\label{fig:yxsz}}
\end{figure}
%==============================================================================================================================================

 \subsection{New clusters in the $z$--$L_{\rm X}$ and $z$--$\Mv$ plane and {\textit Planck} sensitivity}
 \label{sec:yszx}
  
  %%%%%%%%%%%%%%%%%%%%%%%%%%%%%%%%%%%%%%%%%%%%%%%%%%%%%%%%%%%%%%%%%%%%%%%%%%%%%

The present validation sample covers a wide range of redshift, $0.2<z<0.9$, and SZ flux, $2.9\times 10^{-4}\,\arcms< \YSZ<8.8\times 10^{-4}\,\arcms$. As expected from the lower signal-to-noise ratio considered and the deeper sky coverage  (Sec.~\ref{sec:sample}),  the $\YSZ$ range is lower than that of the previous validation sample, $4\times 10^{-4}\,\arcms < \YSZ < 1.4\times 10^{-3}\,\arcms$. Although not perfect, the strategy to preferentially select high-$z$ clusters was successful, with five clusters found at $z>0.5$, including three PHZ candidates.  The full \xmm\  validation sample (single objects only) is shown in the $L_{\rm X}$-$z$ plane in Fig.~\ref{fig:zlx}.  We continue to populate the higher $z$ part of the $L_{\rm X}$-$z$ plane and confirm \planck\ can detect clusters well below the X-ray flux limit of \rass-based catalogues,  ten times lower than \reflex\ at high $z$, and below the limit of the most sensitive \rass\ survey (\macs).  The figure makes obvious the gain in redshift coverage as compared to the \rass-based catalogues.

%==============================================================================================================================================
  \begin{figure}[btp]
 \centering
\includegraphics[width=0.8\columnwidth, clip]{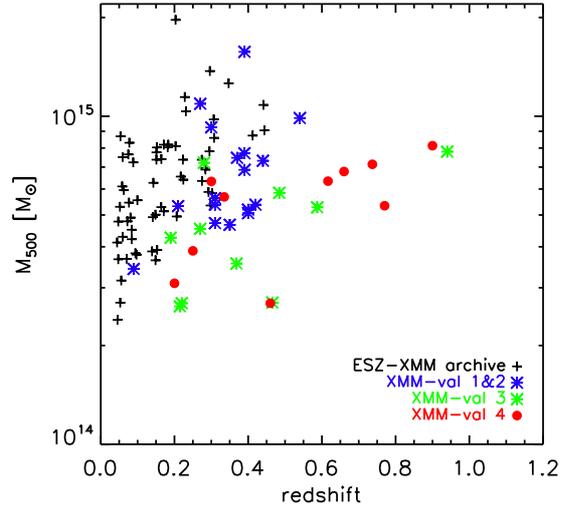}
\caption{ {\footnotesize The new SZ-discovered \planck\ single objects (blue, red and green symbols) in the $z$--$\Mv$ plane. For comparison, black points show known clusters from the ESZ \planck\ catalogue with archival \xmm\  data \citep{planck2011-5.2b}. $\Mv$ are estimated from $\YX$ and the  \MYX\ relation of \citet{arn10}.}} 
\label{fig:zM}
\end{figure}
%==============================================================================================================================================

%==============================================================================================================================================
\begin{table}[tbp]
\caption{{\footnotesize \rass\  information for single confirmed clusters and false candidates.  (1)~Name of the candidate. (2)~signal-to-noise ratio of the \rass\ count rate in the [0.5--2]\,\keV\ band, measured within a region of $5\arcmin$ radius  centred on the SZ candidate position.  (3)~Flux in the  [0.1--2.4]\,\keV\ band as measured with \xmm\  within $\theta_{500}$.  (4)~association with a source from the \rass\ Faint Source Catalogue (F) or Bright Source catalogue (B) published by \citet{vog99,vog00}.  (5)~number of the \xmm\  validation run.  (6)~Confirmed clusters are flagged}.}
\resizebox{\columnwidth}{!} {
\begin{tabular}{l r c l c c}
\toprule
\toprule
\multicolumn{1}{c}{Name} &
\multicolumn{1}{c}{S/N} & 
\multicolumn{1}{c}{$S_{500,\rm {XMM}}$} & 
\multicolumn{1}{c}{Ass.} & 
\multicolumn{1}{c}{Run} & 
\multicolumn{1}{c}{Confirmed} \\
\noalign{\smallskip}
\multicolumn{1}{c}{} &
\multicolumn{1}{c}{\rass} &
\multicolumn{1}{c}{$10^{-12}\,\ergscm$}\\
\midrule
PLCK\,G271.2$-$31.0  &  18.4   & 4.82$\pm$ 0.03  &    B   &2   & Y \\
PLCK\,G286.6$-$31.3  &  10.7   & 3.23$\pm$ 0.07  &    B   &2   & Y \\
PLCK\,G018.7+23.6  &   8.8   & 6.35$\pm$ 0.09  &    B   &2   & Y \\
PLCK\,G305.9$-$44.6  &   7.9   & 2.38$\pm$ 0.05  &    B   &4   & Y \\
PLCK\,G234.2$-$20.5  &   7.2   & 2.45$\pm$ 0.02  &    B   &3   & Y \\
PLCK\,G285.0$-$23.7  &   7.1   & 3.80$\pm$ 0.04  &    B   &2   & Y \\
PLCK\,G060.1+15.6  &   7.0   & 2.86$\pm$ 0.06  &    B   &3   & Y \\
PLCK\,G268.5$-$28.1  &   6.6   & 0.48$\pm$ 0.02  &    F   &3   & Y \\
PLCK\,G171.9$-$40.7  &   6.1   & 5.78$\pm$ 0.06  &    B   &2   & Y \\
PLCK\,G266.6$-$27.3  &   5.6   & 0.84$\pm$ 0.02  &    F   &3   & Y \\
PLCK\,G241.2$-$28.7  &   5.1   & 1.28$\pm$ 0.02  &    B   &2   & Y \\
PLCK\,G019.1+31.2  &   5.1   & 2.93$\pm$ 0.04  &    B   &3   & Y \\
PLCK\,G277.8$-$51.7  &   4.9   & 1.63$\pm$ 0.03  &    F   &1   & Y \\
PLCK\,G208.6$-$74.4  &   4.3   & 0.45$\pm$ 0.02  &    F   &4   & Y \\
PLCK\,G250.0+24.1  &   4.2   & 0.73$\pm$ 0.04  &    F   &1   & Y \\
PLCK\,G286.3$-$38.4  &   4.1   & 1.51$\pm$ 0.04  &    B   &1   & Y \\
PLCK\,G285.6$-$17.2  &   3.8   & 1.24$\pm$ 0.02  &    F   &2   & Y \\
PLCK\,G130.1$-$17.0  &   3.7   & 1.92$\pm$ 0.03  &        &4   & Y \\
PLCK\,G200.9$-$28.2  &   3.7   & 0.77$\pm$ 0.03  &    F   &3   & Y \\
PLCK\,G235.6+23.3  &   3.6   & 0.86$\pm$ 0.02  &    F   &3   & Y \\
PLCK\,G262.2+34.5  &   3.5   & 1.15$\pm$ 0.02  &    F   &3   & Y \\
PLCK\,G004.5$-$19.5  &   3.3   & 2.00$\pm$ 0.03  &    B   &1   & Y \\
PLCK\,G272.9+48.8  &   3.2   & 2.60$\pm$ 0.10  &    F   &2   & Y \\
PLCK\,G205.0$-$63.0  &   3.0   & 1.44$\pm$ 0.02  &    F   &2   & Y \\
PLCK\,G348.4$-$25.5  &   2.9   & 1.72$\pm$ 0.02  &    F   &4   & Y \\
PLCK\,G292.5+22.0  &   2.8   & 2.22$\pm$ 0.04  &        &2   & Y \\
PLCK\,G100.2$-$30.4  &   2.8   & 1.27$\pm$ 0.03  &    F   &2   & Y \\
PLCK\,G226.1$-$16.9   &  2.3   & \ldots   &   F   &1   & \ldots \\
PLCK\,G193.3$-$46.1  &   2.2   & 0.45$\pm$ 0.01  &    F$^\dagger$  &3   & Y \\
PLCK\,G204.7+15.9  &   2.0   & 1.32$\pm$ 0.02  &        &4   & Y \\
PLCK\,G287.0+32.9  &   1.9   & 4.01$\pm$ 0.05  &        &2   & Y \\
PLCK\,G147.3$-$16.6  &   1.8   & 0.59$\pm$ 0.06  &    F   &4   & Y \\
PLCK\,G011.2$-$40.4  &   1.8   & 0.27$\pm$ 0.01  &        &4   & Y \\
PLCK\,G210.6+17.1  &   1.7   & 0.86$\pm$ 0.01  &        &3   & Y \\
PLCK\,G196.4$-$68.3   &  1.7   & \ldots   &       &4   & \ldots \\
PLCK\,G070.8$-$21.5   &  1.6   & \ldots   &       &1   & \ldots \\
PLCK\,G262.7$-$40.9  &   1.3   & 2.26$\pm$ 0.02  &    F   &2   & Y \\
PLCK\,G113.1$-$74.4   &  1.2   & \ldots   &   F   &3   & \ldots \\
PLCK\,G343.4$-$43.4   &  1.2   & \ldots   &   F   &1   & \ldots \\
PLCK\,G239.9$-$40.0  &   1.0   & 0.66$\pm$ 0.01  &    F   &4   & Y \\
PLCK\,G352.1$-$24.0  &   0.7   & 0.52$\pm$ 0.01  &        &4   & Y \\
PLCK\,G219.9$-$34.4  &   0.7   & 0.53$\pm$ 0.02  &        &4   & Y \\
PLCK\,G317.4$-$54.1   & $-$0.4   & \ldots   &       &1   & \ldots \\
PLCK\,G310.5+27.1   & $-$0.9   & \ldots   &       &4   & \ldots \\
PLCK\,G210.6+20.4   & $-$1.0   & \ldots   &       &4   & \ldots \\
\bottomrule
\end{tabular} 
}
\tablefoot{ $^\dagger$: The FSC source is not the cluster.  }
\normalsize
\label{tab:RASS}
\end{table}
%===========================================================================================

We confirm our previous results on the \YSZYX\ relation. Most clusters are consistent with the \rexcess\ prediction:
\begin{equation}
\YSZ  =    0.924\,D_{\rm A}^{-2}\,C_{\rm XSZ}\,\YX
\label{eq:yszx}
\end{equation}
with $C_{\rm XSZ} =1.416 \times 10^{-19}\,{\rm Mpc}^2\,\msol^{-1}\, \keV$.
However, all clusters below a normalised  $\YX\sim5\times 10^{-4}\,{\rm arcmin}^2$ lie above the predicted $\YSZ$--$\YX$ relation and the bin average deviation increases with decreasing $\YX$ (Fig.~\ref{fig:yxsz}).  As noted by \citet{planck2012-I}, this is an indication of Malmquist bias.

Figure~\ref{fig:zM} shows the new  \planck\ clusters confirmed with \xmm\ in the $z$--$\Mv$ plane (single objects only). The minimum mass increases with redshift,  an indication of an increase of the mass detection threshold with $z$. Such an increase is expected from the fact that clusters are not resolved by \planck\ at high $z$; however, we clearly confirm that \planck\ can detect $\Mv > 5 \times 10^{14}\,\msol$ clusters above $z>0.5$. 
Two clear outliers in the $z-M$ plane are evident in Fig.~\ref{fig:zM}. They correspond to the lowest flux clusters PLCK\,G11.2$-$40.4 and  PLCK\,G268.5$-$28.1 at $z=0.46$ and $z=0.47$, respectively (Fig.~\ref{fig:yxsz}), lying in the region most affected by the Malmquist bias.
PLCK\,G11.2$-$40.4 is the cluster mentioned in Sec.~\ref{sec:pos}, which is detected with a large offset between the \planck\ position and the X-ray peak, due to its lying in a region with a noisy background. The blind signal is two times higher than the signal extracted at the X-ray position. This is a clear case of a detection boosted by specific local noise conditions.

\section{Using \rass\ data in  the  construction of the \textit{Planck} cluster catalogue}
\label{sec:rass}

\subsection{Position refinement}

The positions of the associated FSC and BSC source are indicated in the individual \xmm\  image of each candidate in Fig.~\ref{fig:confirmed_gal}, and for previous observations, in Fig.~3 and Fig.~2 published in \citet{planck2011-5.1b} and \citet{planck2012-I}.
Comparing the positions of the SZ candidates and their FSC/BSC counterparts with the X-ray peaks determined from the \xmm\ validation observations, we notice that the FSC/BSC position is a better estimate of the position of the cluster than the position returned by \planck\ alone. Most of the FSC/BSC sources are located within 1\arcmin\ of the \xmm\ position versus 2\arcmin\ for the \planck-SZ position (see Fig.~\ref{fig:disthisto}). Thus, the association with a faint or bright \rass\ source can be used to refine the SZ position estimate.

\subsection{ X-ray flux estimate}
\label{sec:sxcomp} 

\begin{figure*}[t]
\centering
\begin{minipage}[t]{0.9\textwidth}
\resizebox{\hsize}{!} {
\includegraphics[]{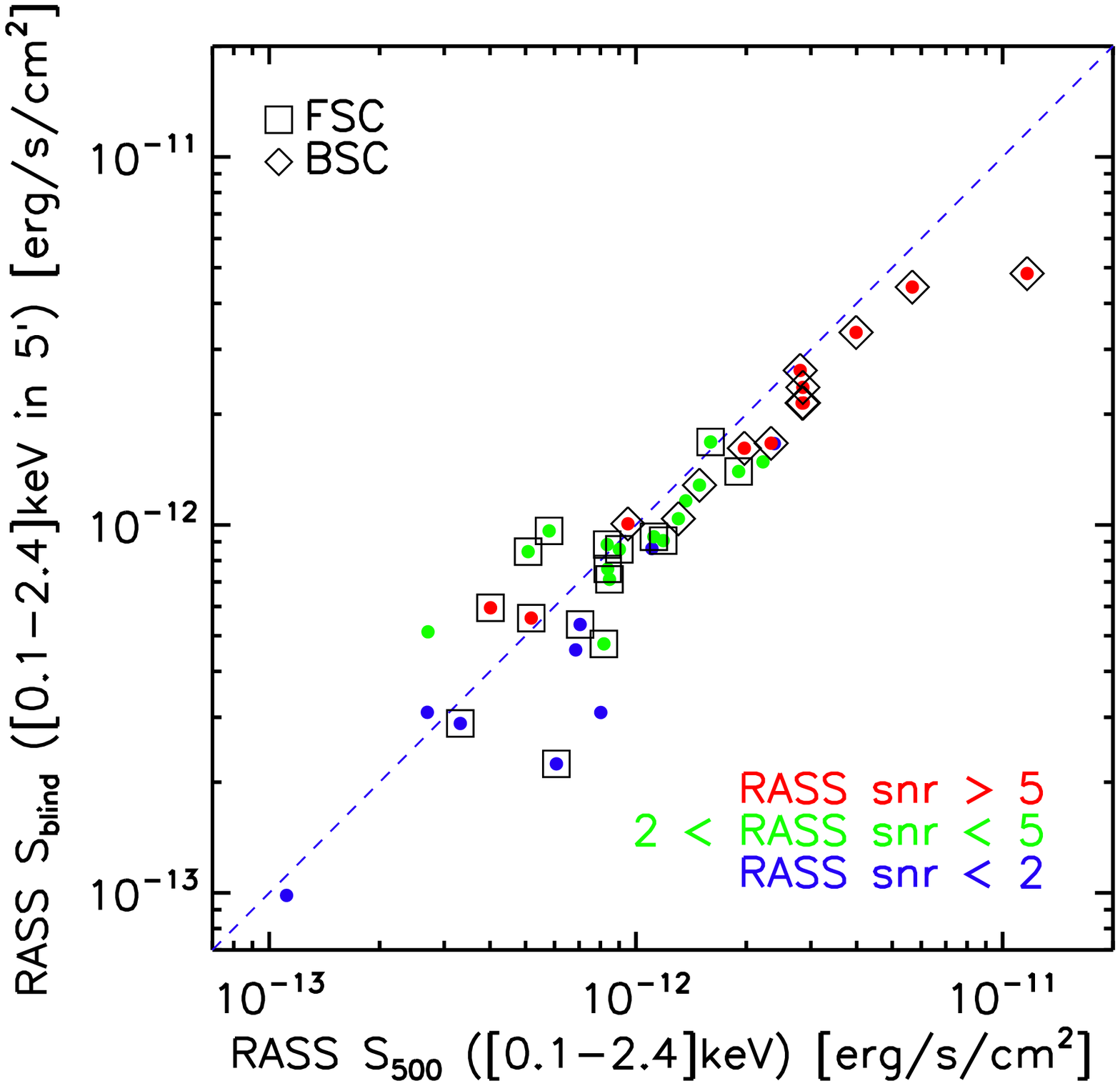}  
\hspace{5mm}
\includegraphics[]{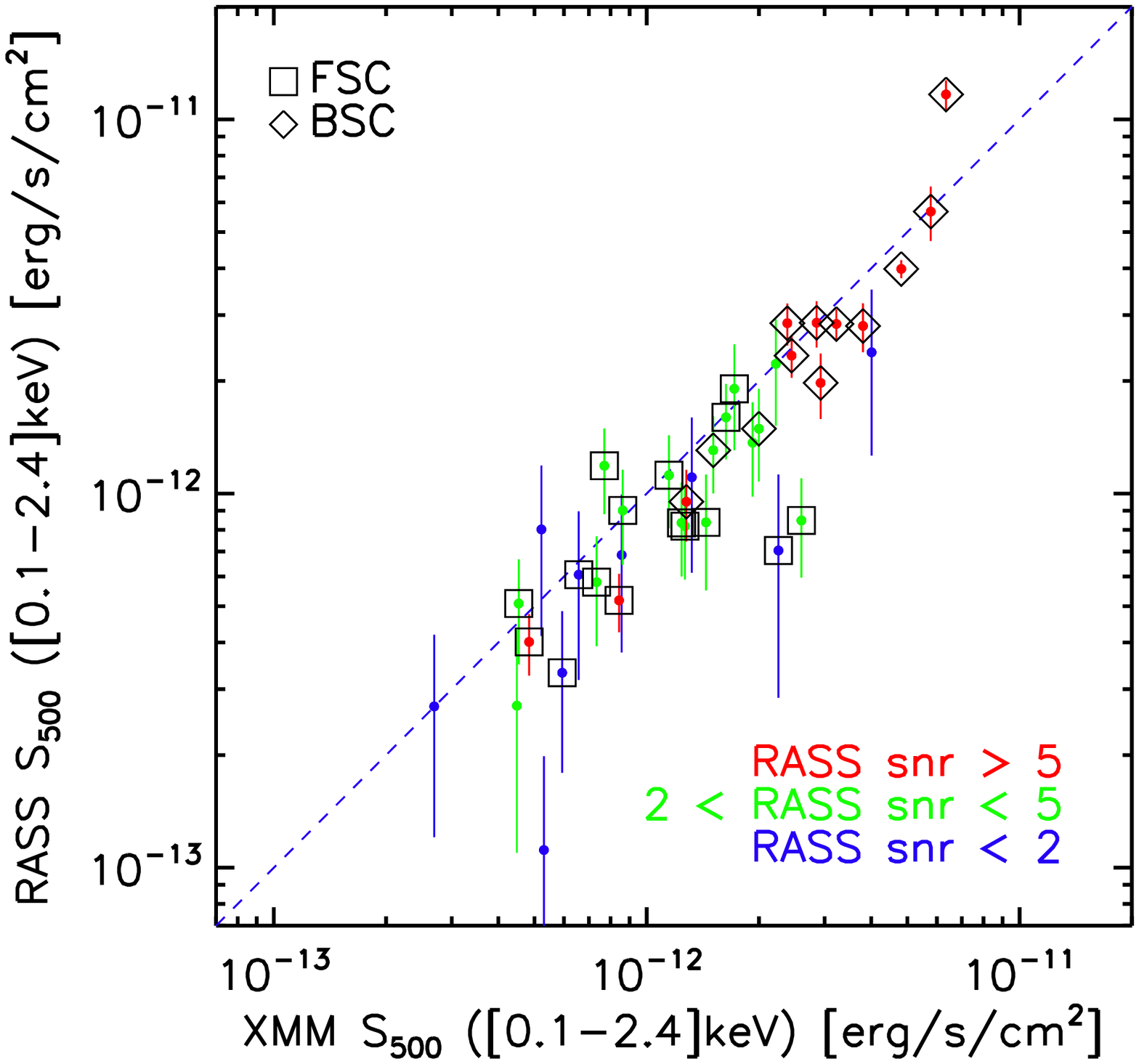} 
\hspace{5mm}
\includegraphics[]{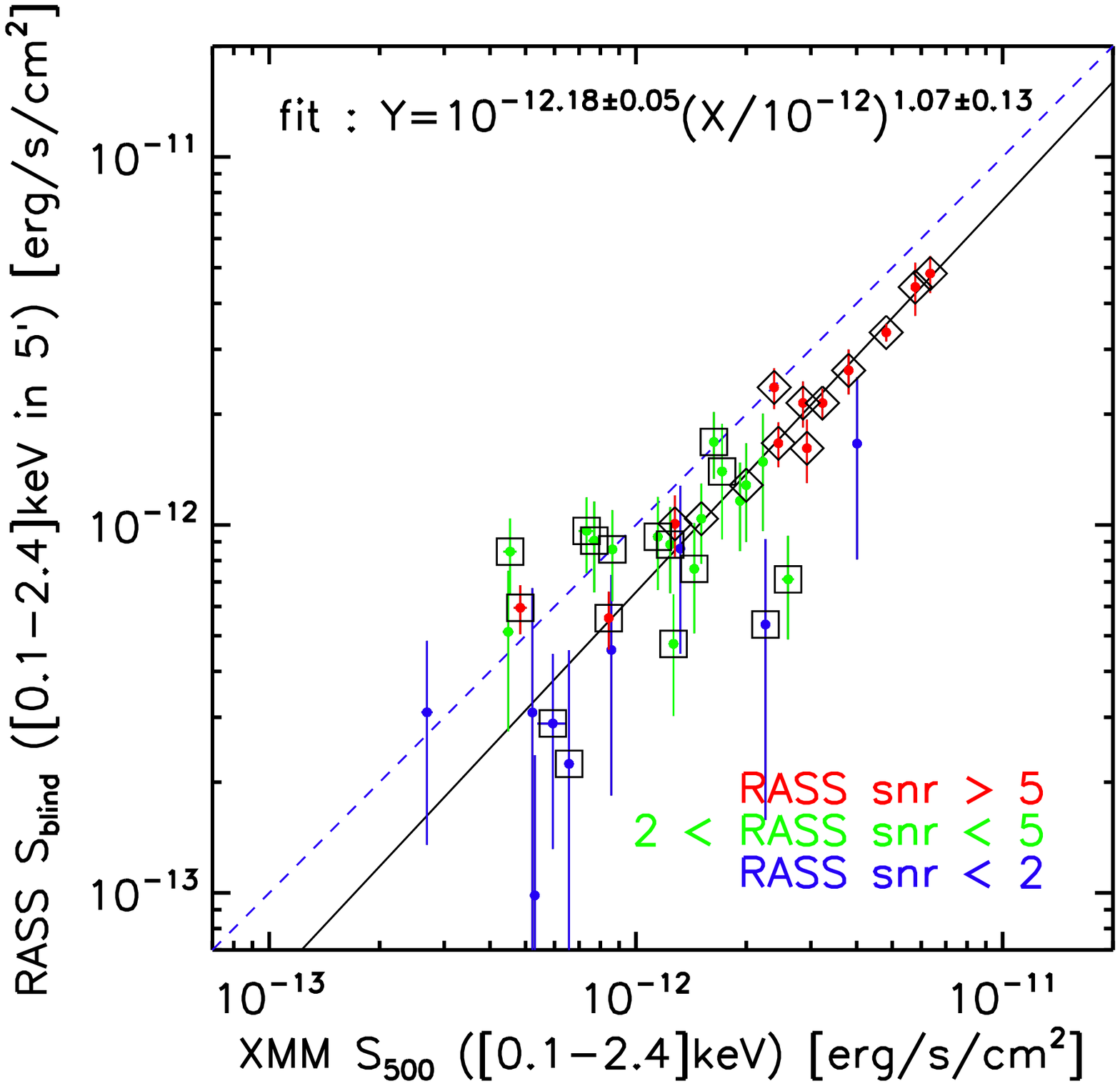} 
}
\end{minipage}
\caption{{\footnotesize Relations between unabsorbed X-ray fluxes measured in the $[0.1$--$2.4]$\,\keV\ band. Blind fluxes are measured in a 5\arcmin\ aperture centred on the \planck\ position; all other fluxes are measured in an aperture corresponding to $\Rv$ centred on the \xmm\ X-ray peak.  
{\it Left panel}: Blind \rass\ flux vs \rass\ flux. {\it Middle panel}: \rass\ flux vs \xmm\ flux. {\it Right panel}: Blind RASS flux vs \xmm\ flux. }}
\label{fig:RASSfx}
\end{figure*}

Figure~\ref{fig:RASSfx} summarises the comparison between \rass\ and \xmm\ unabsorbed fluxes computed in the [0.1--2.4]\,\keV\ band. The \xmm\ flux is given in Table~\ref{tab:RASS}. Fluxes measured in an aperture of 5\arcmin\ centred on the \planck\ candidate position from \rass\ images are referred to as ``blind.'' Here the \rass\ count rate is converted to flux assuming a typical redshift of $z=0.5$, temperature of $\kT=6\,\keV$, and the 21--cm $N_{\rm H}$ value. All other fluxes are recomputed in an aperture corresponding to $\Rv$, centred on the X-ray peak as determined from the \xmm\ validation observations, and using the measured temperature and redshift to convert \xmm\ or \rass\ count rates to  flux.

These figures indicate that the \rass\ blind fluxes and the \rass\ fluxes measured within $\Rv$ are in relatively good agreement, with a slight underestimate at high fluxes (left panel). \rass\ and \xmm\ fluxes measured within $\Rv$ are also in relatively good agreement, although with a slight underestimate together with increased dispersion at low fluxes (middle panel).  As a result, \rass\ blind fluxes slightly underestimate the ``true''  \xmm\ flux measured within $\Rv$, by $\sim 30\,\%$ at $10^{-12}\ergscm$.  The underestimate increases with decreasing S/N (right panel). 

In view of this agreement, we conclude that the \rass\ blind flux  can be used  to  estimate the exposure time required for X-ray follow-up of  a \planck\ candidate, once  confirmed at other wavelengths. The main limitation is the statistical precision on the \rass\ estimate. 
%%==============================================================================================================================================
%%==============================================================================================================================================

%%==============================================================================================================================================
%\begin{figure}[t]
%\centering
%\begin{minipage}[t]{0.6\columnwidth}
%\resizebox{\hsize}{!} {\includegraphics[]{RassBlindRassn.eps}}
%\end{minipage}  
%\begin{minipage}[t]{0.6\columnwidth}
%\resizebox{\hsize}{!} {\includegraphics[]{RassXMMn.eps} }
%\end{minipage}  
%\begin{minipage}[t]{0.6\columnwidth}
%\resizebox{\hsize}{!} {\includegraphics[]{RassBlindXMM_ERR_sanslegFBSC.eps} }
%\end{minipage}  
%\caption{{\footnotesize Relations between unabsorbed X-ray fluxes measured in the [0.1-2.4] keV band. Blind fluxes are measured in a 5\arcmin\ aperture centred on the \planck\ position; all other fluxes are measured in an aperture corresponding to $\Rv$ centred on the \xmm\ X-ray peak.  
%{\it Left panel}: Blind RASS flux vs RASS flux. {\it Middle panel}: RASS flux vs \xmm\ flux. {\it Right panel}: Blind RASS flux vs \xmm\ flux. }}
%\label{fig:RASSfx}
%\end{figure}
%%==============================================================================================================================================

%==============================================================================================================================================
\begin{figure*}[t]
\centering
\includegraphics[width=0.9\textwidth,clip]{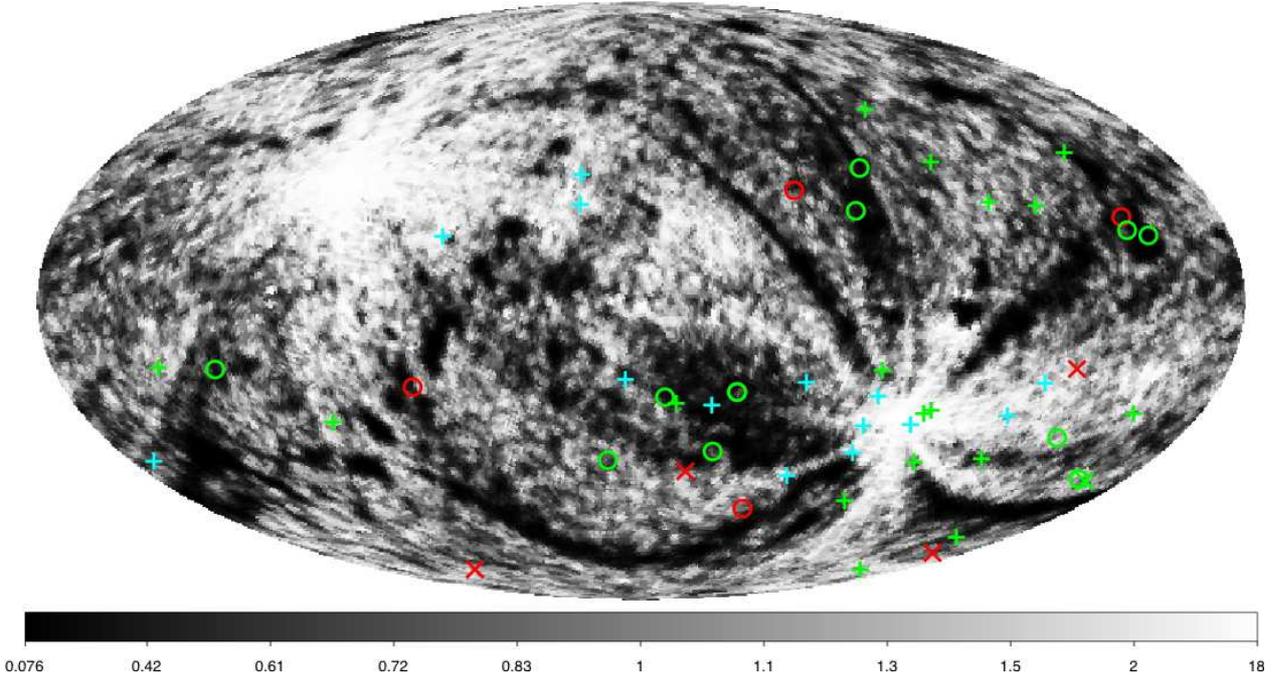} 
\caption{{\footnotesize  Density map of the \rass-Faint Source Catalogue (FSC) with \xmm\ validation results overplotted.  The source density map has been normalised by the median of the pixel density distribution. The source density directly reflects  the \rass\  scanning strategy, with the largest exposure and source density at the Ecliptic poles. Cyan pluses ($+$):  confirmed candidates associated with a BSC source. Other confirmed candidates are plotted in green, and false candidates are plotted in red. Pluses ($+$): good association with a FSC source. Crosses ($\times$): mis-association with an FSC source. Circles ($\bigcirc$): no association with a FSC/BSC source. Confirmed candidates with no association are mostly located in low density regions corresponding to the shallower part of the \rass\ survey.}}
\label{fig:FSCmap}
\end{figure*}
%==============================================================================================================================================

\subsection{Candidate reliability}

The  association of an SZ candidate with a \rass-B/FSC source is neither a necessary nor a sufficient condition for an SZ candidate to be a {\it bona fide} cluster. Putting together the results from all \xmm\  validation observations for a total of 51 \Planck\ cluster candidates, we find that three of the eight false candidates are associated with an FSC source, while eleven candidates are confirmed without association with a \rass-FSC/BSC source. On the other hand, it is striking that PLCK\,G266.6$-$27.3, the most distant cluster of the sample, with a $z=0.97$, is detected at a ${\rm S/N}>5$ in \rass, and is in fact found in the \rass\ Faint Source Catalogue.

\subsubsection{\rass\ source density}
\label{sec:rassdens}

It is important to underline that the \rass\ is not homogeneous, and that neither the BSC nor the FSC are flux-limited or complete in any way. Using the \rass-BSC and FSC, we computed the source density map of each catalogue and the associated probability that a \planck\ candidate will be associated with a B/FSC source within a radius of 5\arcmin. The method is described in Appendix~\ref{appx:rasscal}, and the resulting probabilities are given in Table~\ref{tab:RASSprob}. 

Figure~\ref{fig:FSCmap} and Fig.~\ref{fig:BSCmap} show the \rass-FSC and BSC source density maps with all \xmm\ validation observations overplotted. The  faint source distribution directly reflects  the \rass\ scanning strategy, as evident in Figure~\ref{fig:FSCmap}.  In this context, the probability of chance association is also an indication of how well covered the region is and thus on the depth of the X-ray observation at this position.  We found a mean probability of association with an FSC source of $\mathcal{S(R}\leq 5')\times\bar{\rho} \sim 6$\,\% over the whole sky, where $\mathcal{S(R}\leq 5')$ is the area corresponding to a circle of 5\arcmin\ and $\bar{\rho}$ is the mean density at the position of the candidate, respectively. The corresponding mean probability of association with a BSC source is $\sim 1$\,\%. However, in the best-covered regions of the \rass\  the probability can reach 95\,\% for the FSC and 9\,\% for the BSC, while in the least-covered regions these probabilities drop to 0.4\,\% and 0.2\,\%, respectively.

%==============================================================================================================================================
\begin{figure}[]
 \centering
\includegraphics[width=0.9\columnwidth, clip]{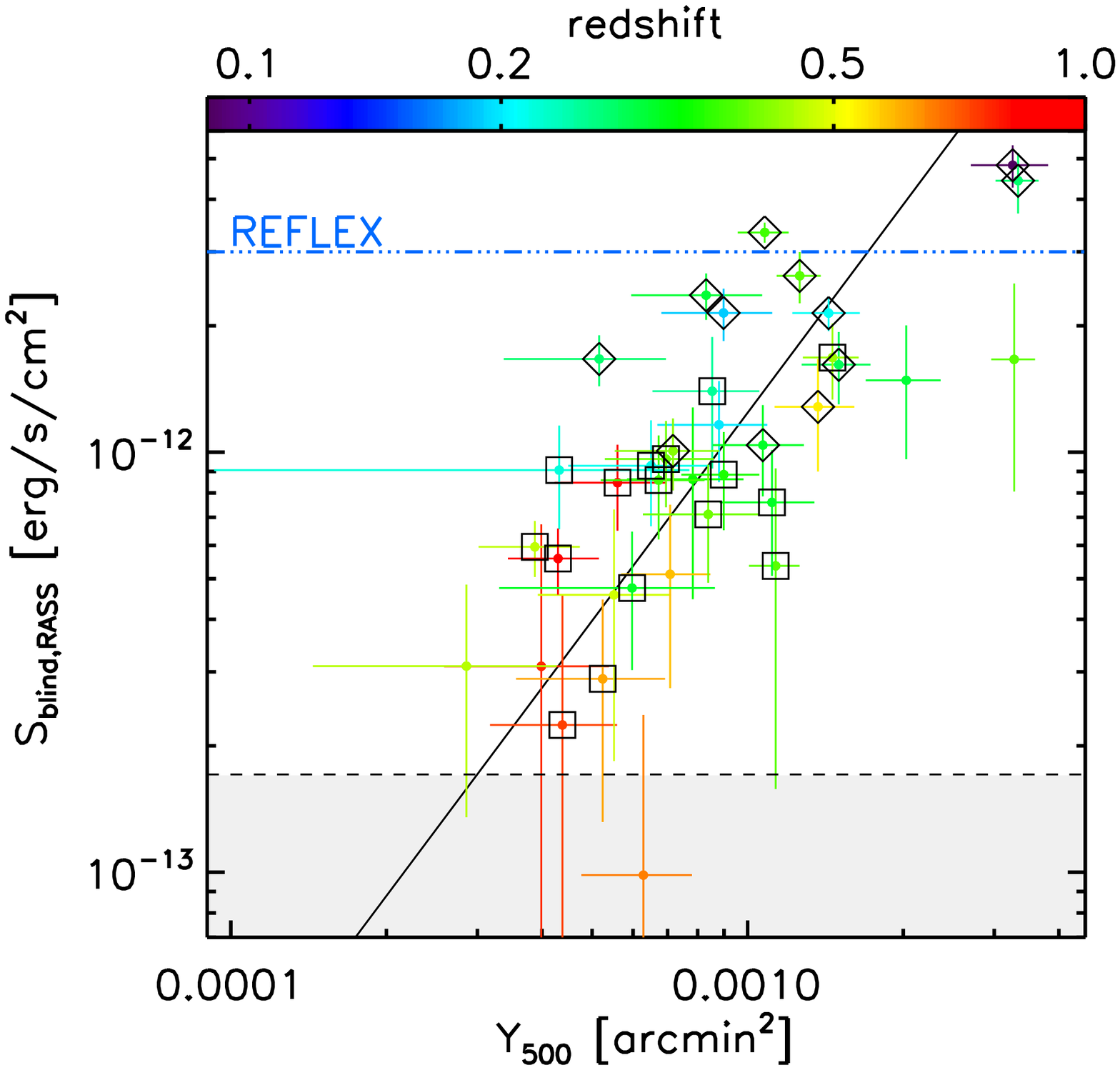}
\caption{ Relation  between \rass\ blind fluxes and SZ fluxes, $\YSZ$,  for single systems confirmed with \xmm\ (all validation observations). The \rass\ flux is the unabsorbed  flux  computed  in the $[0.1$--$2.4]$\,\keV\ band and  measured in a 5\arcmin\ aperture centred on the \planck\ position. The points are colour-coded as a function of redshift.  Squares are  candidates  associated with a FSC source while diamonds are candidates  associated with a BSC source}
\label{fig:Y500SX}
\end{figure}
%==============================================================================================================================================

\subsubsection{BSC source association}

All 12 candidates associated with a BSC source are confirmed.  This is not surprising. For the BSC, the probability of chance association is relatively low, varying from less than 1\,\% to 9\,\%, depending on the sky region. For one cluster, PLCKG305.9$-$44.6, the \xmm\ validation observation  reveals that a point source is located at  the position of the BSC source. However, the source is labelled as extended in the BSC, and in fact the X-ray emission likely corresponds to a blend of the point source and extended cluster emission that was not resolved with the large {\it ROSAT\/} \hbox{PSF}. This is supported by a comparison of the \xmm\ and \rass\ images.

Thus we conclude that the correspondence of a \planck\ SZ candidate with a \rass-BSC source is a very good indication of there being a real cluster at this position.

\subsubsection{FSC source association} 

For the FSC catalogue, on the contrary, the conclusion is more uncertain because of the larger probability of chance association.  Most (18 of 21, i.e., more than 85\,\%) of the candidates associated with a faint source are indeed confirmed. For the triple system PLCK\,G214.6+36.9, the FSC source is classified as extended. Its position as given in the \rass\ catalogue lies between the three clusters and is close to that of a bright \xmm\  point source.  The FSC source is thus in fact a blend of the cluster(s)  and of the point source.  In only one case, PLCK\,G193.3$-$46.1, does the FSC not correspond to the cluster emission. The \xmm\ and \rass\ data shows that it is a point source located $3\farcm3$ away from the cluster centre.  

Taking into account PLCK\,G193.3$-$46.1 and the three false candidates associated with an FSC source, we found four cases of mis-associations out of 51 candidates, i.e., 8\,\%.  This is consistent with the mean probability of chance association of 6\,\% computed above; however, the association with an FSC source is still an indicator of reliability even in the regions of high probability of chance association.  For  instance, the two highest-redshift clusters ($z\approx 0.9$) are correctly associated with a faint source, despite both being in the ecliptic pole region where the probability of false association is high.  The scanning strategies of \planck\ and \rass\ are very similar in that both surveys are deeper in the same regions. In well-covered regions,  the association with the faint source catalogue allows us to probe less massive or higher redshift potential clusters.
A possible indicator of false association might be the distance between the FSC source and the SZ position, although no strict criterion can be applied.  Seventy-five per cent of the false associations correspond to a distance greater than $3\arcmin$, compared to 2 out of 16 (13\,\%) for true associations. 

\subsubsection{No association}
 
Sixteen candidates are not associated with a  B/FSC source. Five of these candidates are false and eleven candidates are true sources with no B/FSC source association. As mentioned above,  the association with a B/FSC is not necessary  for an SZ candidate to be a {\it bona fide} cluster. However, we note that the median probability of FSC chance association, a measure of survey depth as discussed Sec.~\ref{sec:rassdens}, is 2.1\,\% for  clusters without association, to be compared to  6.7\,\% for associated clusters  (see also Fig.~\ref{fig:FSCmap}). These true clusters with no B/FSC counterpart are located in the shallower part of the \rass\ survey, which likely explains why they are not associated.

%\ccor{This shows, together with \citep[][Fig.~9] where we plot the \rass\ blind flux versus the flux measured with XMM-Newton for the whole sample (associated and non-associated), that even if there is no B/FSC identified, the \rass\ signal is representative of the true signal due to the cluster.}
%These are preferentially located in the shallower part of the \rass\ survey, which likely explains why they are not associated. The median probability of FSC chance association, a measure of survey depth as discussed above, is 2.1\,\% for  clusters without association and 6.7\,\% for associated clusters  (see also Fig.~\ref{fig:FSCmap}).  On the other hand, unassociated candidates follow the general correlation observed between the \rass\ blind flux and  the SZ flux (Fig.~\ref{fig:Y500SX}). This correlation presents  some dispersion, due to difference between the blind flux and the true flux (Sec.~\ref{sec:sxcomp}) and the intrinsic dispersion and $z$ dependence of the  $F_{\rm X}/\YSZ$ ratio \citep[][Fig.~9]{planck2012-I}. 
%==============================================================================================================================================
\begin{figure*}[]
\centering
\begin{minipage}[t]{0.8\textwidth}
\resizebox{\hsize}{!} {
\includegraphics{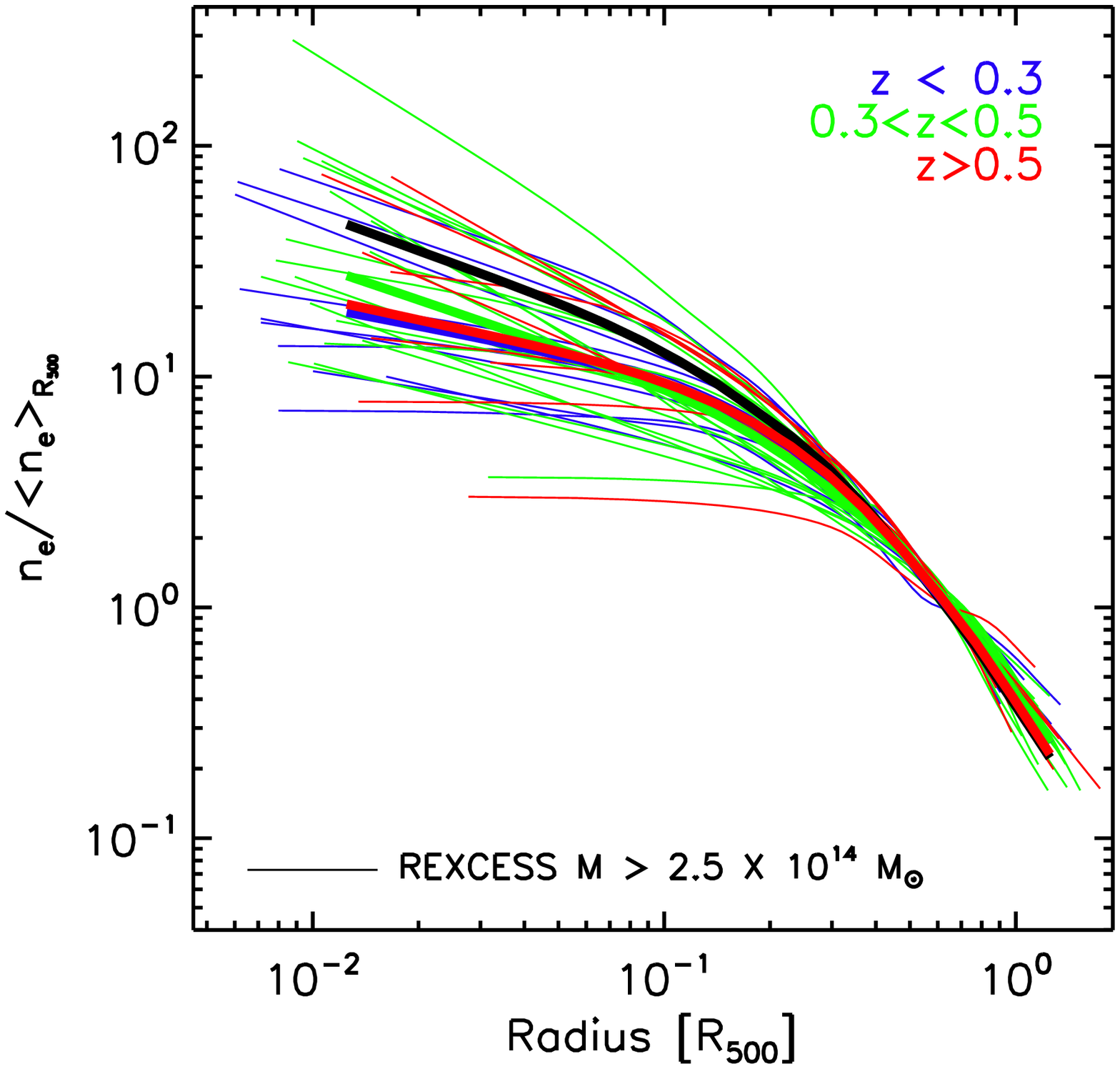}
\hspace{5mm}
\includegraphics{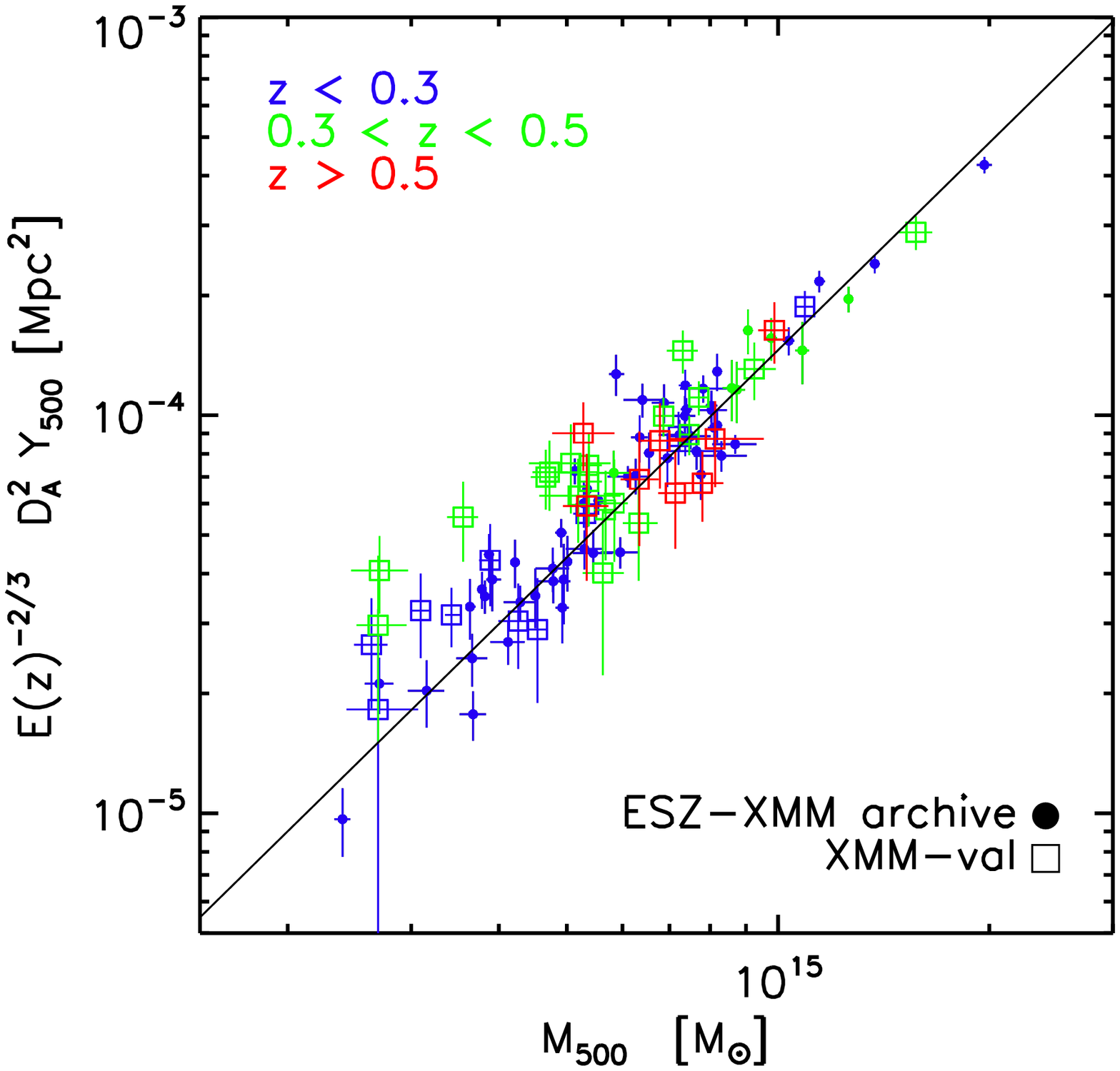}
} 
\vspace{2mm}
\end{minipage}
\begin{minipage}[t]{0.8\textwidth}
\resizebox{\hsize}{!} {
\includegraphics{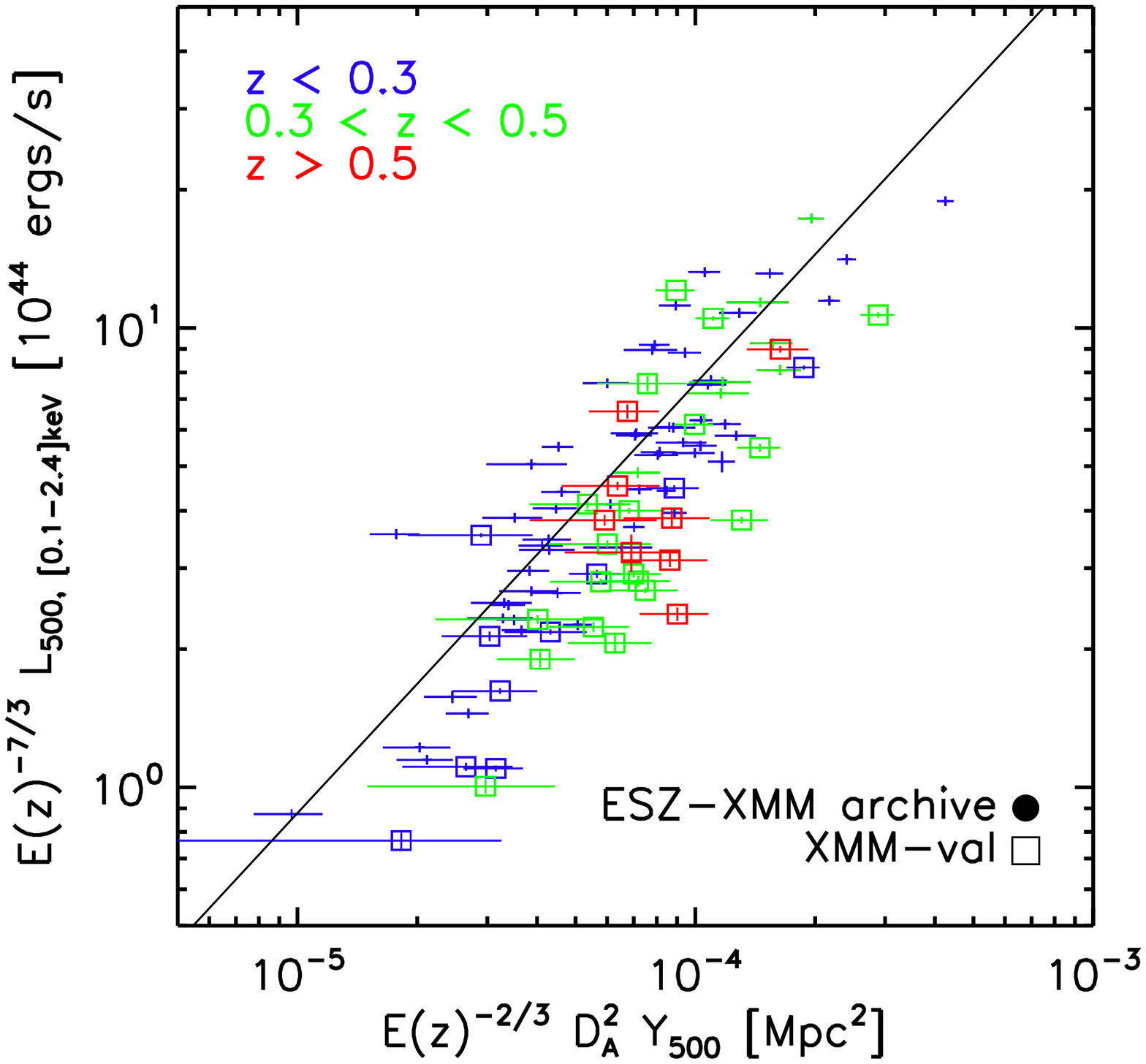}
\hspace{5mm}
\includegraphics{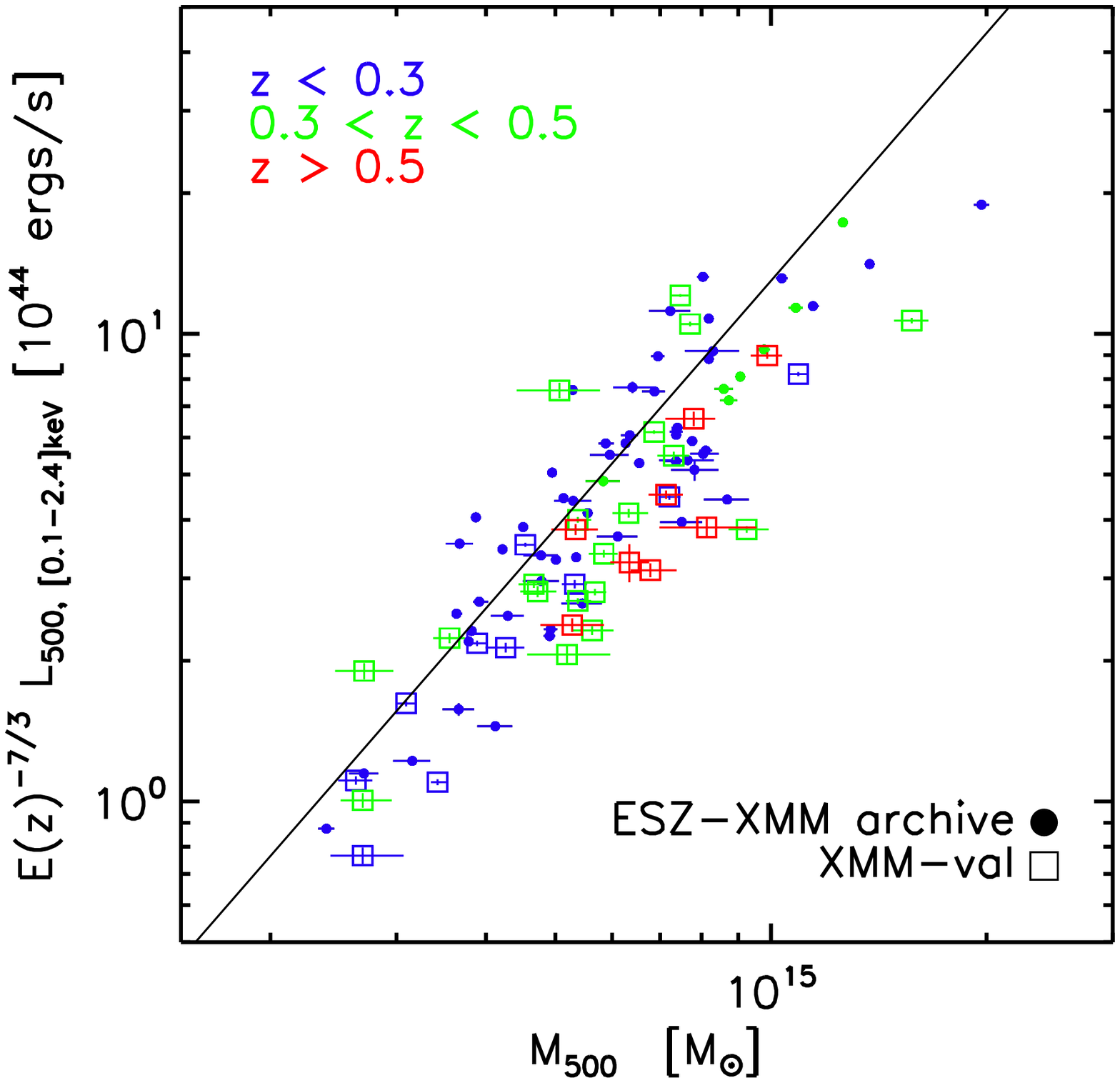}
} 
\end{minipage}
\caption{{\footnotesize 
Scaling properties  of   \planck\ clusters,  colour-coded as a function of redshift.    In all figures, $\Rv$ and $\Mv$ are estimated from the \MYX\ relation of \citet{arn10}. 
{\it Top left panel:}  The scaled density  profiles of the new clusters confirmed with \xmm\  observations. The radii are scaled to $\Rv$. The density is scaled to the mean density within $\Rv$. 
The thick lines denote the mean scaled profile for each sub-sample. The black line is the mean profile of  the  \rexcess\ sample \citep{arn10}. {\it Other panels}:  Scaling relations. Squares show the new clusters confirmed with \xmm\ observations. Points show clusters in the  \planck\--ESZ sample with \xmm\ archival data as presented in \citet{planck2011-5.2b}. 
Relations are plotted between the intrinsic Compton parameter, $D_{\rm A}^2 \YSZ$, and the mass  $\Mv$ ({\it top right panel}),  between the X-ray luminosity and  $Y_{500}$  ({\it bottom left panel}) and between mass and luminosity ({\it bottom right panel}).  Each quantity is scaled with redshift, as expected from standard self-similar evolution. The  lines in the left and middle  panel denotes the predicted $\YSZ$ scaling relations  from the \rexcess\ X-ray observations \citep{arn10}. The line in the right panel is the Malmquist bias corrected $M$--$L$ relation from the \rexcess\ sample \citep{pra09,arn10}. The new clusters are on average less luminous at a given $\YSZ$, or more massive at a given luminosity, than X-ray selected clusters.  There is no evidence of non-standard evolution.}}
\label{fig:xsz}
\end{figure*}
%==============================================================================================================================================
%==============================================================================================================================================
\begin{figure*}[]
\centering
\begin{minipage}[t]{0.8\textwidth}
\resizebox{\hsize}{!} {
\includegraphics{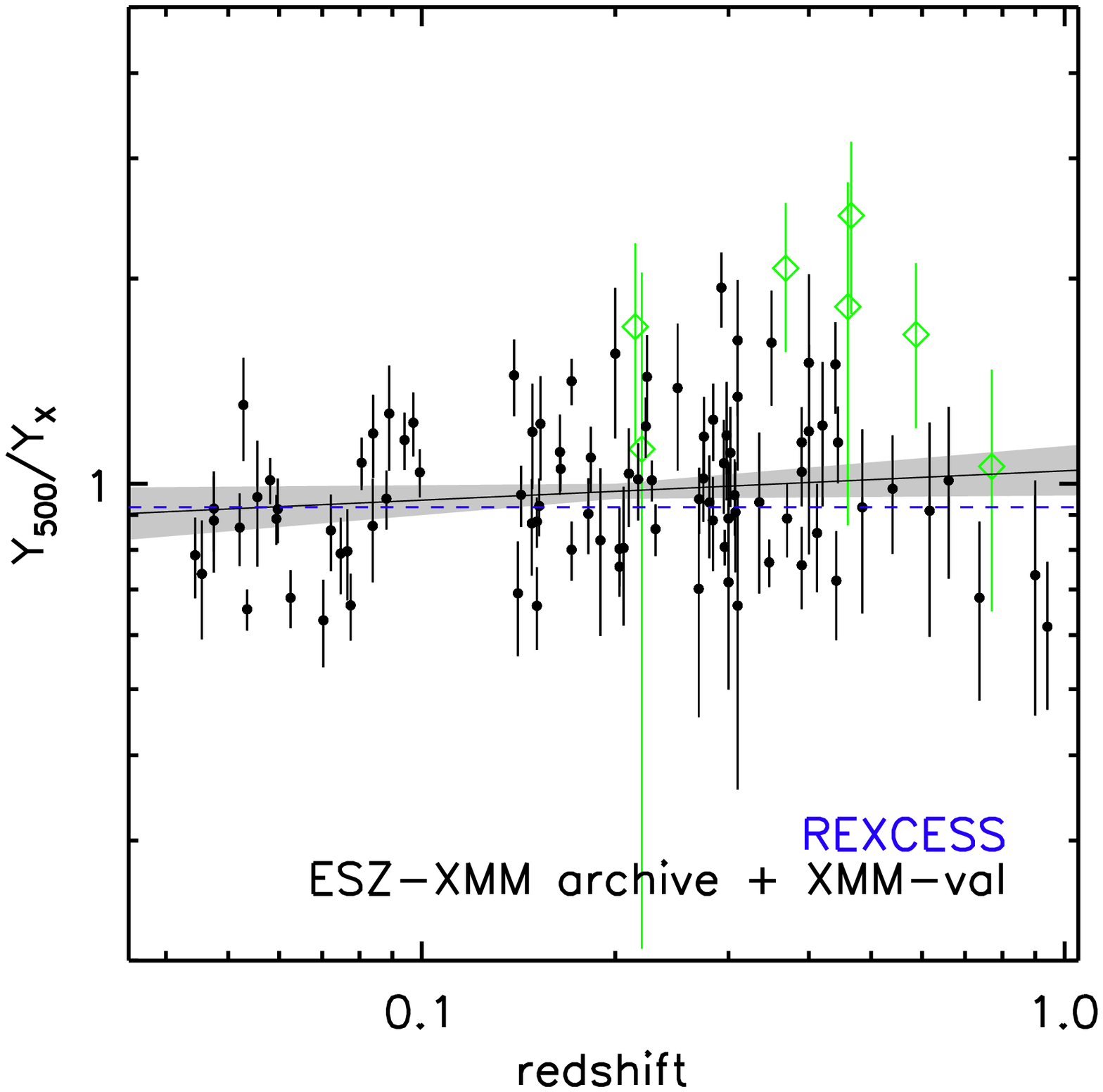}
\hspace{5mm}
\includegraphics{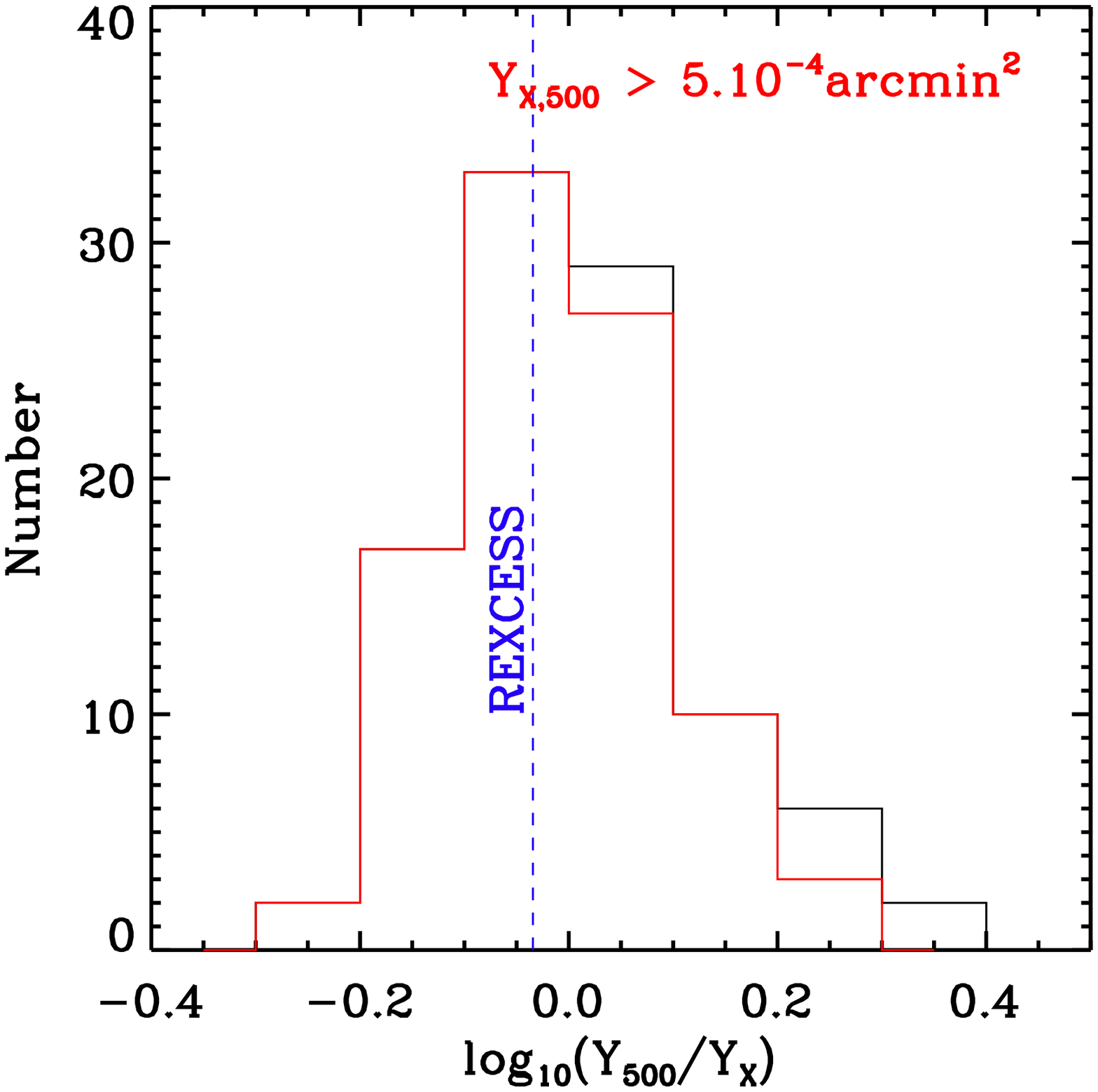}
} 
\end{minipage}
\caption{{\footnotesize   Ratio of the $\YSZ$ Compton parameter to  the normalised $\YX$ parameter. {\it Left panel}: Variation as a function of redshift. The dotted line is the \rexcess\ prediction \citep{arn10}. The full line is the best fit power law and the grey shaded area indicates the $\pm1\sigma$ uncertainty.  Clusters with normalised  $\YX\lesssim5\times 10^{-4}\,{\rm arcmin}^2$ (green points) were excluded from the fit, to minimise Malmquist bias. {\it Right panel}: Histogram of the ratio without and with low flux clusters. 
}
\label{fig:xszevol}}
\end{figure*}
%==============================================================================================================================================

\subsubsection{\rass\ flux and signal-to-noise limit for candidate validation}
 
Unassociated and associated candidates follow the same general correlation between the \rass\ blind flux, $F_{\rm X}$, and  the SZ flux,  $\YSZ$ (Fig.~\ref{fig:Y500SX}). This correlation presents  some dispersion, with deviations from the mean as large as a factor of three. This is expected from the large statistical errors, as well as from the intrinsic dispersion and $z$ dependence of the $F_{\rm X}/\YSZ$ ratio \citep[]{planck2012-I} and the difference between the blind and true X-ray fluxes (Sec.~\ref{sec:sxcomp}).

Because of this large dispersion,  it is not possible to determine a strict  \rass\ flux  (or signal-to-noise ratio) limit  below which a candidate should be discarded. However, we note that all new clusters have an X-ray flux greater than $\sim 2\times 10^{-13}\,\ergscm$ (grey area on Fig.~\ref{fig:Y500SX}). This flux is consistent with the $\YSZ$ threshold $Y_{500, thresh}\approx 2-5\times 10^{-4}\,\mathrm{arcmin^2}$, as defined from the  region  affected by the Malmquist bias (see Fig.~\ref{fig:yxsz}). This \rass\  flux limit is more than 10 times lower than the \reflex\  flux limit of $\sim 2\times 10^{-12}\,\ergscm$, but still detectable with  \rass\footnote{Such clusters, however, could not be identified from \rass\ data alone. They cannot be identified as clusters on the basis of source extent because of the low statistical quality of the signal. Confirmation and identification follow-up is unmanageable in view of the number of sources at such low flux, the vast majority of which are unidentified AGN or noise fluctuations.}.  For the confirmed  candidates, the minimum signal-to-noise ratio computed from \rass\ data is $\sim 0.70$.  Below that limit, all the candidates were false.   All candidates with \rass\ ${\rm S/N}>3$  are confirmed, and only one false candidate is found for \rass\ ${\rm S/N}>2$. The latter is an SZ candidate detected at low \planck\ ${\rm S/N}=4$.

\subsubsection{\rass\ reliability flag}

In view of the above results, we conclude the following regarding the most relevant \rass\ reliability flags:

\begin{itemize}
\item Positional association of a \planck\ SZ candidate with a \rass-BSC source is a very strong indication that the candidate is a cluster; 
\item Positional association of a \planck\ SZ candidate with a \rass-FSC source at ${\rm S/N} > 2$ is a  good indication of a real cluster; 
\item An SZ candidate with no signal at all in \rass\ is false at very high confidence. Obviously, candidates with low signal-to-noise ratio in a well-covered region are particularly likely to be false.
\end{itemize}

\section{A preview of cluster evolution}
\label{sec:evol}

With  this new \xmm\ validation campaign, we have now assembled a sample of 37 new single \planck\ clusters covering a redshift range $0.09<z<0.97$.  With only snapshot \xmm\ observations, the global properties and density profile of each object are measured accurately enough to allow a first assessment of evolution with redshift. The structural and scaling properties of the sample are illustrated in Fig.~\ref{fig:xsz}. We considered three redshift bins, $z<0.3$ (10 clusters), $0.3<z<0.5$ (19 clusters) and $z>0.5$ (8 clusters).  We confirm our previous finding regarding the scaling properties of these new \planck\ selected clusters, and do not find any evidence of departure from standard self-similar evolution.

The average scaled density profile (top left panel of Fig.~\ref{fig:xsz}) is similar for each $z$ bin and is flatter than that of \rexcess, a representative sample of X-ray selected clusters  \citep{arn10}. Once scaled as expected from standard evolution, the new clusters in each redshift bin follow the same trends in scaling properties  (Fig.~\ref{fig:xsz}): they are on average less luminous at a given $\YSZ$, or more massive at a given luminosity, than X-ray selected clusters. On the other hand, they follow  the \YSZYX\ relation predicted from  \rexcess\ data (Eq.~\ref{eq:yszx}). 
 
To study possible evolution with $z$, we plot in Fig.~\ref{fig:xszevol} the $D_{\rm A}^{2}\YSZ/C_{\rm XSZ}\,\YX$ ratio as function of $z$, including the 62 clusters of the \planck\--ESZ sample with \xmm\ archival data \citep{planck2011-5.2b}. We exclude clusters at low flux,  $D_{\rm A}^{-2}\,C_{\rm XSZ}\,\YX<5\times 10^{-4}\,{\rm arcmin}^2$, to minimise possible Malmquist bias (see Sec.~\ref{sec:yszx}). The best fitting power law gives a slope  $\alpha=0.043\pm0.036$, with a normalisation of $0.97\pm0.03$  at $z=0.2$.  The  relation is thus consistent with a constant ratio at the \rexcess\ value of $0.924\pm0.004$.  A histogram of the ratio shows a peak exactly at the \rexcess\ position. The distribution is skewed towards high ratios, the skewness decreasing if low flux clusters are excluded. This skewness might be intrinsic to the cluster population. It might also reflect a residual effect of the  Malmquist bias, clusters with intrinsic high $\YSZ/\YX$ ratio being preferentially detected in SZ surveys.

\section{Conclusions}

We have presented results on the final 15 \planck\ galaxy cluster candidates observed as part of a 500\,ks validation programme undertaken in \xmm\ Director's Discretionary Time. The sample was derived from blind detections in the full 15.5-month nominal \planck\ survey, and includes candidates detected at $4.0\!<\!{\rm S/N}\!<\!6.1$. External flags including \rass\  and DSS detection were used to push the sampling strategy into the low-flux, high-redshift regime and to better assess the use of \rass\ data for candidate validation. This last phase of the follow-up programme yielded 14 clusters from 12 \planck\ candidate detections (two candidates are double systems) with redshifts between 0.2 and 0.9, with six clusters at $z>0.5$. Their masses, estimated using the $\Mv$--$\YX$ relation, range from $2.5 \times 10^{14}$ to $8 \times 10^{14}\, \msol$.  We found an interesting double peaked cluster, PLCK\,G147.3$-$16.6, that is likely an ongoing major merger of two systems of equal mass. Optical observations with NOT, TNG, and Gemini confirmed a redshift of 0.65.

The full \xmm\ validation follow-up programme detailed in this paper and in \citet{planck2011-5.1b, planck2012-I} comprises 51 observations of \planck\ cluster candidates.   The efficiency of validation with \xmm\  stems both from its high sensitivity, allowing easy detection of clusters  in the  \planck\ mass and redshift range,  and from the tight relation between X-ray and SZ properties, which probe the same medium.  The search for  extended \xmm\  emission and a consistency check between the X-ray and SZ flux is then sufficient for  unambiguous discrimination between clusters and false candidates.   We have confirmed the relation between the X-ray flux and the SZ flux, as a function of redshift, and estimated its typical scatter.  This relation is used in the validation procedure.  By contrast, optical validation is hampered by the relatively large \planck\ source position uncertainty and the large scatter between the optical observables (such as galaxy number) and the mass (or SZ signal), both of which increase the chance of false associations. 

 The programme yielded 51 {\it bona fide\/} newly-discovered clusters, including  four double systems and two triple systems. There are eight false candidates. Thirty-two  of the 51 individual clusters  have high quality redshift measurements from the Fe K line. For other cases, the spectral fitting yields several $\chi^2$ minima as a function of $z$, that cannot be distinguished at the $68\,\%$ confidence level. We showed that the relation between the X-ray and SZ properties can be used to further constrain the redshift. The new clusters span the redshift range 0.09 to 0.97 and cover more than one decade in $\YSZ$,  from $2.9\times 10^{-4}$ to $3.0\times 10^{-3}\,\arcms$.  $\Mv$\ of single systems is in the range  ($2.5  \times 10^{14}$--$1.6 \times 10^{15})\,\msol$.  These observations provided a first  characterisation of the new objects that \planck\ is detecting:
\begin{itemize}
\item The newly-detected clusters follow the $\YX$--$\Yv$ relation derived from X-ray selected samples. This is consistent with the prediction that both quantities are tightly related to the cluster mass. 
\item New SZ selected clusters are X-ray underluminous on average compared to X-ray selected clusters, and more morphologically disturbed. The dispersion around the $M$--$L_{\rm X}$ relation may be larger than previously thought and dynamically perturbed (merging) clusters might be under-represented in X-ray surveys. This has implications for statistical studies of X-ray selected samples, either to constrain cosmological models from cluster number counts or to probe the physics of structure formation from the cluster scaling properties.  As discussed in detail by \citet{ang12}, precise knowledge of the actual scatter between the mass and the observable used in the detection is critical in both applications. 
\item We found no indication of departure from standard self-similar evolution in the X-ray versus SZ scaling properties. In particular, there is no significant evolution of the $\YX/\Yv$ ratio. 
\end{itemize}

Beyond new cluster confirmation and characterisation, we checked  the pertinence of  the validation process based on \planck\  internal quality assessments and cross-correlation with ancilliary data. There are eight false candidates in total, all of which were found at ${\rm S/N} < 5$. These failures  underline the importance of the number of methods detecting the clusters and were used to refine our internal quality flag definitions. All candidates with $Q_{\rm SZ}={\rm A}$ are confirmed.  Galaxy overdensity in SDSS data can confirm candidates up to $z\sim0.6$, although it remains difficult to distinguish between massive clusters and  pre-virialised structures at high $z$. The quality of the SZ detection, ancillary data such as significant \rass\ emission, and the offsets between SZ, BCG, and other positions, must all be considered for firm confirmation. Using the full sample of 51 observations, we investigated the use of \rass-based catalogues and maps for \planck\ catalogue construction, finding that:
\begin{itemize}
\item \planck\ clusters appear almost always to be detectable in \rass\ maps, although there is not a one-to-one correspondence between a \rass-BSC or FSC source and the presence of a cluster. 
\item Association of a cluster candidate with a \rass-BSC source is a very strong indication that it is a real cluster. 
\item Whether or not there is a \rass-BSC or FSC source, ${\rm S/N} > 2$ in the \rass\  maps is a good indication of a true candidate, while ${\rm S/N} < 0$ is a good indication of a false candidate. 
\item The association with a faint or bright \rass\ source can be used to refine the SZ position estimate.  The \rass\ blind flux   can be used  to  estimate the exposure time required for X-ray follow-up of  a \planck\ candidate, once  confirmed at other wavelengths. The main limitation is the statistical precision on the \rass\ estimate. 
\end{itemize}

The \xmm\ validation  observations could also be used for the verification of  \planck\ performances, showing that:
\begin{itemize}
\item The mean offset between  the \planck\ position and the cluster position is $1\farcm5$, 
as  expected from \planck\ sky  simulations, and this offset is less  than $2\farcm5$ for $86$\,\% of the clusters.
\item \planck\ can detect clusters well below the X-ray flux limit of \rass\ based catalogues,  ten times lower than \reflex\ at high $z$, and below the limit of the most sensitive \rass\ survey (\macs). 
\item The \planck\ sensitivity threshold for the nominal survey is $\Yv \sim 4 \times 10^{-4}$ arcmin$^2$, with an indication of Malmquist bias in the $\YX$--$\Yv$ relation below this threshold. The corresponding mass threshold depends on redshift, but \planck\ can detect systems with $\Mv > 5 \times 10^{14}\, \msol$ at $z > 0.5$.
\item Overall, there is a high fraction of double/triple systems in the \xmm\ validation follow-up sample, illustrating the problems of confusion in the \planck\ beam. 
\end{itemize}

 These results illustrate the potential of the all-sky \planck\ survey to detect  the most massive clusters in the Universe. Their characterisation, and the determination of their detailed physical properties, depends on a vigorous follow-up programme, which we are currently undertaking.

\begin{acknowledgements}
The \planck\ Collaboration thanks Norbert Schartel for his
 support of the validation process and for granting discretionary time for the observation of \planck\ cluster candidates.  
 The present work is based 
 on observations obtained with \xmm, an ESA science mission with
 instruments and contributions directly funded by ESA Member States
 and the USA (NASA). This research has made use of the following
   databases: SIMBAD, operated at the CDS, Strasbourg, France; the
   NED database, which is operated by the Jet Propulsion Laboratory,
   California Institute of Technology, under contract with the
   National Aeronautics and Space Administration; BAX, which is
   operated by IRAP, under contract with the Centre National d'Etudes Spatiales (CNES); and the  SZ repository operated by
IAS Data and Operation centre (IDOC) under contract with CNES. 
Based on photographic data obtained using The UK Schmidt Telescope. We
further used observations made with the Italian Telescopio Nazionale
Galileo (TNG) 
operated on the island of La Palma by the Fundaci\'on Galileo Galilei
of the INAF 
(Istituto Nazionale di Astrofisica) at the Spanish Observatorio del Roque de los
Muchachos of the Instituto de Astrofisica de Canarias (Science Program ID AOT24/11-A24DDT3), on observations
made with the Nordic Optical Telescope, operated 
on the island of La Palma jointly by Denmark, Finland, Iceland,
Norway, and Sweden, in the Spanish Observatorio del Roque de los
Muchachos of the Instituto de Astrofisica de Canarias (Science Program ID 43-016), observations
obtained at the Gemini Observatory, 
which is operated by the
Association of Universities for Research in Astronomy, Inc., under a
cooperative agreement
with the NSF on behalf of the Gemini partnership: the National Science
Foundation (United
States), the Science and Technology Facilities Council (United Kingdom), the
National Research Council (Canada), CONICYT (Chile), the Australian
Research Council
(Australia), Minist\'erio da Ci\^encia e Tecnologia (Brazil)
and Ministerio de Ciencia, Tecnolog\`ia e Innovaci\`on Productiva
(Argentina), Gemini Science Program ID: GN-2011B-Q-41.
The authors wish to recognize and acknowledge the very
significant cultural role and reverence that the summit of Mauna Kea
has always had within the indigenous Hawaiian community.  We are most
fortunate to have the opportunity to conduct observations from this
mountain.
%We also made use of Digitized Sky Survey southern data. The UK Schmidt Telescope was operated by the Royal Observatory Edinburgh, with funding from the UK Science and Engineering Research Council, until 1988 June, and thereafter by the Anglo-Australian Observatory. Original plate material is copyright (c) of the Royal Observatory Edinburgh and the Anglo-Australian Observatory. The plates were processed into the present compressed digital form with their permission. The Digitized Sky Survey was produced at the Space Telescope Science Institute under US Government grant NAG W-2166.
A description of the \Planck\ Collaboration and a
list of its members, indicating which technical or scientific
activities they have been involved in, can be found at  http://www.rssd.esa.int/Planck\_Collaboration.
The \Planck\ Collaboration acknowledges the support of: ESA; CNES and CNRS/INSU-IN2P3-INP (France); ASI, CNR, and INAF (Italy); NASA and DoE (USA); STFC and UKSA (UK); CSIC, MICINN and JA (Spain); Tekes, AoF and CSC (Finland); DLR and MPG (Germany); CSA (Canada); DTU Space (Denmark); SER/SSO (Switzerland); RCN (Norway); SFI (Ireland); FCT/MCTES (Portugal); and DEISA (EU).\end{acknowledgements}

\bibliographystyle{aa} 

\bibliography{Planck_bib.bib,PIP_22b.bib}

\appendix

\section{Redshift estimates of confirmed candidates}
\subsection{Refinement of the \xmm\ redshift estimate  for $Q_{\rm z}\!\!=\!\!1$ cases }
\label{ap:zx}

The redshift determination from \xmm\ spectral analysis is  uncertain  for five  clusters. There are several $\chi^2$ minima that cannot be distinguished at the $90\,\%$ confidence level ($Q_{\rm z}\!\!=\!\!1$).  As proposed by \citet{planck2012-I}, we estimated the $\YX/\YSZ$ and $F_{\rm X}/\YSZ$ ratios as a function of $z$ and compared them to expected values, to eliminate unphysical solutions. 

Three possible redshifts  were found for PLCK\,G352.1$-$24.0, 0.12, 0.4, and 0.77. The $\YX/\YSZ$ ratio method  enables us to exclude the low redshift $z=0.12$ solution. The $z=0.4$ solution yields a $\YX/\YSZ$  ratio twice higher than expected, at the limit of the observed dispersion. Furthermore, we confirmed that there is no evidence of galaxy concentrations in the DSS red image at the precise \xmm\ cluster location. We thus adopt the highest $z$ value, $z=0.77$, confirming the cluster to be at high $z$. 

The best fitting redshift for  PLCK\,G239.9$-$40.0, $z=0.74$, yields the $\YX/\YSZ$ ratio closest to expectation and is adopted in the further analysis. The lowest $z=0.26$ solution is very unlikely, yielding a $\YX/\YSZ$  ratio twice as high as expected. The other possible solution is $z=0.46$: there are some very faint objects in  the DSS images at the \xmm\  position, although whether those are galaxies is unclear.  

\begin{figure}[t]
\center
\resizebox{\columnwidth}{!} {
\includegraphics[height=0.45\columnwidth,angle=270,origin=br,keepaspectratio]{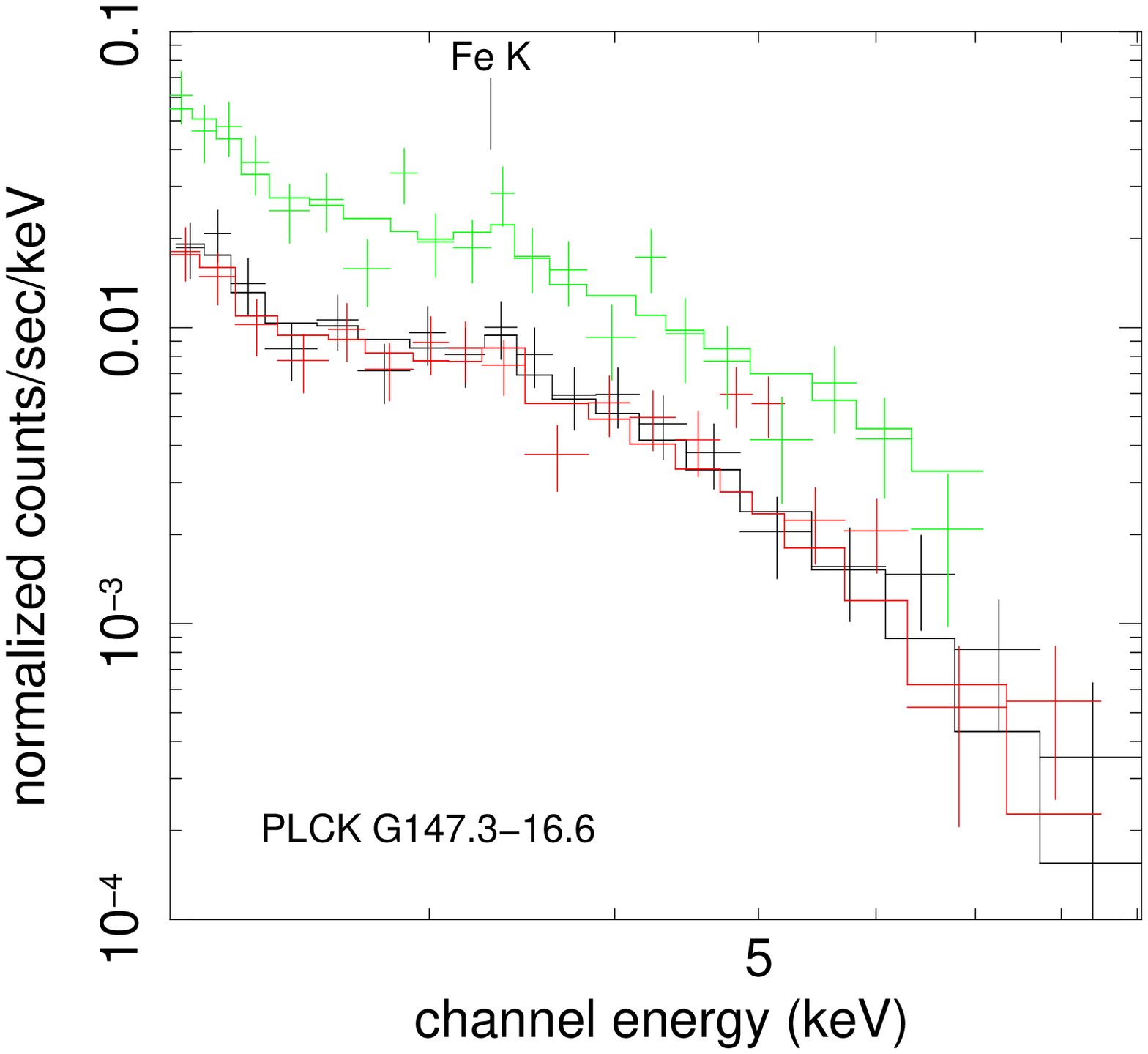}
\hspace{4mm}
\includegraphics[height=0.45\columnwidth,angle=270,origin=br,keepaspectratio]{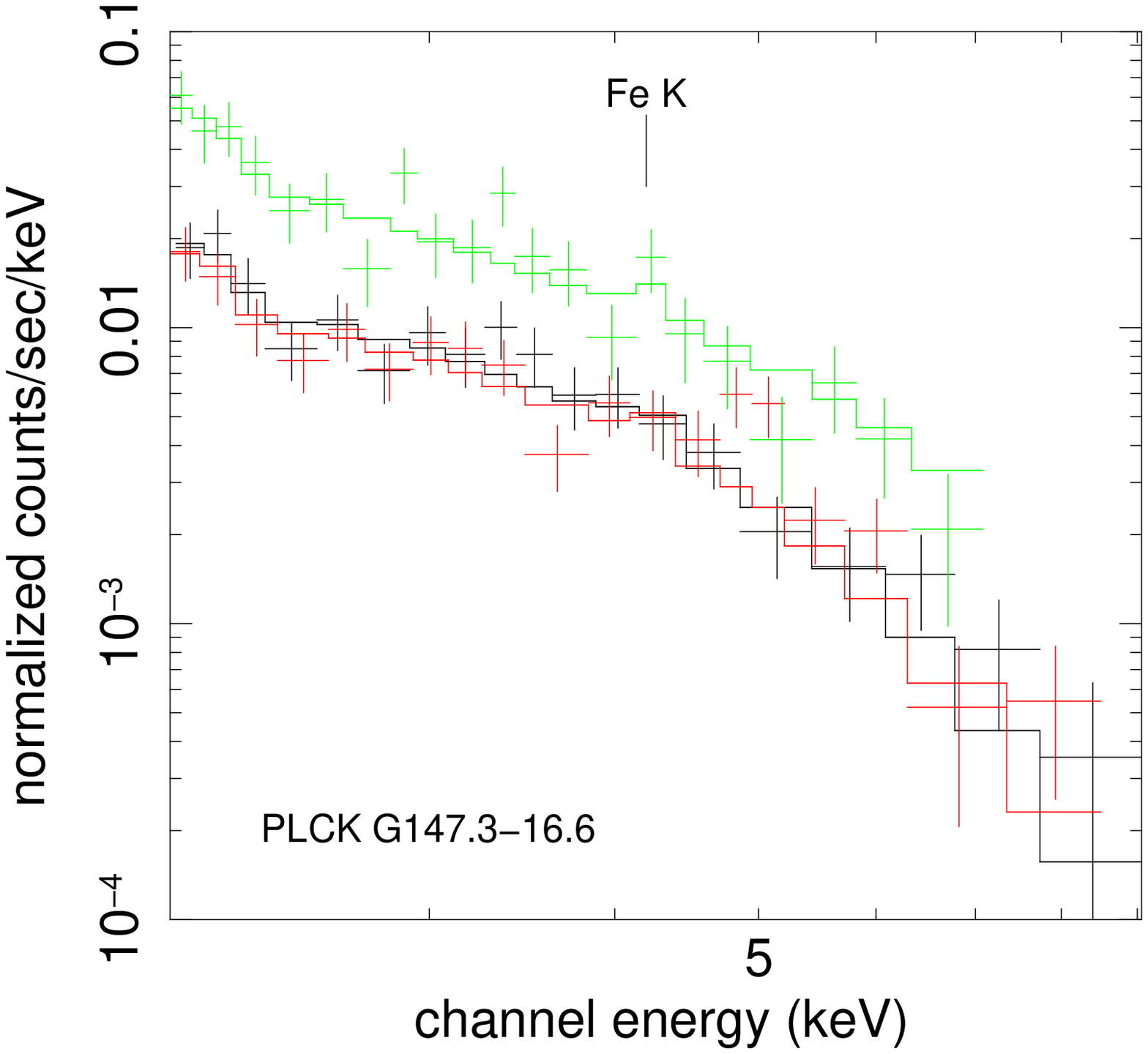}
} 
\caption{{\footnotesize 
EPIC spectra (data points with errors) of PLCK\,G147.3$-$16.6.  Only data points above 2\,keV are shown for clarity, but data down to 0.3\,keV are used in the spectral fitting. The redshift estimate is ambiguous,  with the $\chi^2$ distribution  showing three minima. {\it Left panel}: the best-fitting  thermal model (solid lines) at $z=1.03$ with the position of the redshifted Fe K line marked.  {\it Right panel:} Same for the second best solution at $z=0.62$, consistent with the optical redshift. }
 \label{fig:specz}}
\end{figure}

In the case of  PLCK\,G147.3$-$16.6, all  three redshift solutions, 0.4, 0.62, and 1.03, yield a $\YX/\YSZ$ ratio within the observed dispersion. The best fitting value, $z=1.03$, and the second best solution, $z=0.62$,  are consistent at the $90\,\%$ confidence level, with $\chi^2$ values of 125.9 and 128.7 for 132 degree of freedom, respectively.  The two models are shown in Fig.~\ref{fig:specz}. The optical measurement is described  below (Sec.~\ref{ap:zopt}). 

The redshifts of the two components in PLCK\,G196.7$-$45.5 are uncertain. The $\YX/\YSZ$ and $F_{\rm X}/\YSZ$ ratio methods cannot be used for such double systems, since the individual SZ components are unresolved by \planck.  Of the two solutions, $z=0.57$ and $z=0.87$ for  PLCK\,G196.7$-$45.5A, the latter can be excluded: a clear concentration galaxies at the \xmm\ location is visible in the DSS images, which thus cannot be at such high $z$ (see Sec.~2.2). For PLCK\,G196.7$-$45.5B we adopted the best fitting value, $z=0.42$.

  \begin{figure}[btp]
\centering
\includegraphics[height=0.9\columnwidth, angle=-90,clip]{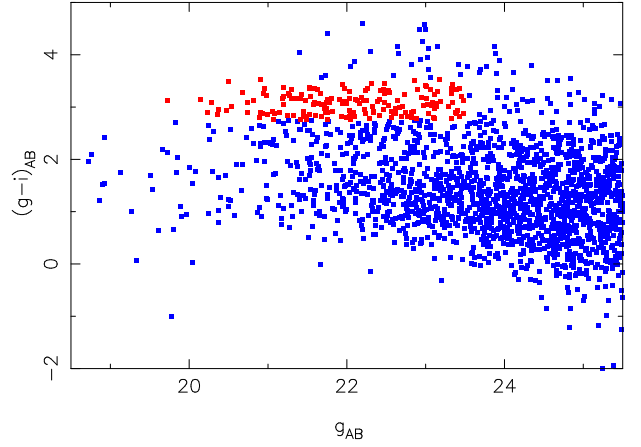}
\caption{ {\footnotesize A $g-i$ vs.\ $g$ colour-magnitude diagram of non-stellar objects 
in the field of PLCK G147.3$-$16.6, observed with NOT/MOSCA. Galaxies plotted 
as red squares, in the region defined by $g-i = 3.15 \pm 0.40$ and $g < 23.5$, 
form the red sequence constituted by early-type galaxies in the cluster.  }}
\label{fig:plckg147rs}
\end{figure}

 \subsection{Optical redshift estimate of  PLCK\,G147.3$-$16.6  }
 \label{ap:zopt}

The optical data for PLCK\,G147.3$-$16.6 were taken using Director's Discretionary Time with DOLORES (Device Optimized for the LOw RESolution), a low resolution spectrograph and imager permanently installed at the TNG telescope (Telescopio Nazionale Galileo La Palma). The camera is equipped with a $2048 \times 2048$ pixel CCD covering a field of view of $8\farcm6\times 8\farcm6$ (pixel scale of 0\parcs252 per pixel).  Exposure times of 3000\,s in the $r$ and $i$ bands were split into 10 single exposures of 300\,s each.  Exposure times of 4000\,s in the $z$ band were split into eight separate exposures. Taking advantage of the dither-offsets between single exposures, no separate sky images were required.  The images were bias and flat field corrected using {\sc IRAF}\footnote{IRAF: \url{http://iraf.noao.edu}}.  For astrometric calibration we used \url{astrometry.net}.  The average seeing derived from the final images is 0\parcs84, 0\parcs85, and 0\parcs84 in the $r$, $i$, and $z$-bands, respectively. In the final images, we reach signal-to-noise ratios (over the PSF area) of 11, 23, and 8 for unresolved sources of 24th magnitude. The colour composite image allows us to pre-identify the cluster members.

The cluster was also observed using the 2.56-m Nordic Optical Telescope with the MOSCA camera,  a $2\times2$ mosaic of $2048\times2048$ pixel CCDs. This camera covers a total field of 7\parcm7 $\times$ 7\parcm7, and was used in $2\times 2$ binned mode.  This gives a pixel scale of 0\farcs217 per binned pixel.  Total exposure times of 900\,s were split into 3 dithered exposures of 300\,s in each of the SDSS $g$- and $i$-bands in photometric conditions.  The telescope was pointed such that the two peaks of the X-ray emission from the cluster would fall in the centreof the mosaic CCD chip that has the best cosmetic quality (named ``CCD7'').     
After standard basic reduction and image registration, the combined images had FWHM of $0\farcs 79$ and $0\farcs65$ in the $g$ and $i$ bands, respectively.  Photometric calibration was based on an ensemble of stars in a field located inside the SDSS footprint, observed at similar airmass immediately following the observations of PLCK G147.3$-$16.6. Stellar objects were removed from the object catalogues based on their location in a size-magnitude diagram. A strong clustering of galaxies with red $g-i$ colours was immediately detected around the position of the X-ray peaks.  The colour-magnitude diagram in Fig.~\ref{fig:plckg147rs}  illustrates the red sequence formed by early-type galaxies at $g-i \simeq 3.15$ in this cluster.  Predicted $g-i$ colours of early-type galaxies as a function of redshift were calculated by convolving the EO template galaxy spectrum of \citet{col80} with the response curves of the SDSS $g$ and $i$ bandpasses. From this, a photometric redshift estimate of $z_{\rm phot} = 0.64 \pm 0.03$ was derived.

The calibrated $g$- and $i$-band photometry from NOT was used to select suitable spectroscopic targets for Gemini North Telescope by choosing galaxies at $g-i \simeq 3.15$. The observations  (Program GN-2011B-Q-41) were made  with GMOS-N, with two exposures of 1800\,s each. The program was in Band~2 service mode, with relaxed observing conditions: the seeing was 1\farcs7 the first night and 0\farcs8 the second night, with cirrus both nights.  The observations were reduced with the standard Gemini IRAF package.  We obtained redshift measurements for 13 objects.  Among those, 10 have redshifts between 0.64 and 0.68, for a cluster redshift measurement of $0.66 \pm 0.05$. If we exclude two objects at $z=0.68$, we obtain $z = 0.645 \pm 0.005$.

\section{Density maps of RASS bright and faint sources\label{appx:rasscal}}
\begin{figure*}[tbp]
\centering
\includegraphics[width=0.9\textwidth,clip]{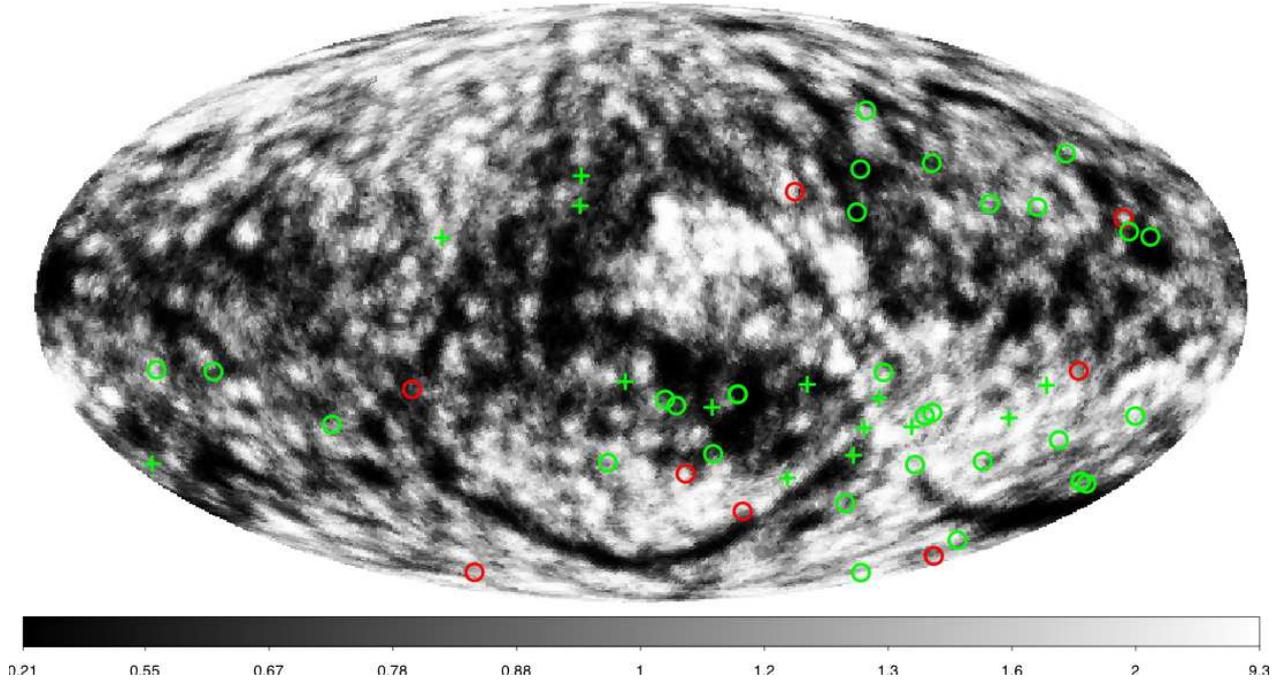} 
\caption{{\footnotesize  \xmm\ validation results overplotted on density map of the \rass-Bright Source Catalogue (BSC). The source density map has been normalised by the median of the pixel density distribution. Confirmed candidates are plotted in green and false candidates are plotted in red. Pluses ($+$): good association with a BSC source. Circles ($\bigcirc$): no association with a BSC  source.}}
\label{fig:BSCmap}
\end{figure*}

\begin{figure*}[tbp]
\centering
\begin{minipage}[t]{0.8\textwidth}
\resizebox{\hsize}{!} {
\includegraphics{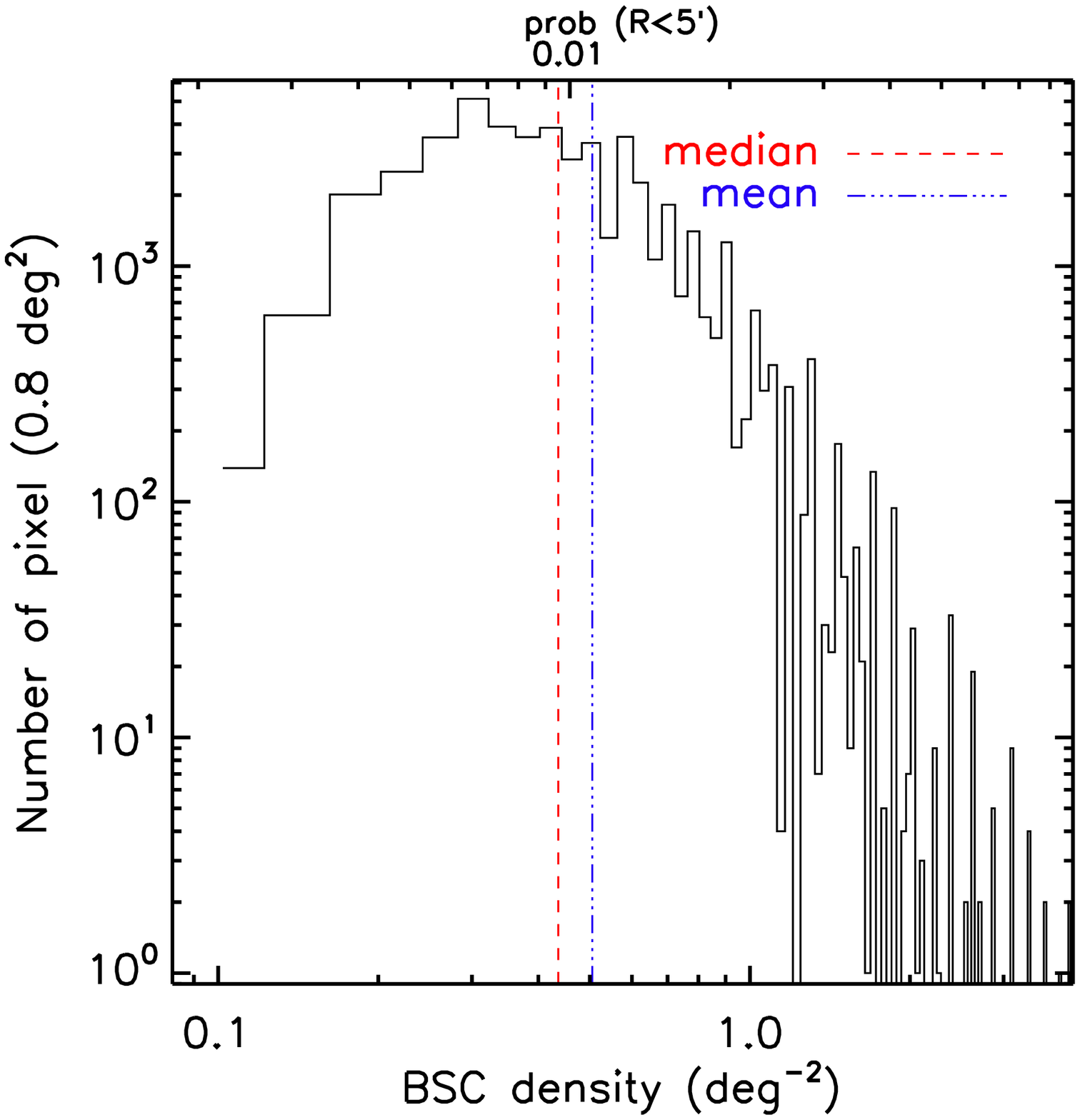}
\hspace{5mm}
\includegraphics{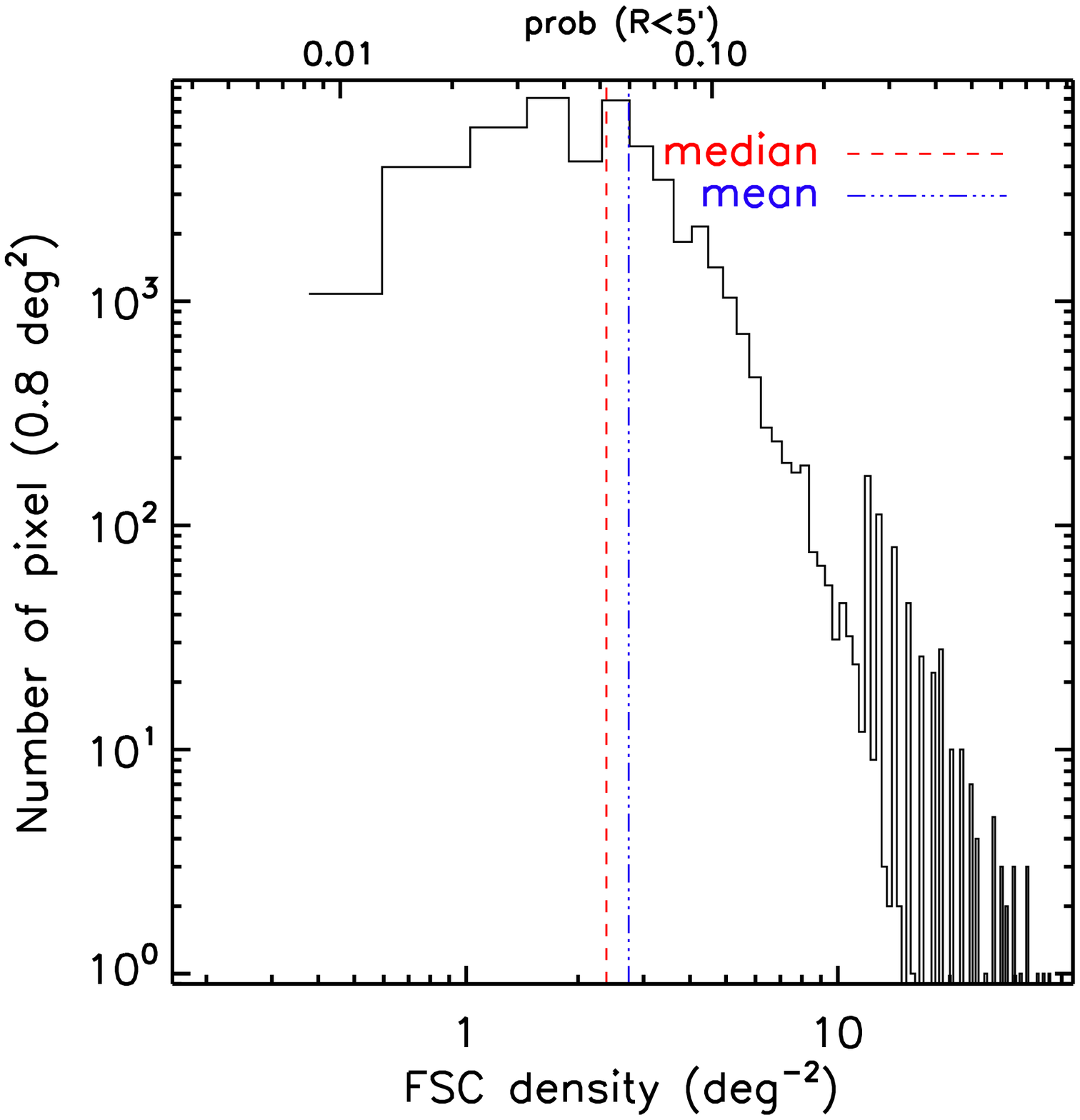}
} 
\end{minipage}
\caption{{\footnotesize Histogram of the source density map of the \rass-BSC ({\it left panel}), and \rass-FSC ({\it right panel}), per square degree. The mean and median source density of each map are plotted in blue dot-dot-dot-dash and in red dashed lines, respectively. The upper $x$--axis shows the associated probability of association within 5\arcmin\ (see text). The sources are drawn from the whole sky so the solid angle is $4\pi$ steradian.}}
\label{fig:histo}
\end{figure*}

In this Appendix we describe the procedure used to calculate the density maps of RASS-BSC and FSC sources, and the associated probability of false association with a \planck\ cluster candidate. We use the catalogues downloaded from Vizier\footnote{\url{http://vizier.u-strasbourg.fr}}. 

\subsection{Source density maps}

To compute the source density maps, we use HEALPix\footnote{\url{http://healpix.jpl.nasa.gov/}} with a resolution of $N_{\rm side} = 64$ (each pixel is 0.8\,$\mathrm{deg^2}$). The HEALPix function {\tt ANG2PIX\_RING} was used to compute the pixel number corresponding to the coordinates of the FSC/BSC sources.

At each pixel, we compute the source density by summing the number of sources in the pixels inside a disc of increasing radius until a threshold number of 10 sources is reached. The source density is then the number of sources found, $N_{\rm src}$, divided by the number of pixels, $N_{\rm pix}$, normalised by the area covered by one pixel:
\begin{equation}
\rho = \frac{N_{\rm src}}{N_{\rm pix}} \times (49,152 / 4\pi) \times \left(\pi/180\right)^2, 
\end{equation}
where 49,152 is the total number of sky pixels for this resolution and $4\pi (180/\pi)^2 \approx 41,000\,\mathrm{deg^2}$ is the total area of the sky.  This gives the mean number of sources per square degree in each pixel. 

The resulting source density maps are plotted in Figs.~\ref{fig:FSCmap} and \ref{fig:BSCmap}. For the FSC density map, the mean source density per square degree ranges from 0.16 to 42.89. There is a clear correspondence between the source density and the depth of the RASS exposure, with regions of maximum source density lying in the regions of maximum RASS exposure at the ecliptic poles (Fig.~\ref{fig:FSCmap}). For the BSC density map, the mean source density per square degree ranges from 0.08 to 4.05, with a much less marked correspondence with the RASS exposure map (Fig.~\ref{fig:FSCmap}).

Figure~\ref{fig:histo} shows the histogram of the number of pixels as a function of mean source density per square degree. We overplot on these histograms the mean ($\bar{\rho}$) and the median ($\rho_{1/2}$) value of the number of sources per square degree. We find $\bar{\rho}\approx \rho_{1/2} \sim 2\,{\rm sources}\,\mathrm{deg^{-2}}$ for the FSC and $\bar{\rho}\approx \rho_{1/2} \sim 0.5\,{\rm sources}\,\mathrm{deg^{-2}}$ for the BSC.

%---------------
\begin{table}
\center
\caption{Summary of the probability of chance association within 5\arcmin\ for the \rass-FSC and the BSC.}
\label{tab:RASSprob}
\resizebox{\columnwidth}{!} {
\begin{tabular}{l ccc c}
\toprule
Catalogue & min prob & max prob & mean prob & median prob \\
\midrule
FSC & 0.004 & 0.936 & 0.060 & 0.052 \\
BSC & 0.002 & 0.088 & 0.011 & 0.010 \\
\bottomrule
\end{tabular}
}
\end{table}
%---------------

\subsection{Probability of association within search radius $\mathcal{R}$}

We can convert the local FSC/BSC source densities   into probabilities of chance association of an SZ candidate with a FSC/BSC source.  The probability of finding a cataloged FSC/BSC source within a search radius  $\mathcal{R}$ of a \planck\  cluster candidate is the product of  the FSC/BSC source density at the candidate location by the search area,  $\mathcal{S(R)}$. This yields a mean probability of association of an SZ candidate with a B/FSC over the full sky of $\mathcal{S(R}\leq 5')\times\bar{\rho} \sim 5\,\%$ for the FSC and 1\,\% for the BSC. However, there is considerable variation depending on how well a given sky region is covered. In the most covered regions, the probability reaches nearly 95\,\% of having an association within 5\arcmin\  for the FSC and 9\,\% for the BSC, while it decreases to 0.4\,\% and 0.2\,\% for the less covered regions for the FSC and BSC catalogues, respectively. We summarise these numbers in Table~\ref{tab:RASSprob}.

\raggedright
\end{document}

%% file: Planck.tex
\def\setsymbol#1#2{\expandafter\def\csname #1\endcsname{#2}}
\def\getsymbol#1{\csname #1\endcsname}

\def\Planck{{\it Planck\/}}

\newbox\tablebox    \newdimen\tablewidth
\def\leaderfil{\leaders\hbox to 5pt{\hss.\hss}\hfil}
\def\endPlancktable{\tablewidth=\columnwidth 
    $$\hss\copy\tablebox\hss$$
    \vskip-\lastskip\vskip -2pt}

\def\tablenote#1 #2\par{\begingroup \parindent=0.8em
    \abovedisplayshortskip=0pt\belowdisplayshortskip=0pt
    \noindent
    $$\hss\vbox{\hsize\tablewidth \hangindent=\parindent \hangafter=1 \noindent
    \hbox to \parindent{\sup{\rm #1}\hss}\strut#2\strut\par}\hss$$
    \endgroup}

\def\L2{\ifmmode L_2\else $L_2$\fi}

\def\DeltaT{\ifmmode \Delta T\else $\Delta T$\fi}
\def\deltat{\ifmmode \Delta t\else $\Delta t$\fi}
\def\fknee{\ifmmode f_{\rm knee}\else $f_{\rm knee}$\fi}
\def\Fmax{\ifmmode F_{\rm max}\else $F_{\rm max}$\fi}
\def\solar{\ifmmode{\rm M}_{\mathord\odot}\else${\rm M}_{\mathord\odot}$\fi}

\def\inv{\ifmmode^{-1}\else$^{-1}$\fi}
\def\mo{\ifmmode^{-1}\else$^{-1}$\fi}
\def\sup#1{\ifmmode ^{\rm #1}\else $^{\rm #1}$\fi}
\def\expo#1{\ifmmode \times 10^{#1}\else $\times 10^{#1}$\fi}
\def\,{\thinspace}
\def\lsim{\mathrel{\raise .4ex\hbox{\rlap{$<$}\lower 1.2ex\hbox{$\sim$}}}}
\def\gsim{\mathrel{\raise .4ex\hbox{\rlap{$>$}\lower 1.2ex\hbox{$\sim$}}}}

\def\simprop{\mathrel{\raise .4ex\hbox{\rlap{$\propto$}\lower 1.2ex\hbox{$\sim$}}}}
\def\deg{\ifmmode^\circ\else$^\circ$\fi}
\def\pdeg{\ifmmode $\setbox0=\hbox{$^{\circ}$}\rlap{\hskip.11\wd0 .}$^{\circ}
          \else \setbox0=\hbox{$^{\circ}$}\rlap{\hskip.11\wd0 .}$^{\circ}$\fi}
\def\arcs{\ifmmode {^{\scriptstyle\prime\prime}}
          \else $^{\scriptstyle\prime\prime}$\fi}
\def\arcm{\ifmmode {^{\scriptstyle\prime}}
          \else $^{\scriptstyle\prime}$\fi}
\newdimen\sa  \newdimen\sb
\def\parcs{\sa=.07em \sb=.03em
     \ifmmode \hbox{\rlap{.}}^{\scriptstyle\prime\kern -\sb\prime}\hbox{\kern -\sa}
     \else \rlap{.}$^{\scriptstyle\prime\kern -\sb\prime}$\kern -\sa\fi}
\def\parcm{\sa=.08em \sb=.03em
     \ifmmode \hbox{\rlap{.}\kern\sa}^{\scriptstyle\prime}\hbox{\kern-\sb}
     \else \rlap{.}\kern\sa$^{\scriptstyle\prime}$\kern-\sb\fi}
\def\ra[#1 #2 #3.#4]{#1\sup{h}#2\sup{m}#3\sup{s}\llap.#4}
\def\dec[#1 #2 #3.#4]{#1\deg#2\arcm#3\arcs\llap.#4}
\def\deco[#1 #2 #3]{#1\deg#2\arcm#3\arcs}
\def\rra[#1 #2]{#1\sup{h}#2\sup{m}}

\def\dots{\relax\ifmmode \ldots\else $\ldots$\fi}
\def\WHzsr{\ifmmode $W\,Hz\mo\,sr\mo$\else W\,Hz\mo\,sr\mo\fi}
\def\mHz{\ifmmode $\,mHz$\else \,mHz\fi}
\def\GHz{\ifmmode $\,GHz$\else \,GHz\fi}
\def\mKs{\ifmmode $\,mK\,s$^{1/2}\else \,mK\,s$^{1/2}$\fi}
\def\muKs{\ifmmode \,\mu$K\,s$^{1/2}\else \,$\mu$K\,s$^{1/2}$\fi}
\def\muKRJs{\ifmmode \,\mu$K$_{\rm RJ}$\,s$^{1/2}\else \,$\mu$K$_{\rm RJ}$\,s$^{1/2}$\fi}
\def\muKHz{\ifmmode \,\mu$K\,Hz$^{-1/2}\else \,$\mu$K\,Hz$^{-1/2}$\fi}
\def\MJysr{\ifmmode \,$MJy\,sr\mo$\else \,MJy\,sr\mo\fi}
\def\MJysrmK{\ifmmode \,$MJy\,sr\mo$\,mK$_{\rm CMB}\mo\else \,MJy\,sr\mo\,mK$_{\rm CMB}\mo$\fi}
\def\microns{\ifmmode \,\mu$m$\else \,$\mu$m\fi}

\def\muK{\ifmmode \,\mu$K$\else \,$\mu$\hbox{K}\fi}
\def\microK{\ifmmode \,\mu$K$\else \,$\mu$\hbox{K}\fi}
\def\muW{\ifmmode \,\mu$W$\else \,$\mu$\hbox{W}\fi}
\def\kms{\ifmmode $\,km\,s$^{-1}\else \,km\,s$^{-1}$\fi}
\def\kmsMpc{\ifmmode $\,\kms\,Mpc\mo$\else \,\kms\,Mpc\mo\fi}
\setsymbol{LFI:center:frequency:70GHz:units}{70.3\,GHz}
\setsymbol{LFI:center:frequency:44GHz:units}{44.1\,GHz}
\setsymbol{LFI:center:frequency:30GHz:units}{28.5\,GHz}

\setsymbol{LFI:center:frequency:70GHz}{70.3}
\setsymbol{LFI:center:frequency:44GHz}{44.1}
\setsymbol{LFI:center:frequency:30GHz}{28.5}

\setsymbol{LFI:center:frequency:LFI18:Rad:M:units}{71.7\GHz}
\setsymbol{LFI:center:frequency:LFI19:Rad:M:units}{67.5\GHz}
\setsymbol{LFI:center:frequency:LFI20:Rad:M:units}{69.2\GHz}
\setsymbol{LFI:center:frequency:LFI21:Rad:M:units}{70.4\GHz}
\setsymbol{LFI:center:frequency:LFI22:Rad:M:units}{71.5\GHz}
\setsymbol{LFI:center:frequency:LFI23:Rad:M:units}{70.8\GHz}
\setsymbol{LFI:center:frequency:LFI24:Rad:M:units}{44.4\GHz}
\setsymbol{LFI:center:frequency:LFI25:Rad:M:units}{44.0\GHz}
\setsymbol{LFI:center:frequency:LFI26:Rad:M:units}{43.9\GHz}
\setsymbol{LFI:center:frequency:LFI27:Rad:M:units}{28.3\GHz}
\setsymbol{LFI:center:frequency:LFI28:Rad:M:units}{28.8\GHz}
\setsymbol{LFI:center:frequency:LFI18:Rad:S:units}{70.1\GHz}
\setsymbol{LFI:center:frequency:LFI19:Rad:S:units}{69.6\GHz}
\setsymbol{LFI:center:frequency:LFI20:Rad:S:units}{69.5\GHz}
\setsymbol{LFI:center:frequency:LFI21:Rad:S:units}{69.5\GHz}
\setsymbol{LFI:center:frequency:LFI22:Rad:S:units}{72.8\GHz}
\setsymbol{LFI:center:frequency:LFI23:Rad:S:units}{71.3\GHz}
\setsymbol{LFI:center:frequency:LFI24:Rad:S:units}{44.1\GHz}
\setsymbol{LFI:center:frequency:LFI25:Rad:S:units}{44.1\GHz}
\setsymbol{LFI:center:frequency:LFI26:Rad:S:units}{44.1\GHz}
\setsymbol{LFI:center:frequency:LFI27:Rad:S:units}{28.5\GHz}
\setsymbol{LFI:center:frequency:LFI28:Rad:S:units}{28.2\GHz}

\setsymbol{LFI:center:frequency:LFI18:Rad:M}{71.7}
\setsymbol{LFI:center:frequency:LFI19:Rad:M}{67.5}
\setsymbol{LFI:center:frequency:LFI20:Rad:M}{69.2}
\setsymbol{LFI:center:frequency:LFI21:Rad:M}{70.4}
\setsymbol{LFI:center:frequency:LFI22:Rad:M}{71.5}
\setsymbol{LFI:center:frequency:LFI23:Rad:M}{70.8}
\setsymbol{LFI:center:frequency:LFI24:Rad:M}{44.4}
\setsymbol{LFI:center:frequency:LFI25:Rad:M}{44.0}
\setsymbol{LFI:center:frequency:LFI26:Rad:M}{43.9}
\setsymbol{LFI:center:frequency:LFI27:Rad:M}{28.3}
\setsymbol{LFI:center:frequency:LFI28:Rad:M}{28.8}
\setsymbol{LFI:center:frequency:LFI18:Rad:S}{70.1}
\setsymbol{LFI:center:frequency:LFI19:Rad:S}{69.6}
\setsymbol{LFI:center:frequency:LFI20:Rad:S}{69.5}
\setsymbol{LFI:center:frequency:LFI21:Rad:S}{69.5}
\setsymbol{LFI:center:frequency:LFI22:Rad:S}{72.8}
\setsymbol{LFI:center:frequency:LFI23:Rad:S}{71.3}
\setsymbol{LFI:center:frequency:LFI24:Rad:S}{44.1}
\setsymbol{LFI:center:frequency:LFI25:Rad:S}{44.1}
\setsymbol{LFI:center:frequency:LFI26:Rad:S}{44.1}
\setsymbol{LFI:center:frequency:LFI27:Rad:S}{28.5}
\setsymbol{LFI:center:frequency:LFI28:Rad:S}{28.2}

\setsymbol{LFI:white:noise:sensitivity:70GHz:units}{152.6\muKs}
\setsymbol{LFI:white:noise:sensitivity:44GHz:units}{173.1\muKs}
\setsymbol{LFI:white:noise:sensitivity:30GHz:units}{146.8\muKs}

\setsymbol{LFI:white:noise:sensitivity:70GHz}{152.6}
\setsymbol{LFI:white:noise:sensitivity:44GHz}{173.1}
\setsymbol{LFI:white:noise:sensitivity:30GHz}{146.8}

\setsymbol{LFI:white:noise:sensitivity:LFI18:Rad:M:units}{512.0\muKs}
\setsymbol{LFI:white:noise:sensitivity:LFI19:Rad:M:units}{581.4\muKs}
\setsymbol{LFI:white:noise:sensitivity:LFI20:Rad:M:units}{590.8\muKs}
\setsymbol{LFI:white:noise:sensitivity:LFI21:Rad:M:units}{455.2\muKs}
\setsymbol{LFI:white:noise:sensitivity:LFI22:Rad:M:units}{492.0\muKs}
\setsymbol{LFI:white:noise:sensitivity:LFI23:Rad:M:units}{507.7\muKs}
\setsymbol{LFI:white:noise:sensitivity:LFI24:Rad:M:units}{462.2\muKs}
\setsymbol{LFI:white:noise:sensitivity:LFI25:Rad:M:units}{413.6\muKs}
\setsymbol{LFI:white:noise:sensitivity:LFI26:Rad:M:units}{478.6\muKs}
\setsymbol{LFI:white:noise:sensitivity:LFI27:Rad:M:units}{277.7\muKs}
\setsymbol{LFI:white:noise:sensitivity:LFI28:Rad:M:units}{312.3\muKs}
\setsymbol{LFI:white:noise:sensitivity:LFI18:Rad:S:units}{465.7\muKs}
\setsymbol{LFI:white:noise:sensitivity:LFI19:Rad:S:units}{555.6\muKs}
\setsymbol{LFI:white:noise:sensitivity:LFI20:Rad:S:units}{623.2\muKs}
\setsymbol{LFI:white:noise:sensitivity:LFI21:Rad:S:units}{564.1\muKs}
\setsymbol{LFI:white:noise:sensitivity:LFI22:Rad:S:units}{534.4\muKs}
\setsymbol{LFI:white:noise:sensitivity:LFI23:Rad:S:units}{542.4\muKs}
\setsymbol{LFI:white:noise:sensitivity:LFI24:Rad:S:units}{399.2\muKs}
\setsymbol{LFI:white:noise:sensitivity:LFI25:Rad:S:units}{392.6\muKs}
\setsymbol{LFI:white:noise:sensitivity:LFI26:Rad:S:units}{418.6\muKs}
\setsymbol{LFI:white:noise:sensitivity:LFI27:Rad:S:units}{302.9\muKs}
\setsymbol{LFI:white:noise:sensitivity:LFI28:Rad:S:units}{285.3\muKs}

\setsymbol{LFI:white:noise:sensitivity:LFI18:Rad:M}{512.0}
\setsymbol{LFI:white:noise:sensitivity:LFI19:Rad:M}{581.4}
\setsymbol{LFI:white:noise:sensitivity:LFI20:Rad:M}{590.8}
\setsymbol{LFI:white:noise:sensitivity:LFI21:Rad:M}{455.2}
\setsymbol{LFI:white:noise:sensitivity:LFI22:Rad:M}{492.0}
\setsymbol{LFI:white:noise:sensitivity:LFI23:Rad:M}{507.7}
\setsymbol{LFI:white:noise:sensitivity:LFI24:Rad:M}{462.2}
\setsymbol{LFI:white:noise:sensitivity:LFI25:Rad:M}{413.6}
\setsymbol{LFI:white:noise:sensitivity:LFI26:Rad:M}{478.6}
\setsymbol{LFI:white:noise:sensitivity:LFI27:Rad:M}{277.7}
\setsymbol{LFI:white:noise:sensitivity:LFI28:Rad:M}{312.3}
\setsymbol{LFI:white:noise:sensitivity:LFI18:Rad:S}{465.7}
\setsymbol{LFI:white:noise:sensitivity:LFI19:Rad:S}{555.6}
\setsymbol{LFI:white:noise:sensitivity:LFI20:Rad:S}{623.2}
\setsymbol{LFI:white:noise:sensitivity:LFI21:Rad:S}{564.1}
\setsymbol{LFI:white:noise:sensitivity:LFI22:Rad:S}{534.4}
\setsymbol{LFI:white:noise:sensitivity:LFI23:Rad:S}{542.4}
\setsymbol{LFI:white:noise:sensitivity:LFI24:Rad:S}{399.2}
\setsymbol{LFI:white:noise:sensitivity:LFI25:Rad:S}{392.6}
\setsymbol{LFI:white:noise:sensitivity:LFI26:Rad:S}{418.6}
\setsymbol{LFI:white:noise:sensitivity:LFI27:Rad:S}{302.9}
\setsymbol{LFI:white:noise:sensitivity:LFI28:Rad:S}{285.3}

\setsymbol{LFI:knee:frequency:70GHz:units}{29.5\mHz}
\setsymbol{LFI:knee:frequency:44GHz:units}{56.2\mHz}
\setsymbol{LFI:knee:frequency:30GHz:units}{113.7\mHz}

\setsymbol{LFI:knee:frequency:70GHz}{29.5}
\setsymbol{LFI:knee:frequency:44GHz}{56.2}
\setsymbol{LFI:knee:frequency:30GHz}{113.7}

\setsymbol{LFI:knee:frequency:LFI18:Rad:M:units}{16.3\mHz}
\setsymbol{LFI:knee:frequency:LFI19:Rad:M:units}{15.1\mHz}
\setsymbol{LFI:knee:frequency:LFI20:Rad:M:units}{18.7\mHz}
\setsymbol{LFI:knee:frequency:LFI21:Rad:M:units}{37.2\mHz}
\setsymbol{LFI:knee:frequency:LFI22:Rad:M:units}{12.7\mHz}
\setsymbol{LFI:knee:frequency:LFI23:Rad:M:units}{34.6\mHz}
\setsymbol{LFI:knee:frequency:LFI24:Rad:M:units}{46.2\mHz}
\setsymbol{LFI:knee:frequency:LFI25:Rad:M:units}{24.9\mHz}
\setsymbol{LFI:knee:frequency:LFI26:Rad:M:units}{67.6\mHz}
\setsymbol{LFI:knee:frequency:LFI27:Rad:M:units}{187.4\mHz}
\setsymbol{LFI:knee:frequency:LFI28:Rad:M:units}{122.2\mHz}
\setsymbol{LFI:knee:frequency:LFI18:Rad:S:units}{17.7\mHz}
\setsymbol{LFI:knee:frequency:LFI19:Rad:S:units}{22.0\mHz}
\setsymbol{LFI:knee:frequency:LFI20:Rad:S:units}{8.7\mHz}
\setsymbol{LFI:knee:frequency:LFI21:Rad:S:units}{25.9\mHz}
\setsymbol{LFI:knee:frequency:LFI22:Rad:S:units}{15.8\mHz}
\setsymbol{LFI:knee:frequency:LFI23:Rad:S:units}{129.8\mHz}
\setsymbol{LFI:knee:frequency:LFI24:Rad:S:units}{100.9\mHz}
\setsymbol{LFI:knee:frequency:LFI25:Rad:S:units}{38.9\mHz}
\setsymbol{LFI:knee:frequency:LFI26:Rad:S:units}{58.9\mHz}
\setsymbol{LFI:knee:frequency:LFI27:Rad:S:units}{104.4\mHz}
\setsymbol{LFI:knee:frequency:LFI28:Rad:S:units}{40.7\mHz}

\setsymbol{LFI:knee:frequency:LFI18:Rad:M}{16.3}
\setsymbol{LFI:knee:frequency:LFI19:Rad:M}{15.1}
\setsymbol{LFI:knee:frequency:LFI20:Rad:M}{18.7}
\setsymbol{LFI:knee:frequency:LFI21:Rad:M}{37.2}
\setsymbol{LFI:knee:frequency:LFI22:Rad:M}{12.7}
\setsymbol{LFI:knee:frequency:LFI23:Rad:M}{34.6}
\setsymbol{LFI:knee:frequency:LFI24:Rad:M}{46.2}
\setsymbol{LFI:knee:frequency:LFI25:Rad:M}{24.9}
\setsymbol{LFI:knee:frequency:LFI26:Rad:M}{67.6}
\setsymbol{LFI:knee:frequency:LFI27:Rad:M}{187.4}
\setsymbol{LFI:knee:frequency:LFI28:Rad:M}{122.2}
\setsymbol{LFI:knee:frequency:LFI18:Rad:S}{17.7}
\setsymbol{LFI:knee:frequency:LFI19:Rad:S}{22.0}
\setsymbol{LFI:knee:frequency:LFI20:Rad:S}{8.7}
\setsymbol{LFI:knee:frequency:LFI21:Rad:S}{25.9}
\setsymbol{LFI:knee:frequency:LFI22:Rad:S}{15.8}
\setsymbol{LFI:knee:frequency:LFI23:Rad:S}{129.8}
\setsymbol{LFI:knee:frequency:LFI24:Rad:S}{100.9}
\setsymbol{LFI:knee:frequency:LFI25:Rad:S}{38.9}
\setsymbol{LFI:knee:frequency:LFI26:Rad:S}{58.9}
\setsymbol{LFI:knee:frequency:LFI27:Rad:S}{104.4}
\setsymbol{LFI:knee:frequency:LFI28:Rad:S}{40.7}

\setsymbol{LFI:slope:70GHz:units}{$-1.03$\mHz}
\setsymbol{LFI:slope:44GHz:units}{$-0.89$\mHz}
\setsymbol{LFI:slope:30GHz:units}{$-0.87$\mHz}

\setsymbol{LFI:slope:70GHz}{$-1.03$}
\setsymbol{LFI:slope:44GHz}{$-0.89$}
\setsymbol{LFI:slope:30GHz}{$-0.87$}

\setsymbol{LFI:slope:LFI18:Rad:M:units}{$-1.04$\mHz}
\setsymbol{LFI:slope:LFI19:Rad:M:units}{$-1.09$\mHz}
\setsymbol{LFI:slope:LFI20:Rad:M:units}{$-0.69$\mHz}
\setsymbol{LFI:slope:LFI21:Rad:M:units}{$-1.56$\mHz}
\setsymbol{LFI:slope:LFI22:Rad:M:units}{$-1.01$\mHz}
\setsymbol{LFI:slope:LFI23:Rad:M:units}{$-0.96$\mHz}
\setsymbol{LFI:slope:LFI24:Rad:M:units}{$-0.83$\mHz}
\setsymbol{LFI:slope:LFI25:Rad:M:units}{$-0.91$\mHz}
\setsymbol{LFI:slope:LFI26:Rad:M:units}{$-0.95$\mHz}
\setsymbol{LFI:slope:LFI27:Rad:M:units}{$-0.87$\mHz}
\setsymbol{LFI:slope:LFI28:Rad:M:units}{$-0.88$\mHz}
\setsymbol{LFI:slope:LFI18:Rad:S:units}{$-1.15$\mHz}
\setsymbol{LFI:slope:LFI19:Rad:S:units}{$-1.00$\mHz}
\setsymbol{LFI:slope:LFI20:Rad:S:units}{$-0.95$\mHz}
\setsymbol{LFI:slope:LFI21:Rad:S:units}{$-0.92$\mHz}
\setsymbol{LFI:slope:LFI22:Rad:S:units}{$-1.01$\mHz}
\setsymbol{LFI:slope:LFI23:Rad:S:units}{$-0.95$\mHz}
\setsymbol{LFI:slope:LFI24:Rad:S:units}{$-0.73$\mHz}
\setsymbol{LFI:slope:LFI25:Rad:S:units}{$-1.16$\mHz}
\setsymbol{LFI:slope:LFI26:Rad:S:units}{$-0.79$\mHz}
\setsymbol{LFI:slope:LFI27:Rad:S:units}{$-0.82$\mHz}
\setsymbol{LFI:slope:LFI28:Rad:S:units}{$-0.91$\mHz}

\setsymbol{LFI:slope:LFI18:Rad:M}{$-1.04$}
\setsymbol{LFI:slope:LFI19:Rad:M}{$-1.09$}
\setsymbol{LFI:slope:LFI20:Rad:M}{$-0.69$}
\setsymbol{LFI:slope:LFI21:Rad:M}{$-1.56$}
\setsymbol{LFI:slope:LFI22:Rad:M}{$-1.01$}
\setsymbol{LFI:slope:LFI23:Rad:M}{$-0.96$}
\setsymbol{LFI:slope:LFI24:Rad:M}{$-0.83$}
\setsymbol{LFI:slope:LFI25:Rad:M}{$-0.91$}
\setsymbol{LFI:slope:LFI26:Rad:M}{$-0.95$}
\setsymbol{LFI:slope:LFI27:Rad:M}{$-0.87$}
\setsymbol{LFI:slope:LFI28:Rad:M}{$-0.88$}
\setsymbol{LFI:slope:LFI18:Rad:S}{$-1.15$}
\setsymbol{LFI:slope:LFI19:Rad:S}{$-1.00$}
\setsymbol{LFI:slope:LFI20:Rad:S}{$-0.95$}
\setsymbol{LFI:slope:LFI21:Rad:S}{$-0.92$}
\setsymbol{LFI:slope:LFI22:Rad:S}{$-1.01$}
\setsymbol{LFI:slope:LFI23:Rad:S}{$-0.95$}
\setsymbol{LFI:slope:LFI24:Rad:S}{$-0.73$}
\setsymbol{LFI:slope:LFI25:Rad:S}{$-1.16$}
\setsymbol{LFI:slope:LFI26:Rad:S}{$-0.79$}
\setsymbol{LFI:slope:LFI27:Rad:S}{$-0.82$}
\setsymbol{LFI:slope:LFI28:Rad:S}{$-0.91$}

\setsymbol{LFI:FWHM:70GHz:units}{13\parcm01}
\setsymbol{LFI:FWHM:44GHz:units}{27\parcm92}
\setsymbol{LFI:FWHM:30GHz:units}{32\parcm65}

\setsymbol{LFI:FWHM:70GHz}{13.01}
\setsymbol{LFI:FWHM:44GHz}{27.92}
\setsymbol{LFI:FWHM:30GHz}{32.65}

\setsymbol{LFI:FWHM:LFI18:units}{13\parcm39}
\setsymbol{LFI:FWHM:LFI19:units}{13\parcm01}
\setsymbol{LFI:FWHM:LFI20:units}{12\parcm75}
\setsymbol{LFI:FWHM:LFI21:units}{12\parcm74}
\setsymbol{LFI:FWHM:LFI22:units}{12\parcm87}
\setsymbol{LFI:FWHM:LFI23:units}{13\parcm27}
\setsymbol{LFI:FWHM:LFI24:units}{22\parcm98}
\setsymbol{LFI:FWHM:LFI25:units}{30\parcm46}
\setsymbol{LFI:FWHM:LFI26:units}{30\parcm31}
\setsymbol{LFI:FWHM:LFI27:units}{32\parcm65}
\setsymbol{LFI:FWHM:LFI28:units}{32\parcm66}

\setsymbol{LFI:FWHM:LFI18}{13.39}
\setsymbol{LFI:FWHM:LFI19}{13.01}
\setsymbol{LFI:FWHM:LFI20}{12.75}
\setsymbol{LFI:FWHM:LFI21}{12.74}
\setsymbol{LFI:FWHM:LFI22}{12.87}
\setsymbol{LFI:FWHM:LFI23}{13.27}
\setsymbol{LFI:FWHM:LFI24}{22.98}
\setsymbol{LFI:FWHM:LFI25}{30.46}
\setsymbol{LFI:FWHM:LFI26}{30.31}
\setsymbol{LFI:FWHM:LFI27}{32.65}
\setsymbol{LFI:FWHM:LFI28}{32.66}

\setsymbol{LFI:FWHM:uncertainty:LFI18:units}{0.170\arcm}
\setsymbol{LFI:FWHM:uncertainty:LFI19:units}{0.174\arcm}
\setsymbol{LFI:FWHM:uncertainty:LFI20:units}{0.170\arcm}
\setsymbol{LFI:FWHM:uncertainty:LFI21:units}{0.156\arcm}
\setsymbol{LFI:FWHM:uncertainty:LFI22:units}{0.164\arcm}
\setsymbol{LFI:FWHM:uncertainty:LFI23:units}{0.171\arcm}
\setsymbol{LFI:FWHM:uncertainty:LFI24:units}{0.652\arcm}
\setsymbol{LFI:FWHM:uncertainty:LFI25:units}{1.075\arcm}
\setsymbol{LFI:FWHM:uncertainty:LFI26:units}{1.131\arcm}
\setsymbol{LFI:FWHM:uncertainty:LFI27:units}{1.266\arcm}
\setsymbol{LFI:FWHM:uncertainty:LFI28:units}{1.287\arcm}

\setsymbol{LFI:FWHM:uncertainty:LFI18}{0.170}
\setsymbol{LFI:FWHM:uncertainty:LFI19}{0.174}
\setsymbol{LFI:FWHM:uncertainty:LFI20}{0.170}
\setsymbol{LFI:FWHM:uncertainty:LFI21}{0.156}
\setsymbol{LFI:FWHM:uncertainty:LFI22}{0.164}
\setsymbol{LFI:FWHM:uncertainty:LFI23}{0.171}
\setsymbol{LFI:FWHM:uncertainty:LFI24}{0.652}
\setsymbol{LFI:FWHM:uncertainty:LFI25}{1.075}
\setsymbol{LFI:FWHM:uncertainty:LFI26}{1.131}
\setsymbol{LFI:FWHM:uncertainty:LFI27}{1.266}
\setsymbol{LFI:FWHM:uncertainty:LFI28}{1.287}

\setsymbol{HFI:center:frequency:100GHz:units}{100\,GHz}
\setsymbol{HFI:center:frequency:143GHz:units}{143\,GHz}
\setsymbol{HFI:center:frequency:217GHz:units}{217\,GHz}
\setsymbol{HFI:center:frequency:353GHz:units}{353\,GHz}
\setsymbol{HFI:center:frequency:545GHz:units}{545\,GHz}
\setsymbol{HFI:center:frequency:857GHz:units}{857\,GHz}

\setsymbol{HFI:center:frequency:100GHz}{100}
\setsymbol{HFI:center:frequency:143GHz}{143}
\setsymbol{HFI:center:frequency:217GHz}{217}
\setsymbol{HFI:center:frequency:353GHz}{353}
\setsymbol{HFI:center:frequency:545GHz}{545}
\setsymbol{HFI:center:frequency:857GHz}{857}

\setsymbol{HFI:Ndetectors:100GHz}{8}
\setsymbol{HFI:Ndetectors:143GHz}{11}
\setsymbol{HFI:Ndetectors:217GHz}{12}
\setsymbol{HFI:Ndetectors:353GHz}{12}
\setsymbol{HFI:Ndetectors:545GHz}{3}
\setsymbol{HFI:Ndetectors:857GHz}{4}

\setsymbol{HFI:FWHM:Maps:100GHz:units}{9\parcm88}
\setsymbol{HFI:FWHM:Maps:143GHz:units}{7\parcm18}
\setsymbol{HFI:FWHM:Maps:217GHz:units}{4\parcm87}
\setsymbol{HFI:FWHM:Maps:353GHz:units}{4\parcm65}
\setsymbol{HFI:FWHM:Maps:545GHz:units}{4\parcm72}
\setsymbol{HFI:FWHM:Maps:857GHz:units}{4\parcm39}
\setsymbol{HFI:FWHM:Maps:100GHz}{9.88}
\setsymbol{HFI:FWHM:Maps:143GHz}{7.18}
\setsymbol{HFI:FWHM:Maps:217GHz}{4.87}
\setsymbol{HFI:FWHM:Maps:353GHz}{4.65}
\setsymbol{HFI:FWHM:Maps:545GHz}{4.72}
\setsymbol{HFI:FWHM:Maps:857GHz}{4.39}

\setsymbol{HFI:beam:ellipticity:Maps:100GHz}{1.15}
\setsymbol{HFI:beam:ellipticity:Maps:143GHz}{1.01}
\setsymbol{HFI:beam:ellipticity:Maps:217GHz}{1.06}
\setsymbol{HFI:beam:ellipticity:Maps:353GHz}{1.05}
\setsymbol{HFI:beam:ellipticity:Maps:545GHz}{1.14}
\setsymbol{HFI:beam:ellipticity:Maps:857GHz}{1.19}

\setsymbol{HFI:FWHM:Mars:100GHz:units}{9\parcm37}
\setsymbol{HFI:FWHM:Mars:143GHz:units}{7\parcm04}
\setsymbol{HFI:FWHM:Mars:217GHz:units}{4\parcm68}
\setsymbol{HFI:FWHM:Mars:353GHz:units}{4\parcm43}
\setsymbol{HFI:FWHM:Mars:545GHz:units}{3\parcm80}
\setsymbol{HFI:FWHM:Mars:857GHz:units}{3\parcm67}

\setsymbol{HFI:FWHM:Mars:100GHz}{9.37}
\setsymbol{HFI:FWHM:Mars:143GHz}{7.04}
\setsymbol{HFI:FWHM:Mars:217GHz}{4.68}
\setsymbol{HFI:FWHM:Mars:353GHz}{4.43}
\setsymbol{HFI:FWHM:Mars:545GHz}{3.80}
\setsymbol{HFI:FWHM:Mars:857GHz}{3.67}

\setsymbol{HFI:beam:ellipticity:Mars:100GHz}{1.18}
\setsymbol{HFI:beam:ellipticity:Mars:143GHz}{1.03}
\setsymbol{HFI:beam:ellipticity:Mars:217GHz}{1.14}
\setsymbol{HFI:beam:ellipticity:Mars:353GHz}{1.09}
\setsymbol{HFI:beam:ellipticity:Mars:545GHz}{1.25}
\setsymbol{HFI:beam:ellipticity:Mars:857GHz}{1.03}

\setsymbol{HFI:CMB:relative:calibration:100GHz}{$\lsim 1\%$}
\setsymbol{HFI:CMB:relative:calibration:143GHz}{$\lsim 1\%$}
\setsymbol{HFI:CMB:relative:calibration:217GHz}{$\lsim 1\%$}
\setsymbol{HFI:CMB:relative:calibration:353GHz}{$\lsim 1\%$}
\setsymbol{HFI:CMB:relative:calibration:545GHz}{}
\setsymbol{HFI:CMB:relative:calibration:857GHz}{}

\setsymbol{HFI:CMB:absolute:calibration:100GHz}{$\lsim 2\%$}
\setsymbol{HFI:CMB:absolute:calibration:143GHz}{$\lsim 2\%$}
\setsymbol{HFI:CMB:absolute:calibration:217GHz}{$\lsim 2\%$}
\setsymbol{HFI:CMB:absolute:calibration:353GHz}{$\lsim 2\%$}
\setsymbol{HFI:CMB:absolute:calibration:545GHz}{}
\setsymbol{HFI:CMB:absolute:calibration:857GHz}{}

\setsymbol{HFI:FIRAS:gain:calibration:accuracy:statistical:100GHz}{}
\setsymbol{HFI:FIRAS:gain:calibration:accuracy:statistical:143GHz}{}
\setsymbol{HFI:FIRAS:gain:calibration:accuracy:statistical:217GHz}{}
\setsymbol{HFI:FIRAS:gain:calibration:accuracy:statistical:353GHz}{2.5\%}
\setsymbol{HFI:FIRAS:gain:calibration:accuracy:statistical:545GHz}{1\%}
\setsymbol{HFI:FIRAS:gain:calibration:accuracy:statistical:857GHz}{0.5\%}

\setsymbol{HFI:FIRAS:gain:calibration:accuracy:systematic:100GHz}{}
\setsymbol{HFI:FIRAS:gain:calibration:accuracy:systematic:143GHz}{}
\setsymbol{HFI:FIRAS:gain:calibration:accuracy:systematic:217GHz}{}
\setsymbol{HFI:FIRAS:gain:calibration:accuracy:systematic:353GHz}{}
\setsymbol{HFI:FIRAS:gain:calibration:accuracy:systematic:545GHz}{7\%}
\setsymbol{HFI:FIRAS:gain:calibration:accuracy:systematic:857GHz}{7\%}

\setsymbol{HFI:FIRAS:zero:point:accuracy:100GHz:units}{0.8\MJysr}
\setsymbol{HFI:FIRAS:zero:point:accuracy:143GHz:units}{}
\setsymbol{HFI:FIRAS:zero:point:accuracy:217GHz:units}{}
\setsymbol{HFI:FIRAS:zero:point:accuracy:353GHz:units}{1.4\MJysr}
\setsymbol{HFI:FIRAS:zero:point:accuracy:545GHz:units}{2.2\MJysr}
\setsymbol{HFI:FIRAS:zero:point:accuracy:857GHz:units}{1.7\MJysr}

\setsymbol{HFI:FIRAS:zero:point:accuracy:100GHz}{0.8}
\setsymbol{HFI:FIRAS:zero:point:accuracy:143GHz}{}
\setsymbol{HFI:FIRAS:zero:point:accuracy:217GHz}{}
\setsymbol{HFI:FIRAS:zero:point:accuracy:353GHz}{1.4}
\setsymbol{HFI:FIRAS:zero:point:accuracy:545GHz}{2.2}
\setsymbol{HFI:FIRAS:zero:point:accuracy:857GHz}{1.7}

\setsymbol{HFI:unit:conversion:100GHz:units}{0.2415\MJysrmK}
\setsymbol{HFI:unit:conversion:143GHz:units}{0.3694\MJysrmK}
\setsymbol{HFI:unit:conversion:217GHz:units}{0.4811\MJysrmK}
\setsymbol{HFI:unit:conversion:353GHz:units}{0.2883\MJysrmK}
\setsymbol{HFI:unit:conversion:545GHz:units}{0.05826\MJysrmK}
\setsymbol{HFI:unit:conversion:857GHz:units}{0.002238\MJysrmK}

\setsymbol{HFI:unit:conversion:100GHz}{0.2415}
\setsymbol{HFI:unit:conversion:143GHz}{0.3694}
\setsymbol{HFI:unit:conversion:217GHz}{0.4811}
\setsymbol{HFI:unit:conversion:353GHz}{0.2883}
\setsymbol{HFI:unit:conversion:545GHz}{0.05826}
\setsymbol{HFI:unit:conversion:857GHz}{0.002238}

\setsymbol{HFI:colour:correction:alpha=-2:V1.01:100GHz}{0.9893}
\setsymbol{HFI:colour:correction:alpha=-2:V1.01:143GHz}{0.9759}
\setsymbol{HFI:colour:correction:alpha=-2:V1.01:217GHz}{1.0007}
\setsymbol{HFI:colour:correction:alpha=-2:V1.01:353GHz}{1.0028}
\setsymbol{HFI:colour:correction:alpha=-2:V1.01:545GHz}{1.0019}
\setsymbol{HFI:colour:correction:alpha=-2:V1.01:857GHz}{0.9889}

\setsymbol{HFI:colour:correction:alpha=0:V1.01:100GHz}{1.0008}
\setsymbol{HFI:colour:correction:alpha=0:V1.01:143GHz}{1.0148}
\setsymbol{HFI:colour:correction:alpha=0:V1.01:217GHz}{0.9909}
\setsymbol{HFI:colour:correction:alpha=0:V1.01:353GHz}{0.9888}
\setsymbol{HFI:colour:correction:alpha=0:V1.01:545GHz}{0.9878}
\setsymbol{HFI:colour:correction:alpha=0:V1.01:857GHz}{1.0014}

%% file: PIP_22b_authors_and_institutes.tex
%This author list corresponds to \title{Author list for PIP 22b, Proj. Ref. 5.1/5.2/5.5: Planck intermediate results. IV. The XMM-Newton validation programme for new Planck clusters}
%Prepared by R. Leonardi (rleonardi@sciops.esa.int), ESAC/ESA
%This version is from Wed Jul 11 15:42:59 2012 CET
%\subtitle{There are 193 co-authors in this list}
\author{\small
Planck Collaboration:
P.~A.~R.~Ade\inst{85}
\and
N.~Aghanim\inst{59}
\and
M.~Arnaud\inst{74}
\and
M.~Ashdown\inst{71, 6}
\and
J.~Aumont\inst{59}
\and
C.~Baccigalupi\inst{83}
\and
A.~Balbi\inst{35}
\and
A.~J.~Banday\inst{91, 9}
\and
R.~B.~Barreiro\inst{67}
\and
J.~G.~Bartlett\inst{1, 69}
% ...
}
\institute{\small
APC, AstroParticule et Cosmologie, Universit\'{e} Paris Diderot, CNRS/IN2P3, CEA/lrfu, Observatoire de Paris, Sorbonne Paris Cit\'{e}, 10, rue Alice Domon et L\'{e}onie Duquet, 75205 Paris Cedex 13, France\\
\and
Aalto University Mets\"{a}hovi Radio Observatory, Mets\"{a}hovintie 114, FIN-02540 Kylm\"{a}l\"{a}, Finland\\
\and
Academy of Sciences of Tatarstan, Bauman Str., 20, Kazan, 420111, Republic of Tatarstan, Russia\\
\and
Agenzia Spaziale Italiana Science Data Center, c/o ESRIN, via Galileo Galilei, Frascati, Italy\\
\and
Agenzia Spaziale Italiana, Viale Liegi 26, Roma, Italy\\
\and
Astrophysics Group, Cavendish Laboratory, University of Cambridge, J J Thomson Avenue, Cambridge CB3 0HE, U.K.\\
\and
Atacama Large Millimeter/submillimeter Array, ALMA Santiago Central Offices, Alonso de Cordova 3107, Vitacura, Casilla 763 0355, Santiago, Chile\\
\and
CITA, University of Toronto, 60 St. George St., Toronto, ON M5S 3H8, Canada\\
\and
CNRS, IRAP, 9 Av. colonel Roche, BP 44346, F-31028 Toulouse cedex 4, France\\
\and
California Institute of Technology, Pasadena, California, U.S.A.\\
\and
Centre of Mathematics for Applications, University of Oslo, Blindern, Oslo, Norway\\
\and
Centro de Astrof\'{\i}sica, Universidade do Porto, Rua das Estrelas, 4150-762 Porto, Portugal\\
\and
Centro de Estudios de F\'{i}sica del Cosmos de Arag\'{o}n (CEFCA), Plaza San Juan, 1, planta 2, E-44001, Teruel, Spain\\
\and
Computational Cosmology Center, Lawrence Berkeley National Laboratory, Berkeley, California, U.S.A.\\
\and
DSM/Irfu/SPP, CEA-Saclay, F-91191 Gif-sur-Yvette Cedex, France\\
\and
DTU Space, National Space Institute, Juliane Mariesvej 30, Copenhagen, Denmark\\
\and
D\'{e}partement de Physique Th\'{e}orique, Universit\'{e} de Gen\`{e}ve, 24, Quai E. Ansermet,1211 Gen\`{e}ve 4, Switzerland\\
\and
Departamento de F\'{\i}sica, Universidad de Oviedo, Avda. Calvo Sotelo s/n, Oviedo, Spain\\
\and
Department of Astronomy and Geodesy, Kazan Federal University,  Kremlevskaya Str., 18, Kazan, 420008, Russia\\
\and
Department of Physics \& Astronomy, University of British Columbia, 6224 Agricultural Road, Vancouver, British Columbia, Canada\\
\and
Department of Physics and Astronomy, Dana and David Dornsife College of Letter, Arts and Sciences, University of Southern California, Los Angeles, CA 90089, U.S.A.\\
\and
Department of Physics and Astronomy, University of Sussex, Brighton BN1 9QH, U.K.\\
\and
Department of Physics, Gustaf H\"{a}llstr\"{o}min katu 2a, University of Helsinki, Helsinki, Finland\\
\and
Department of Physics, Princeton University, Princeton, New Jersey, U.S.A.\\
\and
Department of Physics, University of California, Berkeley, California, U.S.A.\\
\and
Department of Physics, University of California, One Shields Avenue, Davis, California, U.S.A.\\
\and
Department of Physics, University of California, Santa Barbara, California, U.S.A.\\
\and
Department of Physics, University of Illinois at Urbana-Champaign, 1110 West Green Street, Urbana, Illinois, U.S.A.\\
\and
Department of Statistics, Purdue University, 250 N. University Street, West Lafayette, Indiana, U.S.A.\\
\and
Dipartimento di Fisica e Astronomia G. Galilei, Universit\`{a} degli Studi di Padova, via Marzolo 8, 35131 Padova, Italy\\
\and
Dipartimento di Fisica, Universit\`{a} La Sapienza, P. le A. Moro 2, Roma, Italy\\
\and
Dipartimento di Fisica, Universit\`{a} degli Studi di Milano, Via Celoria, 16, Milano, Italy\\
\and
Dipartimento di Fisica, Universit\`{a} degli Studi di Trieste, via A. Valerio 2, Trieste, Italy\\
\and
Dipartimento di Fisica, Universit\`{a} di Ferrara, Via Saragat 1, 44122 Ferrara, Italy\\
\and
Dipartimento di Fisica, Universit\`{a} di Roma Tor Vergata, Via della Ricerca Scientifica, 1, Roma, Italy\\
\and
Dipartimento di Matematica, Universit\`{a} di Roma Tor Vergata, Via della Ricerca Scientifica, 1, Roma, Italy\\
\and
Discovery Center, Niels Bohr Institute, Blegdamsvej 17, Copenhagen, Denmark\\
\and
Dpto. Astrof\'{i}sica, Universidad de La Laguna (ULL), E-38206 La Laguna, Tenerife, Spain\\
\and
European Southern Observatory, ESO Vitacura, Alonso de Cordova 3107, Vitacura, Casilla 19001, Santiago, Chile\\
\and
European Space Agency, ESAC, Camino bajo del Castillo, s/n, Urbanizaci\'{o}n Villafranca del Castillo, Villanueva de la Ca\~{n}ada, Madrid, Spain\\
\and
European Space Agency, ESAC, Planck Science Office, Camino bajo del Castillo, s/n, Urbanizaci\'{o}n Villafranca del Castillo, Villanueva de la Ca\~{n}ada, Madrid, Spain\\
\and
European Space Agency, ESTEC, Keplerlaan 1, 2201 AZ Noordwijk, The Netherlands\\
\and
GEPI, Observatoire de Paris, Section de Meudon, 5 Place J. Janssen, 92195 Meudon Cedex, France\\
\and
Helsinki Institute of Physics, Gustaf H\"{a}llstr\"{o}min katu 2, University of Helsinki, Helsinki, Finland\\
\and
INAF - Osservatorio Astronomico di Padova, Vicolo dell'Osservatorio 5, Padova, Italy\\
\and
INAF - Osservatorio Astronomico di Roma, via di Frascati 33, Monte Porzio Catone, Italy\\
\and
INAF - Osservatorio Astronomico di Trieste, Via G.B. Tiepolo 11, Trieste, Italy\\
\and
INAF Istituto di Radioastronomia, Via P. Gobetti 101, 40129 Bologna, Italy\\
\and
INAF/IASF Bologna, Via Gobetti 101, Bologna, Italy\\
\and
INAF/IASF Milano, Via E. Bassini 15, Milano, Italy\\
\and
INFN, Sezione di Roma 1, Universit`{a} di Roma Sapienza, Piazzale Aldo Moro 2, 00185, Roma, Italy\\
\and
INRIA, Laboratoire de Recherche en Informatique, Universit\'{e} Paris-Sud 11, B\^{a}timent 490, 91405 Orsay Cedex, France\\
\and
IPAG: Institut de Plan\'{e}tologie et d'Astrophysique de Grenoble, Universit\'{e} Joseph Fourier, Grenoble 1 / CNRS-INSU, UMR 5274, Grenoble, F-38041, France\\
\and
IUCAA, Post Bag 4, Ganeshkhind, Pune University Campus, Pune 411 007, India\\
\and
Imperial College London, Astrophysics group, Blackett Laboratory, Prince Consort Road, London, SW7 2AZ, U.K.\\
\and
Infrared Processing and Analysis Center, California Institute of Technology, Pasadena, CA 91125, U.S.A.\\
\and
Institut N\'{e}el, CNRS, Universit\'{e} Joseph Fourier Grenoble I, 25 rue des Martyrs, Grenoble, France\\
\and
Institut Universitaire de France, 103, bd Saint-Michel, 75005, Paris, France\\
\and
Institut d'Astrophysique Spatiale, CNRS (UMR8617) Universit\'{e} Paris-Sud 11, B\^{a}timent 121, Orsay, France\\
\and
Institut d'Astrophysique de Paris, CNRS (UMR7095), 98 bis Boulevard Arago, F-75014, Paris, France\\
\and
Institute for Space Sciences, Bucharest-Magurale, Romania\\
\and
Institute of Astro and Particle Physics, Technikerstrasse 25/8, University of Innsbruck, A-6020, Innsbruck, Austria\\
\and
Institute of Astronomy and Astrophysics, Academia Sinica, Taipei, Taiwan\\
\and
Institute of Astronomy, University of Cambridge, Madingley Road, Cambridge CB3 0HA, U.K.\\
\and
Institute of Theoretical Astrophysics, University of Oslo, Blindern, Oslo, Norway\\
\and
Instituto de Astrof\'{\i}sica de Canarias, C/V\'{\i}a L\'{a}ctea s/n, La Laguna, Tenerife, Spain\\
\and
Instituto de F\'{\i}sica de Cantabria (CSIC-Universidad de Cantabria), Avda. de los Castros s/n, Santander, Spain\\
\and
Istituto di Fisica del Plasma, CNR-ENEA-EURATOM Association, Via R. Cozzi 53, Milano, Italy\\
\and
Jet Propulsion Laboratory, California Institute of Technology, 4800 Oak Grove Drive, Pasadena, California, U.S.A.\\
\and
Jodrell Bank Centre for Astrophysics, Alan Turing Building, School of Physics and Astronomy, The University of Manchester, Oxford Road, Manchester, M13 9PL, U.K.\\
\and
Kavli Institute for Cosmology Cambridge, Madingley Road, Cambridge, CB3 0HA, U.K.\\
\and
LAL, Universit\'{e} Paris-Sud, CNRS/IN2P3, Orsay, France\\
\and
LERMA, CNRS, Observatoire de Paris, 61 Avenue de l'Observatoire, Paris, France\\
\and
Laboratoire AIM, IRFU/Service d'Astrophysique - CEA/DSM - CNRS - Universit\'{e} Paris Diderot, B\^{a}t. 709, CEA-Saclay, F-91191 Gif-sur-Yvette Cedex, France\\
\and
Laboratoire de Physique Subatomique et de Cosmologie, Universit\'{e} Joseph Fourier Grenoble I, CNRS/IN2P3, Institut National Polytechnique de Grenoble, 53 rue des Martyrs, 38026 Grenoble cedex, France\\
\and
Laboratoire de Physique Th\'{e}orique, Universit\'{e} Paris-Sud 11 \& CNRS, B\^{a}timent 210, 91405 Orsay, France\\
\and
Lawrence Berkeley National Laboratory, Berkeley, California, U.S.A.\\
\and
Max-Planck-Institut f\"{u}r Astrophysik, Karl-Schwarzschild-Str. 1, 85741 Garching, Germany\\
\and
Max-Planck-Institut f\"{u}r Extraterrestrische Physik, Giessenbachstra{\ss}e, 85748 Garching, Germany\\
\and
Niels Bohr Institute, Blegdamsvej 17, Copenhagen, Denmark\\
\and
Observational Cosmology, Mail Stop 367-17, California Institute of Technology, Pasadena, CA, 91125, U.S.A.\\
\and
Optical Science Laboratory, University College London, Gower Street, London, U.K.\\
\and
SISSA, Astrophysics Sector, via Bonomea 265, 34136, Trieste, Italy\\
\and
SUPA, Institute for Astronomy, University of Edinburgh, Royal Observatory, Blackford Hill, Edinburgh EH9 3HJ, U.K.\\
\and
School of Physics and Astronomy, Cardiff University, Queens Buildings, The Parade, Cardiff, CF24 3AA, U.K.\\
\and
Space Research Institute (IKI), Russian Academy of Sciences, Profsoyuznaya Str, 84/32, Moscow, 117997, Russia\\
\and
Space Sciences Laboratory, University of California, Berkeley, California, U.S.A.\\
\and
Stanford University, Dept of Physics, Varian Physics Bldg, 382 Via Pueblo Mall, Stanford, California, U.S.A.\\
\and
UPMC Univ Paris 06, UMR7095, 98 bis Boulevard Arago, F-75014, Paris, France\\
\and
Universit\'{e} Denis Diderot (Paris 7), 75205 Paris Cedex 13, France\\
\and
Universit\'{e} de Toulouse, UPS-OMP, IRAP, F-31028 Toulouse cedex 4, France\\
% ...
}